\documentclass[longauth]{aa}

\usepackage{graphicx}
\usepackage{natbib}
\usepackage{scalerel}

\usepackage[table]{xcolor}

\usepackage{txfonts}
\usepackage[pdfencoding=auto,psdextra]{hyperref}
\hypersetup{
    colorlinks=true,
    linkcolor=blue,
    filecolor=magenta,      
    urlcolor=blue,
    citecolor=blue
}
\makeatletter
\renewcommand*\aa@pageof{, page \thepage{} of \pageref*{LastPage}}
\makeatother

\usepackage[utf8]{inputenc}

\usepackage[switch, modulo]{lineno}

\usepackage{euclid}

\begin{document}
%
%

%
%
   \title{\Euclid preparation. Optical emission-line predictions of intermediate-$z$ galaxy populations in GAEA for the Euclid Deep and Wide Surveys}

\newcommand{\orcid}[1]{} 
\author{Euclid Collaboration: L.~Scharr\'e\orcid{0000-0003-2551-4430}\thanks{\email{lucie.scharre@epfl.ch}}\inst{\ref{aff2}}
\and M.~Hirschmann\orcid{0000-0002-3301-3321}\inst{\ref{aff2},\ref{aff3}}
\and G.~De~Lucia\orcid{0000-0002-6220-9104}\inst{\ref{aff3}}
\and S.~Charlot\orcid{0000-0003-3458-2275}\inst{\ref{aff4}}
\and F.~Fontanot\orcid{0000-0003-4744-0188}\inst{\ref{aff3},\ref{aff5}}
\and M.~Spinelli\orcid{0000-0003-0148-3254}\inst{\ref{aff6},\ref{aff3},\ref{aff7}}
\and L.~Xie\inst{\ref{aff8}}
\and A.~Feltre\orcid{0000-0003-0492-4924}\inst{\ref{aff9}}
\and V.~Allevato\orcid{0000-0001-7232-5152}\inst{\ref{aff10}}
\and A.~Plat\inst{\ref{aff2}}
\and M.~N.~Bremer\inst{\ref{aff11}}
\and S.~Fotopoulou\orcid{0000-0002-9686-254X}\inst{\ref{aff11}}
\and L.~Gabarra\orcid{0000-0002-8486-8856}\inst{\ref{aff12}}
\and B.~R.~Granett\orcid{0000-0003-2694-9284}\inst{\ref{aff13}}
\and M.~Moresco\orcid{0000-0002-7616-7136}\inst{\ref{aff14},\ref{aff15}}
\and C.~Scarlata\orcid{0000-0002-9136-8876}\inst{\ref{aff16}}
\and L.~Pozzetti\orcid{0000-0001-7085-0412}\inst{\ref{aff15}}
\and L.~Spinoglio\orcid{0000-0001-8840-1551}\inst{\ref{aff17}}
\and M.~Talia\orcid{0000-0003-4352-2063}\inst{\ref{aff14},\ref{aff15}}
\and G.~Zamorani\orcid{0000-0002-2318-301X}\inst{\ref{aff15}}
\and B.~Altieri\orcid{0000-0003-3936-0284}\inst{\ref{aff18}}
\and A.~Amara\inst{\ref{aff19}}
\and S.~Andreon\orcid{0000-0002-2041-8784}\inst{\ref{aff13}}
\and N.~Auricchio\orcid{0000-0003-4444-8651}\inst{\ref{aff15}}
\and M.~Baldi\orcid{0000-0003-4145-1943}\inst{\ref{aff20},\ref{aff15},\ref{aff21}}
\and S.~Bardelli\orcid{0000-0002-8900-0298}\inst{\ref{aff15}}
\and D.~Bonino\orcid{0000-0002-3336-9977}\inst{\ref{aff22}}
\and E.~Branchini\orcid{0000-0002-0808-6908}\inst{\ref{aff23},\ref{aff24},\ref{aff13}}
\and M.~Brescia\orcid{0000-0001-9506-5680}\inst{\ref{aff25},\ref{aff10},\ref{aff26}}
\and J.~Brinchmann\orcid{0000-0003-4359-8797}\inst{\ref{aff27}}
\and S.~Camera\orcid{0000-0003-3399-3574}\inst{\ref{aff28},\ref{aff29},\ref{aff22}}
\and V.~Capobianco\orcid{0000-0002-3309-7692}\inst{\ref{aff22}}
\and C.~Carbone\orcid{0000-0003-0125-3563}\inst{\ref{aff30}}
\and J.~Carretero\orcid{0000-0002-3130-0204}\inst{\ref{aff31},\ref{aff32}}
\and S.~Casas\orcid{0000-0002-4751-5138}\inst{\ref{aff33}}
\and F.~J.~Castander\orcid{0000-0001-7316-4573}\inst{\ref{aff34},\ref{aff35}}
\and M.~Castellano\orcid{0000-0001-9875-8263}\inst{\ref{aff36}}
\and S.~Cavuoti\orcid{0000-0002-3787-4196}\inst{\ref{aff10},\ref{aff26}}
\and A.~Cimatti\inst{\ref{aff37}}
\and G.~Congedo\orcid{0000-0003-2508-0046}\inst{\ref{aff38}}
\and C.~J.~Conselice\inst{\ref{aff39}}
\and L.~Conversi\orcid{0000-0002-6710-8476}\inst{\ref{aff40},\ref{aff18}}
\and Y.~Copin\orcid{0000-0002-5317-7518}\inst{\ref{aff41}}
\and L.~Corcione\orcid{0000-0002-6497-5881}\inst{\ref{aff22}}
\and F.~Courbin\orcid{0000-0003-0758-6510}\inst{\ref{aff42}}
\and H.~M.~Courtois\orcid{0000-0003-0509-1776}\inst{\ref{aff43}}
\and A.~Da~Silva\orcid{0000-0002-6385-1609}\inst{\ref{aff44},\ref{aff45}}
\and H.~Degaudenzi\orcid{0000-0002-5887-6799}\inst{\ref{aff46}}
\and J.~Dinis\inst{\ref{aff44},\ref{aff45}}
\and M.~Douspis\inst{\ref{aff47}}
\and F.~Dubath\orcid{0000-0002-6533-2810}\inst{\ref{aff46}}
\and X.~Dupac\inst{\ref{aff18}}
\and S.~Dusini\orcid{0000-0002-1128-0664}\inst{\ref{aff48}}
\and M.~Farina\orcid{0000-0002-3089-7846}\inst{\ref{aff17}}
\and S.~Farrens\orcid{0000-0002-9594-9387}\inst{\ref{aff49}}
\and S.~Ferriol\inst{\ref{aff41}}
\and M.~Frailis\orcid{0000-0002-7400-2135}\inst{\ref{aff3}}
\and E.~Franceschi\orcid{0000-0002-0585-6591}\inst{\ref{aff15}}
\and S.~Galeotta\orcid{0000-0002-3748-5115}\inst{\ref{aff3}}
\and B.~Garilli\orcid{0000-0001-7455-8750}\inst{\ref{aff30}}
\and B.~Gillis\orcid{0000-0002-4478-1270}\inst{\ref{aff38}}
\and C.~Giocoli\orcid{0000-0002-9590-7961}\inst{\ref{aff15},\ref{aff50}}
\and A.~Grazian\orcid{0000-0002-5688-0663}\inst{\ref{aff51}}
\and F.~Grupp\inst{\ref{aff52},\ref{aff53}}
\and L.~Guzzo\orcid{0000-0001-8264-5192}\inst{\ref{aff54},\ref{aff13},\ref{aff55}}
\and S.~V.~H.~Haugan\orcid{0000-0001-9648-7260}\inst{\ref{aff56}}
\and W.~Holmes\inst{\ref{aff57}}
\and I.~Hook\orcid{0000-0002-2960-978X}\inst{\ref{aff58}}
\and F.~Hormuth\inst{\ref{aff59}}
\and A.~Hornstrup\orcid{0000-0002-3363-0936}\inst{\ref{aff60},\ref{aff61}}
\and K.~Jahnke\orcid{0000-0003-3804-2137}\inst{\ref{aff62}}
\and E.~Keih\"anen\orcid{0000-0003-1804-7715}\inst{\ref{aff63}}
\and S.~Kermiche\orcid{0000-0002-0302-5735}\inst{\ref{aff64}}
\and A.~Kiessling\orcid{0000-0002-2590-1273}\inst{\ref{aff57}}
\and T.~Kitching\orcid{0000-0002-4061-4598}\inst{\ref{aff65}}
\and B.~Kubik\inst{\ref{aff41}}
\and M.~K\"ummel\orcid{0000-0003-2791-2117}\inst{\ref{aff66}}
\and M.~Kunz\orcid{0000-0002-3052-7394}\inst{\ref{aff67}}
\and H.~Kurki-Suonio\orcid{0000-0002-4618-3063}\inst{\ref{aff68},\ref{aff69}}
\and S.~Ligori\orcid{0000-0003-4172-4606}\inst{\ref{aff22}}
\and P.~B.~Lilje\orcid{0000-0003-4324-7794}\inst{\ref{aff56}}
\and V.~Lindholm\orcid{0000-0003-2317-5471}\inst{\ref{aff68},\ref{aff69}}
\and I.~Lloro\inst{\ref{aff70}}
\and D.~Maino\inst{\ref{aff54},\ref{aff30},\ref{aff55}}
\and E.~Maiorano\orcid{0000-0003-2593-4355}\inst{\ref{aff15}}
\and O.~Mansutti\orcid{0000-0001-5758-4658}\inst{\ref{aff3}}
\and O.~Marggraf\orcid{0000-0001-7242-3852}\inst{\ref{aff71}}
\and K.~Markovic\orcid{0000-0001-6764-073X}\inst{\ref{aff57}}
\and N.~Martinet\orcid{0000-0003-2786-7790}\inst{\ref{aff72}}
\and F.~Marulli\orcid{0000-0002-8850-0303}\inst{\ref{aff14},\ref{aff15},\ref{aff21}}
\and R.~Massey\orcid{0000-0002-6085-3780}\inst{\ref{aff73}}
\and S.~Maurogordato\inst{\ref{aff6}}
\and H.~J.~McCracken\orcid{0000-0002-9489-7765}\inst{\ref{aff74}}
\and E.~Medinaceli\orcid{0000-0002-4040-7783}\inst{\ref{aff15}}
\and S.~Mei\orcid{0000-0002-2849-559X}\inst{\ref{aff75}}
\and Y.~Mellier\inst{\ref{aff76},\ref{aff74}}
\and M.~Meneghetti\orcid{0000-0003-1225-7084}\inst{\ref{aff15},\ref{aff21}}
\and E.~Merlin\orcid{0000-0001-6870-8900}\inst{\ref{aff36}}
\and G.~Meylan\inst{\ref{aff42}}
\and L.~Moscardini\orcid{0000-0002-3473-6716}\inst{\ref{aff14},\ref{aff15},\ref{aff21}}
\and E.~Munari\orcid{0000-0002-1751-5946}\inst{\ref{aff3}}
\and S.-M.~Niemi\inst{\ref{aff77}}
\and C.~Padilla\orcid{0000-0001-7951-0166}\inst{\ref{aff31}}
\and S.~Paltani\orcid{0000-0002-8108-9179}\inst{\ref{aff46}}
\and F.~Pasian\orcid{0000-0002-4869-3227}\inst{\ref{aff3}}
\and K.~Pedersen\inst{\ref{aff78}}
\and V.~Pettorino\inst{\ref{aff79},\ref{aff77}}
\and G.~Polenta\orcid{0000-0003-4067-9196}\inst{\ref{aff80}}
\and M.~Poncet\inst{\ref{aff81}}
\and L.~A.~Popa\inst{\ref{aff82}}
\and F.~Raison\orcid{0000-0002-7819-6918}\inst{\ref{aff52}}
\and A.~Renzi\orcid{0000-0001-9856-1970}\inst{\ref{aff83},\ref{aff48}}
\and J.~Rhodes\inst{\ref{aff57}}
\and G.~Riccio\inst{\ref{aff10}}
\and E.~Romelli\orcid{0000-0003-3069-9222}\inst{\ref{aff3}}
\and M.~Roncarelli\orcid{0000-0001-9587-7822}\inst{\ref{aff15}}
\and E.~Rossetti\inst{\ref{aff20}}
\and R.~Saglia\orcid{0000-0003-0378-7032}\inst{\ref{aff66},\ref{aff52}}
\and D.~Sapone\orcid{0000-0001-7089-4503}\inst{\ref{aff84}}
\and B.~Sartoris\inst{\ref{aff66},\ref{aff3}}
\and M.~Schirmer\orcid{0000-0003-2568-9994}\inst{\ref{aff62}}
\and P.~Schneider\orcid{0000-0001-8561-2679}\inst{\ref{aff71}}
\and A.~Secroun\orcid{0000-0003-0505-3710}\inst{\ref{aff64}}
\and G.~Seidel\orcid{0000-0003-2907-353X}\inst{\ref{aff62}}
\and S.~Serrano\orcid{0000-0002-0211-2861}\inst{\ref{aff35},\ref{aff34},\ref{aff85}}
\and C.~Sirignano\orcid{0000-0002-0995-7146}\inst{\ref{aff83},\ref{aff48}}
\and G.~Sirri\orcid{0000-0003-2626-2853}\inst{\ref{aff21}}
\and L.~Stanco\orcid{0000-0002-9706-5104}\inst{\ref{aff48}}
\and C.~Surace\orcid{0000-0003-2592-0113}\inst{\ref{aff72}}
\and P.~Tallada-Cresp\'{i}\orcid{0000-0002-1336-8328}\inst{\ref{aff86},\ref{aff32}}
\and A.~N.~Taylor\inst{\ref{aff38}}
\and H.~I.~Teplitz\orcid{0000-0002-7064-5424}\inst{\ref{aff87}}
\and I.~Tereno\inst{\ref{aff44},\ref{aff88}}
\and R.~Toledo-Moreo\orcid{0000-0002-2997-4859}\inst{\ref{aff89}}
\and F.~Torradeflot\orcid{0000-0003-1160-1517}\inst{\ref{aff32},\ref{aff86}}
\and I.~Tutusaus\orcid{0000-0002-3199-0399}\inst{\ref{aff90}}
\and L.~Valenziano\orcid{0000-0002-1170-0104}\inst{\ref{aff15},\ref{aff91}}
\and T.~Vassallo\orcid{0000-0001-6512-6358}\inst{\ref{aff66},\ref{aff3}}
\and A.~Veropalumbo\orcid{0000-0003-2387-1194}\inst{\ref{aff13},\ref{aff24}}
\and Y.~Wang\orcid{0000-0002-4749-2984}\inst{\ref{aff87}}
\and J.~Weller\orcid{0000-0002-8282-2010}\inst{\ref{aff66},\ref{aff52}}
\and J.~Zoubian\inst{\ref{aff64}}
\and E.~Zucca\orcid{0000-0002-5845-8132}\inst{\ref{aff15}}
\and A.~Biviano\orcid{0000-0002-0857-0732}\inst{\ref{aff3},\ref{aff5}}
\and M.~Bolzonella\orcid{0000-0003-3278-4607}\inst{\ref{aff15}}
\and E.~Bozzo\orcid{0000-0002-8201-1525}\inst{\ref{aff46}}
\and C.~Burigana\orcid{0000-0002-3005-5796}\inst{\ref{aff92},\ref{aff91}}
\and C.~Colodro-Conde\inst{\ref{aff93}}
\and D.~Di~Ferdinando\inst{\ref{aff21}}
\and R.~Farinelli\inst{\ref{aff15}}
\and J.~Graci\'{a}-Carpio\inst{\ref{aff52}}
\and G.~Mainetti\inst{\ref{aff94}}
\and M.~Martinelli\orcid{0000-0002-6943-7732}\inst{\ref{aff36},\ref{aff95}}
\and N.~Mauri\orcid{0000-0001-8196-1548}\inst{\ref{aff37},\ref{aff21}}
\and C.~Neissner\orcid{0000-0001-8524-4968}\inst{\ref{aff31},\ref{aff32}}
\and A.~A.~Nucita\inst{\ref{aff96},\ref{aff97},\ref{aff98}}
\and Z.~Sakr\orcid{0000-0002-4823-3757}\inst{\ref{aff99},\ref{aff90},\ref{aff100}}
\and V.~Scottez\inst{\ref{aff76},\ref{aff101}}
\and M.~Tenti\orcid{0000-0002-4254-5901}\inst{\ref{aff21}}
\and M.~Viel\orcid{0000-0002-2642-5707}\inst{\ref{aff5},\ref{aff3},\ref{aff102},\ref{aff103},\ref{aff104}}
\and M.~Wiesmann\orcid{0009-0000-8199-5860}\inst{\ref{aff56}}
\and Y.~Akrami\orcid{0000-0002-2407-7956}\inst{\ref{aff105},\ref{aff106}}
\and S.~Anselmi\orcid{0000-0002-3579-9583}\inst{\ref{aff48},\ref{aff83},\ref{aff107}}
\and C.~Baccigalupi\orcid{0000-0002-8211-1630}\inst{\ref{aff102},\ref{aff3},\ref{aff103},\ref{aff5}}
\and M.~Ballardini\orcid{0000-0003-4481-3559}\inst{\ref{aff108},\ref{aff109},\ref{aff15}}
\and M.~Bethermin\orcid{0000-0002-3915-2015}\inst{\ref{aff110},\ref{aff72}}
\and A.~Blanchard\orcid{0000-0001-8555-9003}\inst{\ref{aff90}}
\and S.~Borgani\orcid{0000-0001-6151-6439}\inst{\ref{aff111},\ref{aff5},\ref{aff3},\ref{aff103}}
\and A.~S.~Borlaff\orcid{0000-0003-3249-4431}\inst{\ref{aff112},\ref{aff113}}
\and S.~Bruton\orcid{0000-0002-6503-5218}\inst{\ref{aff16}}
\and R.~Cabanac\orcid{0000-0001-6679-2600}\inst{\ref{aff90}}
\and A.~Calabro\orcid{0000-0003-2536-1614}\inst{\ref{aff36}}
\and G.~Ca\~{n}as-Herrera\orcid{0000-0003-2796-2149}\inst{\ref{aff77},\ref{aff114}}
\and A.~Cappi\inst{\ref{aff15},\ref{aff6}}
\and C.~S.~Carvalho\inst{\ref{aff88}}
\and G.~Castignani\orcid{0000-0001-6831-0687}\inst{\ref{aff14},\ref{aff15}}
\and T.~Castro\orcid{0000-0002-6292-3228}\inst{\ref{aff3},\ref{aff103},\ref{aff5},\ref{aff104}}
\and K.~C.~Chambers\orcid{0000-0001-6965-7789}\inst{\ref{aff115}}
\and S.~Contarini\orcid{0000-0002-9843-723X}\inst{\ref{aff52},\ref{aff14}}
\and T.~Contini\orcid{0000-0003-0275-938X}\inst{\ref{aff90}}
\and A.~R.~Cooray\orcid{0000-0002-3892-0190}\inst{\ref{aff116}}
\and J.~Coupon\inst{\ref{aff46}}
\and O.~Cucciati\orcid{0000-0002-9336-7551}\inst{\ref{aff15}}
\and G.~Desprez\inst{\ref{aff117}}
\and S.~Di~Domizio\orcid{0000-0003-2863-5895}\inst{\ref{aff23},\ref{aff24}}
\and H.~Dole\orcid{0000-0002-9767-3839}\inst{\ref{aff47}}
\and A.~D\'{i}az-S\'{a}nchez\orcid{0000-0003-0748-4768}\inst{\ref{aff118}}
\and J.~A.~Escartin~Vigo\inst{\ref{aff52}}
\and S.~Escoffier\orcid{0000-0002-2847-7498}\inst{\ref{aff64}}
\and I.~Ferrero\orcid{0000-0002-1295-1132}\inst{\ref{aff56}}
\and K.~Ganga\orcid{0000-0001-8159-8208}\inst{\ref{aff75}}
\and J.~Garc\'ia-Bellido\orcid{0000-0002-9370-8360}\inst{\ref{aff105}}
\and E.~Gaztanaga\orcid{0000-0001-9632-0815}\inst{\ref{aff34},\ref{aff35},\ref{aff119}}
\and K.~George\orcid{0000-0002-1734-8455}\inst{\ref{aff66}}
\and F.~Giacomini\orcid{0000-0002-3129-2814}\inst{\ref{aff21}}
\and G.~Gozaliasl\orcid{0000-0002-0236-919X}\inst{\ref{aff120},\ref{aff68}}
\and A.~Gregorio\orcid{0000-0003-4028-8785}\inst{\ref{aff111},\ref{aff3},\ref{aff103}}
\and A.~Hall\orcid{0000-0002-3139-8651}\inst{\ref{aff38}}
\and H.~Hildebrandt\orcid{0000-0002-9814-3338}\inst{\ref{aff121}}
\and J.~J.~E.~Kajava\orcid{0000-0002-3010-8333}\inst{\ref{aff122},\ref{aff123}}
\and V.~Kansal\orcid{0000-0002-4008-6078}\inst{\ref{aff124},\ref{aff125},\ref{aff126}}
\and C.~C.~Kirkpatrick\inst{\ref{aff63}}
\and L.~Legrand\orcid{0000-0003-0610-5252}\inst{\ref{aff127}}
\and A.~Loureiro\orcid{0000-0002-4371-0876}\inst{\ref{aff128},\ref{aff129}}
\and J.~Macias-Perez\orcid{0000-0002-5385-2763}\inst{\ref{aff130}}
\and M.~Magliocchetti\orcid{0000-0001-9158-4838}\inst{\ref{aff17}}
\and C.~Mancini\orcid{0000-0002-4297-0561}\inst{\ref{aff30}}
\and F.~Mannucci\orcid{0000-0002-4803-2381}\inst{\ref{aff9}}
\and R.~Maoli\orcid{0000-0002-6065-3025}\inst{\ref{aff131},\ref{aff36}}
\and C.~J.~A.~P.~Martins\orcid{0000-0002-4886-9261}\inst{\ref{aff132},\ref{aff27}}
\and S.~Matthew\inst{\ref{aff38}}
\and L.~Maurin\orcid{0000-0002-8406-0857}\inst{\ref{aff47}}
\and R.~B.~Metcalf\orcid{0000-0003-3167-2574}\inst{\ref{aff14},\ref{aff15}}
\and M.~Migliaccio\inst{\ref{aff133},\ref{aff134}}
\and P.~Monaco\orcid{0000-0003-2083-7564}\inst{\ref{aff111},\ref{aff3},\ref{aff103},\ref{aff5}}
\and G.~Morgante\inst{\ref{aff15}}
\and Nicholas~A.~Walton\orcid{0000-0003-3983-8778}\inst{\ref{aff135}}
\and M.~P{\"o}ntinen\orcid{0000-0001-5442-2530}\inst{\ref{aff68}}
\and V.~Popa\inst{\ref{aff82}}
\and C.~Porciani\orcid{0000-0002-7797-2508}\inst{\ref{aff71}}
\and D.~Potter\orcid{0000-0002-0757-5195}\inst{\ref{aff136}}
\and I.~Risso\orcid{0000-0003-2525-7761}\inst{\ref{aff137}}
\and P.-F.~Rocci\inst{\ref{aff47}}
\and M.~Sahl\'en\orcid{0000-0003-0973-4804}\inst{\ref{aff138}}
\and A.~G.~S\'anchez\orcid{0000-0003-1198-831X}\inst{\ref{aff52}}
\and A.~Schneider\orcid{0000-0001-7055-8104}\inst{\ref{aff136}}
\and M.~Schultheis\inst{\ref{aff6}}
\and M.~Sereno\orcid{0000-0003-0302-0325}\inst{\ref{aff15},\ref{aff21}}
\and P.~Simon\inst{\ref{aff71}}
\and J.~Steinwagner\inst{\ref{aff52}}
\and G.~Testera\inst{\ref{aff24}}
\and M.~Tewes\orcid{0000-0002-1155-8689}\inst{\ref{aff71}}
\and R.~Teyssier\orcid{0000-0001-7689-0933}\inst{\ref{aff139}}
\and S.~Toft\orcid{0000-0003-3631-7176}\inst{\ref{aff61},\ref{aff140},\ref{aff141}}
\and S.~Tosi\orcid{0000-0002-7275-9193}\inst{\ref{aff23},\ref{aff24},\ref{aff13}}
\and A.~Troja\orcid{0000-0003-0239-4595}\inst{\ref{aff83},\ref{aff48}}
\and M.~Tucci\inst{\ref{aff46}}
\and J.~Valiviita\orcid{0000-0001-6225-3693}\inst{\ref{aff68},\ref{aff69}}
\and D.~Vergani\orcid{0000-0003-0898-2216}\inst{\ref{aff15}}
\and G.~Verza\orcid{0000-0002-1886-8348}\inst{\ref{aff142},\ref{aff143}}
\and I.~A.~Zinchenko\inst{\ref{aff66}}}
										   
\institute{
Institute of Physics, Laboratory for Galaxy Evolution, Ecole Polytechnique F\'ed\'erale de Lausanne, Observatoire de Sauverny, CH-1290 Versoix, Switzerland\label{aff2}
\and
INAF-Osservatorio Astronomico di Trieste, Via G. B. Tiepolo 11, 34143 Trieste, Italy\label{aff3}
\and
Sorbonne Universit{\'e}s, UPMC Univ Paris 6 et CNRS, UMR 7095, Institut d'Astrophysique de Paris, 98 bis bd Arago, 75014 Paris, France\label{aff4}
\and
IFPU, Institute for Fundamental Physics of the Universe, via Beirut 2, 34151 Trieste, Italy\label{aff5}
\and
Universit\'e C\^{o}te d'Azur, Observatoire de la C\^{o}te d'Azur, CNRS, Laboratoire Lagrange, Bd de l'Observatoire, CS 34229, 06304 Nice cedex 4, France\label{aff6}
\and
Department of Physics and Astronomy, University of the Western Cape, Bellville, Cape Town, 7535, South Africa\label{aff7}
\and
Tianjin Normal University, Binshuixidao 393, Tianjin 300387, China\label{aff8}
\and
INAF-Osservatorio Astrofisico di Arcetri, Largo E. Fermi 5, 50125, Firenze, Italy\label{aff9}
\and
INAF-Osservatorio Astronomico di Capodimonte, Via Moiariello 16, 80131 Napoli, Italy\label{aff10}
\and
School of Physics, HH Wills Physics Laboratory, University of Bristol, Tyndall Avenue, Bristol, BS8 1TL, UK\label{aff11}
\and
Department of Physics, Oxford University, Keble Road, Oxford OX1 3RH, UK\label{aff12}
\and
INAF-Osservatorio Astronomico di Brera, Via Brera 28, 20122 Milano, Italy\label{aff13}
\and
Dipartimento di Fisica e Astronomia "Augusto Righi" - Alma Mater Studiorum Universit\`a di Bologna, via Piero Gobetti 93/2, 40129 Bologna, Italy\label{aff14}
\and
INAF-Osservatorio di Astrofisica e Scienza dello Spazio di Bologna, Via Piero Gobetti 93/3, 40129 Bologna, Italy\label{aff15}
\and
Minnesota Institute for Astrophysics, University of Minnesota, 116 Church St SE, Minneapolis, MN 55455, USA\label{aff16}
\and
INAF-Istituto di Astrofisica e Planetologia Spaziali, via del Fosso del Cavaliere, 100, 00100 Roma, Italy\label{aff17}
\and
ESAC/ESA, Camino Bajo del Castillo, s/n., Urb. Villafranca del Castillo, 28692 Villanueva de la Ca\~nada, Madrid, Spain\label{aff18}
\and
School of Mathematics and Physics, University of Surrey, Guildford, Surrey, GU2 7XH, UK\label{aff19}
\and
Dipartimento di Fisica e Astronomia, Universit\`a di Bologna, Via Gobetti 93/2, 40129 Bologna, Italy\label{aff20}
\and
INFN-Sezione di Bologna, Viale Berti Pichat 6/2, 40127 Bologna, Italy\label{aff21}
\and
INAF-Osservatorio Astrofisico di Torino, Via Osservatorio 20, 10025 Pino Torinese (TO), Italy\label{aff22}
\and
Dipartimento di Fisica, Universit\`a di Genova, Via Dodecaneso 33, 16146, Genova, Italy\label{aff23}
\and
INFN-Sezione di Genova, Via Dodecaneso 33, 16146, Genova, Italy\label{aff24}
\and
Department of Physics "E. Pancini", University Federico II, Via Cinthia 6, 80126, Napoli, Italy\label{aff25}
\and
INFN section of Naples, Via Cinthia 6, 80126, Napoli, Italy\label{aff26}
\and
Instituto de Astrof\'isica e Ci\^encias do Espa\c{c}o, Universidade do Porto, CAUP, Rua das Estrelas, PT4150-762 Porto, Portugal\label{aff27}
\and
Dipartimento di Fisica, Universit\`a degli Studi di Torino, Via P. Giuria 1, 10125 Torino, Italy\label{aff28}
\and
INFN-Sezione di Torino, Via P. Giuria 1, 10125 Torino, Italy\label{aff29}
\and
INAF-IASF Milano, Via Alfonso Corti 12, 20133 Milano, Italy\label{aff30}
\and
Institut de F\'{i}sica d'Altes Energies (IFAE), The Barcelona Institute of Science and Technology, Campus UAB, 08193 Bellaterra (Barcelona), Spain\label{aff31}
\and
Port d'Informaci\'{o} Cient\'{i}fica, Campus UAB, C. Albareda s/n, 08193 Bellaterra (Barcelona), Spain\label{aff32}
\and
Institute for Theoretical Particle Physics and Cosmology (TTK), RWTH Aachen University, 52056 Aachen, Germany\label{aff33}
\and
Institute of Space Sciences (ICE, CSIC), Campus UAB, Carrer de Can Magrans, s/n, 08193 Barcelona, Spain\label{aff34}
\and
Institut d'Estudis Espacials de Catalunya (IEEC), Carrer Gran Capit\'a 2-4, 08034 Barcelona, Spain\label{aff35}
\and
INAF-Osservatorio Astronomico di Roma, Via Frascati 33, 00078 Monteporzio Catone, Italy\label{aff36}
\and
Dipartimento di Fisica e Astronomia "Augusto Righi" - Alma Mater Studiorum Universit\`a di Bologna, Viale Berti Pichat 6/2, 40127 Bologna, Italy\label{aff37}
\and
Institute for Astronomy, University of Edinburgh, Royal Observatory, Blackford Hill, Edinburgh EH9 3HJ, UK\label{aff38}
\and
Jodrell Bank Centre for Astrophysics, Department of Physics and Astronomy, University of Manchester, Oxford Road, Manchester M13 9PL, UK\label{aff39}
\and
European Space Agency/ESRIN, Largo Galileo Galilei 1, 00044 Frascati, Roma, Italy\label{aff40}
\and
University of Lyon, Univ Claude Bernard Lyon 1, CNRS/IN2P3, IP2I Lyon, UMR 5822, 69622 Villeurbanne, France\label{aff41}
\and
Institute of Physics, Laboratory of Astrophysics, Ecole Polytechnique F\'ed\'erale de Lausanne (EPFL), Observatoire de Sauverny, 1290 Versoix, Switzerland\label{aff42}
\and
UCB Lyon 1, CNRS/IN2P3, IUF, IP2I Lyon, 4 rue Enrico Fermi, 69622 Villeurbanne, France\label{aff43}
\and
Departamento de F\'isica, Faculdade de Ci\^encias, Universidade de Lisboa, Edif\'icio C8, Campo Grande, PT1749-016 Lisboa, Portugal\label{aff44}
\and
Instituto de Astrof\'isica e Ci\^encias do Espa\c{c}o, Faculdade de Ci\^encias, Universidade de Lisboa, Campo Grande, 1749-016 Lisboa, Portugal\label{aff45}
\and
Department of Astronomy, University of Geneva, ch. d'Ecogia 16, 1290 Versoix, Switzerland\label{aff46}
\and
Universit\'e Paris-Saclay, CNRS, Institut d'astrophysique spatiale, 91405, Orsay, France\label{aff47}
\and
INFN-Padova, Via Marzolo 8, 35131 Padova, Italy\label{aff48}
\and
Universit\'e Paris-Saclay, Universit\'e Paris Cit\'e, CEA, CNRS, AIM, 91191, Gif-sur-Yvette, France\label{aff49}
\and
Istituto Nazionale di Fisica Nucleare, Sezione di Bologna, Via Irnerio 46, 40126 Bologna, Italy\label{aff50}
\and
INAF-Osservatorio Astronomico di Padova, Via dell'Osservatorio 5, 35122 Padova, Italy\label{aff51}
\and
Max Planck Institute for Extraterrestrial Physics, Giessenbachstr. 1, 85748 Garching, Germany\label{aff52}
\and
University Observatory, Faculty of Physics, Ludwig-Maximilians-Universit{\"a}t, Scheinerstr. 1, 81679 Munich, Germany\label{aff53}
\and
Dipartimento di Fisica "Aldo Pontremoli", Universit\`a degli Studi di Milano, Via Celoria 16, 20133 Milano, Italy\label{aff54}
\and
INFN-Sezione di Milano, Via Celoria 16, 20133 Milano, Italy\label{aff55}
\and
Institute of Theoretical Astrophysics, University of Oslo, P.O. Box 1029 Blindern, 0315 Oslo, Norway\label{aff56}
\and
Jet Propulsion Laboratory, California Institute of Technology, 4800 Oak Grove Drive, Pasadena, CA, 91109, USA\label{aff57}
\and
Department of Physics, Lancaster University, Lancaster, LA1 4YB, UK\label{aff58}
\and
von Hoerner \& Sulger GmbH, Schlo{\ss}Platz 8, 68723 Schwetzingen, Germany\label{aff59}
\and
Technical University of Denmark, Elektrovej 327, 2800 Kgs. Lyngby, Denmark\label{aff60}
\and
Cosmic Dawn Center (DAWN), Denmark\label{aff61}
\and
Max-Planck-Institut f\"ur Astronomie, K\"onigstuhl 17, 69117 Heidelberg, Germany\label{aff62}
\and
Department of Physics and Helsinki Institute of Physics, Gustaf H\"allstr\"omin katu 2, 00014 University of Helsinki, Finland\label{aff63}
\and
Aix-Marseille Universit\'e, CNRS/IN2P3, CPPM, Marseille, France\label{aff64}
\and
Mullard Space Science Laboratory, University College London, Holmbury St Mary, Dorking, Surrey RH5 6NT, UK\label{aff65}
\and
Universit\"ats-Sternwarte M\"unchen, Fakult\"at f\"ur Physik, Ludwig-Maximilians-Universit\"at M\"unchen, Scheinerstrasse 1, 81679 M\"unchen, Germany\label{aff66}
\and
Universit\'e de Gen\`eve, D\'epartement de Physique Th\'eorique and Centre for Astroparticle Physics, 24 quai Ernest-Ansermet, CH-1211 Gen\`eve 4, Switzerland\label{aff67}
\and
Department of Physics, P.O. Box 64, 00014 University of Helsinki, Finland\label{aff68}
\and
Helsinki Institute of Physics, Gustaf H{\"a}llstr{\"o}min katu 2, University of Helsinki, Helsinki, Finland\label{aff69}
\and
NOVA optical infrared instrumentation group at ASTRON, Oude Hoogeveensedijk 4, 7991PD, Dwingeloo, The Netherlands\label{aff70}
\and
Universit\"at Bonn, Argelander-Institut f\"ur Astronomie, Auf dem H\"ugel 71, 53121 Bonn, Germany\label{aff71}
\and
Aix-Marseille Universit\'e, CNRS, CNES, LAM, Marseille, France\label{aff72}
\and
Department of Physics, Institute for Computational Cosmology, Durham University, South Road, DH1 3LE, UK\label{aff73}
\and
Institut d'Astrophysique de Paris, UMR 7095, CNRS, and Sorbonne Universit\'e, 98 bis boulevard Arago, 75014 Paris, France\label{aff74}
\and
Universit\'e Paris Cit\'e, CNRS, Astroparticule et Cosmologie, 75013 Paris, France\label{aff75}
\and
Institut d'Astrophysique de Paris, 98bis Boulevard Arago, 75014, Paris, France\label{aff76}
\and
European Space Agency/ESTEC, Keplerlaan 1, 2201 AZ Noordwijk, The Netherlands\label{aff77}
\and
Department of Physics and Astronomy, University of Aarhus, Ny Munkegade 120, DK-8000 Aarhus C, Denmark\label{aff78}
\and
Universit\'e Paris-Saclay, Universit\'e Paris Cit\'e, CEA, CNRS, Astrophysique, Instrumentation et Mod\'elisation Paris-Saclay, 91191 Gif-sur-Yvette, France\label{aff79}
\and
Space Science Data Center, Italian Space Agency, via del Politecnico snc, 00133 Roma, Italy\label{aff80}
\and
Centre National d'Etudes Spatiales -- Centre spatial de Toulouse, 18 avenue Edouard Belin, 31401 Toulouse Cedex 9, France\label{aff81}
\and
Institute of Space Science, Str. Atomistilor, nr. 409 M\u{a}gurele, Ilfov, 077125, Romania\label{aff82}
\and
Dipartimento di Fisica e Astronomia "G. Galilei", Universit\`a di Padova, Via Marzolo 8, 35131 Padova, Italy\label{aff83}
\and
Departamento de F\'isica, FCFM, Universidad de Chile, Blanco Encalada 2008, Santiago, Chile\label{aff84}
\and
Satlantis, University Science Park, Sede Bld 48940, Leioa-Bilbao, Spain\label{aff85}
\and
Centro de Investigaciones Energ\'eticas, Medioambientales y Tecnol\'ogicas (CIEMAT), Avenida Complutense 40, 28040 Madrid, Spain\label{aff86}
\and
Infrared Processing and Analysis Center, California Institute of Technology, Pasadena, CA 91125, USA\label{aff87}
\and
Instituto de Astrof\'isica e Ci\^encias do Espa\c{c}o, Faculdade de Ci\^encias, Universidade de Lisboa, Tapada da Ajuda, 1349-018 Lisboa, Portugal\label{aff88}
\and
Universidad Polit\'ecnica de Cartagena, Departamento de Electr\'onica y Tecnolog\'ia de Computadoras,  Plaza del Hospital 1, 30202 Cartagena, Spain\label{aff89}
\and
Institut de Recherche en Astrophysique et Plan\'etologie (IRAP), Universit\'e de Toulouse, CNRS, UPS, CNES, 14 Av. Edouard Belin, 31400 Toulouse, France\label{aff90}
\and
INFN-Bologna, Via Irnerio 46, 40126 Bologna, Italy\label{aff91}
\and
INAF, Istituto di Radioastronomia, Via Piero Gobetti 101, 40129 Bologna, Italy\label{aff92}
\and
Instituto de Astrof\'isica de Canarias, Calle V\'ia L\'actea s/n, 38204, San Crist\'obal de La Laguna, Tenerife, Spain\label{aff93}
\and
Centre de Calcul de l'IN2P3/CNRS, 21 avenue Pierre de Coubertin 69627 Villeurbanne Cedex, France\label{aff94}
\and
INFN-Sezione di Roma, Piazzale Aldo Moro, 2 - c/o Dipartimento di Fisica, Edificio G. Marconi, 00185 Roma, Italy\label{aff95}
\and
Department of Mathematics and Physics E. De Giorgi, University of Salento, Via per Arnesano, CP-I93, 73100, Lecce, Italy\label{aff96}
\and
INAF-Sezione di Lecce, c/o Dipartimento Matematica e Fisica, Via per Arnesano, 73100, Lecce, Italy\label{aff97}
\and
INFN, Sezione di Lecce, Via per Arnesano, CP-193, 73100, Lecce, Italy\label{aff98}
\and
Institut f\"ur Theoretische Physik, University of Heidelberg, Philosophenweg 16, 69120 Heidelberg, Germany\label{aff99}
\and
Universit\'e St Joseph; Faculty of Sciences, Beirut, Lebanon\label{aff100}
\and
Junia, EPA department, 41 Bd Vauban, 59800 Lille, France\label{aff101}
\and
SISSA, International School for Advanced Studies, Via Bonomea 265, 34136 Trieste TS, Italy\label{aff102}
\and
INFN, Sezione di Trieste, Via Valerio 2, 34127 Trieste TS, Italy\label{aff103}
\and
ICSC - Centro Nazionale di Ricerca in High Performance Computing, Big Data e Quantum Computing, Via Magnanelli 2, Bologna, Italy\label{aff104}
\and
Instituto de F\'isica Te\'orica UAM-CSIC, Campus de Cantoblanco, 28049 Madrid, Spain\label{aff105}
\and
CERCA/ISO, Department of Physics, Case Western Reserve University, 10900 Euclid Avenue, Cleveland, OH 44106, USA\label{aff106}
\and
Laboratoire Univers et Th\'eorie, Observatoire de Paris, Universit\'e PSL, Universit\'e Paris Cit\'e, CNRS, 92190 Meudon, France\label{aff107}
\and
Dipartimento di Fisica e Scienze della Terra, Universit\`a degli Studi di Ferrara, Via Giuseppe Saragat 1, 44122 Ferrara, Italy\label{aff108}
\and
Istituto Nazionale di Fisica Nucleare, Sezione di Ferrara, Via Giuseppe Saragat 1, 44122 Ferrara, Italy\label{aff109}
\and
Universit\'e de Strasbourg, CNRS, Observatoire astronomique de Strasbourg, UMR 7550, 67000 Strasbourg, France\label{aff110}
\and
Dipartimento di Fisica - Sezione di Astronomia, Universit\`a di Trieste, Via Tiepolo 11, 34131 Trieste, Italy\label{aff111}
\and
NASA Ames Research Center, Moffett Field, CA 94035, USA\label{aff112}
\and
Bay Area Environmental Research Institute, Moffett Field, California 94035, USA\label{aff113}
\and
Institute Lorentz, Leiden University, PO Box 9506, Leiden 2300 RA, The Netherlands\label{aff114}
\and
Institute for Astronomy, University of Hawaii, 2680 Woodlawn Drive, Honolulu, HI 96822, USA\label{aff115}
\and
Department of Physics \& Astronomy, University of California Irvine, Irvine CA 92697, USA\label{aff116}
\and
Department of Astronomy \& Physics and Institute for Computational Astrophysics, Saint Mary's University, 923 Robie Street, Halifax, Nova Scotia, B3H 3C3, Canada\label{aff117}
\and
Departamento F\'isica Aplicada, Universidad Polit\'ecnica de Cartagena, Campus Muralla del Mar, 30202 Cartagena, Murcia, Spain\label{aff118}
\and
Institute of Cosmology and Gravitation, University of Portsmouth, Portsmouth PO1 3FX, UK\label{aff119}
\and
Department of Computer Science, Aalto University, PO Box 15400, Espoo, FI-00 076, Finland\label{aff120}
\and
Ruhr University Bochum, Faculty of Physics and Astronomy, Astronomical Institute (AIRUB), German Centre for Cosmological Lensing (GCCL), 44780 Bochum, Germany\label{aff121}
\and
Department of Physics and Astronomy, Vesilinnantie 5, 20014 University of Turku, Finland\label{aff122}
\and
Serco for European Space Agency (ESA), Camino bajo del Castillo, s/n, Urbanizacion Villafranca del Castillo, Villanueva de la Ca\~nada, 28692 Madrid, Spain\label{aff123}
\and
ARC Centre of Excellence for Dark Matter Particle Physics, Melbourne, Australia\label{aff124}
\and
Centre for Astrophysics \& Supercomputing, Swinburne University of Technology, Victoria 3122, Australia\label{aff125}
\and
W.M. Keck Observatory, 65-1120 Mamalahoa Hwy, Kamuela, HI, USA\label{aff126}
\and
ICTP South American Institute for Fundamental Research, Instituto de F\'{\i}sica Te\'orica, Universidade Estadual Paulista, S\~ao Paulo, Brazil\label{aff127}
\and
Oskar Klein Centre for Cosmoparticle Physics, Department of Physics, Stockholm University, Stockholm, SE-106 91, Sweden\label{aff128}
\and
Astrophysics Group, Blackett Laboratory, Imperial College London, London SW7 2AZ, UK\label{aff129}
\and
Univ. Grenoble Alpes, CNRS, Grenoble INP, LPSC-IN2P3, 53, Avenue des Martyrs, 38000, Grenoble, France\label{aff130}
\and
Dipartimento di Fisica, Sapienza Universit\`a di Roma, Piazzale Aldo Moro 2, 00185 Roma, Italy\label{aff131}
\and
Centro de Astrof\'{\i}sica da Universidade do Porto, Rua das Estrelas, 4150-762 Porto, Portugal\label{aff132}
\and
Dipartimento di Fisica, Universit\`a di Roma Tor Vergata, Via della Ricerca Scientifica 1, Roma, Italy\label{aff133}
\and
INFN, Sezione di Roma 2, Via della Ricerca Scientifica 1, Roma, Italy\label{aff134}
\and
Institute of Astronomy, University of Cambridge, Madingley Road, Cambridge CB3 0HA, UK\label{aff135}
\and
Institute for Computational Science, University of Zurich, Winterthurerstrasse 190, 8057 Zurich, Switzerland\label{aff136}
\and
Dipartimento di Fisica, Universit\`a degli studi di Genova, and INFN-Sezione di Genova, via Dodecaneso 33, 16146, Genova, Italy\label{aff137}
\and
Theoretical astrophysics, Department of Physics and Astronomy, Uppsala University, Box 515, 751 20 Uppsala, Sweden\label{aff138}
\and
Department of Astrophysical Sciences, Peyton Hall, Princeton University, Princeton, NJ 08544, USA\label{aff139}
\and
Niels Bohr Institute, University of Copenhagen, Jagtvej 128, 2200 Copenhagen, Denmark\label{aff140}
\and
Cosmic Dawn Center (DAWN)\label{aff141}
\and
Center for Cosmology and Particle Physics, Department of Physics, New York University, New York, NY 10003, USA\label{aff142}
\and
Center for Computational Astrophysics, Flatiron Institute, 162 5th Avenue, 10010, New York, NY, USA\label{aff143}}    
 \date{\today}

%
%
   \abstract{In anticipation of the upcoming Euclid Wide and Deep Surveys, we present optical emission-line predictions at intermediate redshifts from 0.4 to 2.5. Our approach combines a mock light cone from the \textsc{Gaea} semi-analytic model with advanced photoionisation models to construct emission-line catalogues. This allows us to self-consistently model nebular emission from \ion{H}{ii} regions around young stars, and, for the first time with a semi-analytic model, narrow-line regions of active galactic nuclei (AGN) and evolved stellar populations. \textsc{Gaea}, with a box size of 500\,$h^{-1} \,\mathrm{Mpc}$, marks the largest volume this set of models has been applied to. We validate our methodology against observational and theoretical data at low redshift. 
   Our analysis focuses on seven optical emission lines: H$\alpha$, H$\beta$, [\ion{S}{ii}]$\lambda\lambda 6717, 6731$, [\ion{N}{ii}]$\lambda 6584$,  [\ion{O}{i}]$\lambda 6300$, [\ion{O}{iii}]$\lambda 5007$, and [\ion{O}{ii}]$\lambda\lambda 3727, 3729$. In assessing \Euclid's selection bias, we find that it will predominantly observe line-emitting galaxies, which are massive (stellar mass $\gtrsim 10^{9}\si{\solarmass}$), star-forming (specific star-formation rate $> 10^{-10}\mathrm{yr^{-1}}$), and metal-rich (oxygen-to-hydrogen abundance $\mathrm{\logten(O/H)+12} > 8$). We provide \Euclid-observable percentages of emission-line populations in our underlying \textsc{Gaea} sample with a mass resolution limit of $10^{9}\si{\solarmass}$ and an $H$-band magnitude cut of 25. 
   We compare results with and without an estimate of interstellar dust attenuation, which we model using a Calzetti law with a mass-dependent scaling. According to this estimate, the presence of dust may decrease observable percentages by a further 20-30\% with respect to the overall population, which presents challenges for detecting intrinsically fainter lines. We predict \Euclid to observe around 30--70\% of H$\alpha$-, [\ion{N}{ii}]-, [\ion{S}{ii}]-, and [\ion{O}{iii}]-emitting galaxies at redshift below 1. At higher redshift, these percentages decrease below 10\%. H$\beta$, [\ion{O}{ii}], and [\ion{O}{i}] emission are expected to appear relatively faint, thus limiting observability to at most 5\% at the lower end of their detectable redshift range, and below 1\% at the higher end. This is the case both for these lines individually and in combination with other lines. 
   For galaxies with line emission above the flux threshold in the Euclid Deep Survey, we find that BPT diagrams can effectively distinguish between different galaxy types up to around redshift 1.8, attributed to the bias toward metal-rich systems. Moreover, we show that the relationships of H$\alpha$ and [O{\sc iii}]+H$\beta$ to the star-formation rate, as well as the [\ion{O}{iii}]-AGN luminosity relation, exhibit minimal, if any, changes with increasing redshift when compared to local calibrations. Based on the line ratios $\rm [\ion{N}{ii}]/H\alpha$, $[\ion{N}{ii}]/[\ion{O}{ii}]$, and $[\ion{N}{ii}]/[\ion{S}{ii}]$, we further propose novel redshift-invariant tracers for the black hole accretion rate-to-star formation rate ratio. Lastly, we find that commonly used metallicity estimators display gradual shifts in normalisations with increasing redshift, while maintaining the overall shape of local calibrations. This is in tentative agreement with recent JWST data. 
}
%
%
\keywords{surveys -- Galaxies: abundances -- Galaxies: evolution -- Galaxies: general -- Methods: numerical -- Techniques: spectroscopic
}
%
%
   \titlerunning{\Euclid\/: Emission-line forecasts}
   \authorrunning{L. Scharr\'e et al.}
   
   \maketitle
%
%
%
%
   
\section{Introduction}
\label{sec:intro}
During its six-year mission to constrain the dark Universe, \Euclid \citep{Laureijs11,Racca2016TheDesign} will catalogue billions of galaxies and collect an unprecedented abundance of highly-accurate photometric and spectroscopic data. Using weak lensing and galaxy clustering as cosmological probes, it will study the growth of cosmic structures and the Universe's accelerated expansion over the past 10 billion years. For the purpose of recovering accurate distance measurements, \Euclid's near-infrared spectrometer and photometer \citep[NISP,][]{Maciaszek22,Schirmer-EP18} was designed to probe redshifted optical emission from galaxies out to redshift 2 by observing in the near-infrared range. Crucially, the regime around redshift 2 represents the peak of star formation \citep{Madau2014CosmicHistory}, black hole growth, and quasar activity \citep{Richards2006TheThree}, making the resulting data set ideal for studying galaxy formation and evolution. 

In the Euclid Wide Survey \citep[EWS,][]{Scaramella-EP1}, \Euclid is set to observe roughly $15\,000\, \mathrm{deg}^2$ of extragalactic sky. The NISP spectrometer has been tuned to measure $\mathrm{H\alpha}$ line emission at redshift 0.9--1.8 and is expected to recover spectra for about 35 million galaxies. It contains three `red' grisms (RGS, resolving power $\mathcal{R}$ > 380) oriented at different angles, each covering rest-frame 1.21--1.89\,\si{\micron} to a flux limit of $2 \times 10^{-16}\, \mathrm{erg} \, \mathrm{s}^{-1}\, \mathrm{cm}^{-2}$. \Euclid's initial specification forecasts a number density of $1700\, \mathrm{deg}^{-2}$ H$\alpha$ emitters; however, this estimate strongly depends on the uncertain intrinsic H$\alpha$ luminosity function in this redshift range \citep[see][]{Pozzetti2016ModellingMissions,Bagley2020HSTSurveys}.

In addition to the EWS, \Euclid will cover selected fields of 50\,$\mathrm{deg}^2$ at two magnitudes deeper in the Deep Survey (EDS; Scaramella et al., in prep.). In this mode, the NISP spectrometer is able to measure emission lines at fluxes greater than $6 \times 10^{-17}\,\mathrm{erg} \, \mathrm{s}^{-1}\, \mathrm{cm}^{-2}$ and observations will be made in a second `blue' grism (BGS, $\mathcal{R}$ > 250), covering rest-frame 0.93--1.37\,\si{\micron}. These capabilities make the EDS ideal for performing detailed sample characterisations, as it allows the detection of fainter emission overall, as well as the simultaneous recovery of the most useful rest-frame optical emission lines for galaxies at redshift 0.4--2.5.

Strong optical emission lines, like [\ion{N}{ii}]$\lambda 6584$, H$\alpha$, [\ion{O}{i}]$\lambda 6300$, [\ion{O}{iii}]$\lambda 5007$, H$\beta$, and the doublets [\ion{O}{ii}]$\lambda\lambda 3727, 3729$ and [\ion{S}{ii}]$\lambda\lambda 6717, 6731$, have long been known to be particularly sensitive probes of both the local conditions of the ionised gas in the interstellar medium (ISM), as well as the nature of the ionising radiation (\citealt{Ferland1983Are,Osterbrock2006AstrophysicsNuclei}, see \citealt{Kewley2019UnderstandingLines} for a recent review). As a result, emission-line intensities can be used in spectroscopic diagnostics to trace various galaxy properties.

Diagnostic diagrams combining two emission-line ratios are widely used to determine whether the ionising radiation in a galaxy is dominated by young, massive stars (produced in recent star formation, SF) or by an active galactic nucleus (AGN). The standard Baldwin--Phillips--Terlevich \citep[BPT,][]{Baldwin1981ClassificationObjects., Veilleux1987SpectralGalaxies} diagrams, which connect the [\ion{O}{iii}]$\lambda 5700$/H$\beta$ ratio to the [\ion{N}{ii}]$\lambda 6584$/H$\alpha$, [\ion{S}{ii}]$\lambda 6724$/H$\alpha$, and [\ion{O}{i}]$\lambda 6300$/H$\alpha$ ratios, have proved successful at distinguishing between ionising sources in local galaxies \citep{Kewley2001OpticalGalaxies,Kauffmann2003TheAGN}. With large-scale spectroscopic surveys like \Euclid collecting high-quality spectra in the more distant Universe, it remains unclear whether their use can be extended to higher redshifts. 

In fact, theoretical works utilising photoionisation models have indicated that in metal-poor galaxies, which are more prevalent at high redshift \citep[see][]{Maiolino2008AMAZE3}, AGN produce similar optical emission-line strengths to young stars in SF galaxies, leading them to overlap on the [\ion{O}{iii}]$\lambda 5700$/H$\beta$ versus [\ion{N}{ii}]$\lambda 6584$/H$\alpha$ BPT diagram as early as redshift 1 \citep{Groves2006Emission-lineAGN,Feltre2016NuclearWavelengths,Hirschmann2019SyntheticSources}. Recently, \citet{Kocevski2022CEERSJWST} and \citet{Harikane2023JWST/NIRSpecProperties} seemingly confirmed this using JWST NIRSpec spectroscopy, as their emission-line measurements sample of faint AGN above redshift 5 and 4, respectively, are indistinguishable from SF galaxies in the BPT diagram. However, at redshift 2.3, \citet{Coil2014TheZ2.3} found that their sample of 50 SF galaxies and 10 confirmed AGNs is still robustly separable in the standard BPT diagram, indicating that its breakdown as spectral diagnostic might only occur beyond intermediate redshifts. Considering the limited sample size and redshift coverage, it is nevertheless necessary to verify if EDS-like galaxy populations conform to BPT selection criteria and can thus be classified according to their ionising sources in upcoming data releases.

Optical emission lines have also been used to estimate properties of ionising sources, such as the SFR for SF-dominated galaxies and the intrinsic AGN luminosity $L_{\rm AGN}$ for AGN-dominated galaxies. H$\alpha$ is a particularly appealing tracer for the SFR, as it is well-calibrated for local galaxies \citep[see][]{Kennicutt1998STARSEQUENCE,Hopkins2003StarSurvey,Kennicutt2012} and could potentially be used to derive SFRs for galaxies observed in the EWS. In the absence of H$\alpha$ measurements, the [\ion{O}{iii}]$\lambda 5007$ line luminosity (often combined with $\mathrm{H}\beta$ into [\ion{O}{iii}]$\lambda 5007 + \mathrm{H}\beta$, as they are inseparable in photometric narrow-band surveys like HiZELS, see \citealt{Geach2008HiZELS:Z=2.23,Sobral2009HiZELS:Z=0.84}) has been used as a tracer for the SFR \citep[e.g.][]{Teplitz2000MeasurementGalaxies,Moustakas2006OPTICALINDICATORS,Osterbrock2006AstrophysicsNuclei,Sobral2015CF-HiZELS2.2}. However, some studies \citep[e.g.][]{ Kennicutt1992TheSpectra,Sobral2015CF-HiZELS2.2} have warned about potential biasing due to dust and AGN contributions, indicating that [\ion{O}{iii}]$\lambda 5007$ is an unreliable SFR proxy, especially at high redshift. Despite this, there is tentative evidence from line-emitting galaxies at redshift 0.84, 1.42, 2.23, and 3.24 in the HiZELS survey that both H$\alpha$ and [\ion{O}{iii}]$\lambda 5007( + \mathrm{H}\beta)$ can be used to estimate the SFR at moderate and high redshift \citep{Sobral2013AHiZELS,Sobral2015CF-HiZELS2.2,Khostovan2015EvolutionHiZELS,Suzuki2016OIIIHiZELS}. 

Across various AGN types in nearby galaxies, the [\ion{O}{iii}]$\lambda 5007$ luminosity has also been found to correlate with the 2--10\,keV X-ray AGN luminosity $L_{\rm X}$ \citep[e.g.][]{Netzer2006TheNuclei,Panessa2006OnGalaxies,Lamastra2009TheSources, Georgantopoulos2010ComparisonAGN,Feltre2023OpticalGalaxies}, which is itself used as a proxy for the bolometric AGN luminosity $L_{\rm AGN}$. Thus far, there has been no work done to verify its applicability to redshifts greater than 1. Given the reported evolution of the $\rm [\ion{O}{iii}]\lambda 5700/H\beta$ for SF galaxies, it is unclear whether similar effects could pollute the correlation in AGN-dominated galaxies. 

While line emission in AGN-dominated galaxies is mainly driven by the central AGN, there may still be a significant identifiable contribution from the star-forming component. One way to quantify the relative influence of the AGN is via the ratio of the black hole accretion rate (BHAR) and SFR. The BHAR-SFR relationship has been studied extensively to constrain the co-evolution of the black hole and its host galaxy, with values of $\mathrm{\logten(BHAR/SFR)}$ ranging from $-4$ to $-1$ (see \citealt{McDonald2021ObservationalGalaxies} and references therein). As a result of varying methods and assumptions, combined uncertainties from separate BHAR and SFR estimates are large. Thus, for the purposes of source characterisation in the EDS, a direct and consistently-measured estimate of this ratio from emission-line intensities should place such measurements on to a more secure footing.

The gas-phase metallicity (often expressed as the oxygen-to-hydrogen abundance O/H) is another key property which imprints onto the emission from ionised gas in galaxies. Various calibrations for local galaxies, derived from both direct temperature ($T_{\rm e}$) estimates and photoionisation models, relate intensity ratios of strong emission lines to the O/H abundance (early works by  \citealt{Jensen1976COMPOSITIONTHEORY,Pagel1979On1365.}, for recent reviews see \citealt{Kewley2019UnderstandingLines,Maiolino2019DeGalaxies}). These relations exhibit a significant scatter at low redshift  and recent JWST/NIRSpec observations of galaxies at redshift 2--9 show that O/H estimates derived from low-redshift calibrations can differ significantly from more robust direct $T_{\rm e}$ estimates \citep{Curti2022The8,Sanders2023DirectNoon}. This may indicate a significant difference in the metallicity-related properties of the ISM of low- and high-redshift galaxies. Using photoionisation models coupled to the cosmological IllustrisTNG simulations, \citet{Hirschmann2023High-redshiftSimulations} also found that some line ratio-metallicity relations evolve by up to 1\,dex between redshift 2 and 0.
It remains to be clarified how far exactly different calibrations for the gas-phase metallicity relations can be extended from the local Universe before starting to break down.

In summary, rest-frame optical emission lines are powerful probes with which to characterise galaxies and, consequently, the large number of upcoming \Euclid spectra at intermediate redshifts will help constrain one of the most important regimes for galaxy evolution. However, as outlined, many locally-used spectroscopic diagnostics and emission line-based calibrations are yet to be validated in this domain. Recently, \citet{Gabarra-EP31} assessed the performance of the NISP red grisms using mock-spectra constructed by combining galaxy properties from SED fits of star-forming galaxies between redshift 0.3 and 2.5 and some of the calibrations detailed above, thus explicitly assuming their validity at intermediate redshifts. 
In order to strengthen these pre-launch forecasts and guide observers in their analysis of future \Euclid data releases, it is vital to complement calculations based on empirical relations with self-consistent theoretical frameworks, which allow for the study of emission-line properties across cosmic time and make targeted forecasts for specific surveys and instruments. Predicting these emission lines from first principles in a self-consistent and robust manner has been a long-standing challenge, precisely because of the scarcity of spectroscopic data at intermediate redshifts, in addition to the complex interplay of various physical processes.

Past studies have demonstrated success in coupling nebular emission-line models to cosmological simulations and semi-analytic models. These have thus far been limited to modelling only the line emission due to young stars
\citep[e.g.][]{Orsi2014TheUniverse,Shimizu2016NebularLines,Wilkins2020Nebular-lineReionization, Pellegrini2020WARPFIELD-EMP:Clouds,Garg2022TheRedshift,Baugh2022ModellingGalaxies}, or, if including the contribution from AGN narrow-line regions, are limited in statistics \citep{Hirschmann2017SyntheticRatios,Hirschmann2019SyntheticSources} or focus their predictions on specific emission-line properties \citep{Hirschmann2023Emission-lineJWST} and high-redshift galaxies \citep{Hirschmann2023High-redshiftSimulations}. Consequently, a lack of comprehensive theoretical guidance for intermediate redshifts persists, which ideally would account for emission-line contribution from AGN and provide adequate statistics.

In this paper, we aim to close this gap by adopting a \Euclid-like mock light cone constructed from the \textsc{Gaea} semi-analytic model \citep{DeLucia2014ElementalModel,Hirschmann2016GalaxyModel,Fontanot2020TheModel}, which we couple to photoionisation models used in previous works by \citet{Hirschmann2017SyntheticRatios,Hirschmann2019SyntheticSources,Hirschmann2023Emission-lineJWST,Hirschmann2023High-redshiftSimulations}. Our framework is uniquely successful in its self-consistent modelling of emission lines originating not only from young stars \citep{Gutkin2016ModellingGalaxies}, but also from AGN narrow-line regions \citep{Feltre2016NuclearWavelengths}, and evolved post-asymptotic giant branch stars \citep{Hirschmann2017SyntheticRatios}. We focus our analysis on different redshift intervals between 0.4 and 2.5, in which \Euclid will recover various combinations of the brightest and most useful emission lines, such as H$\alpha$, H$\beta$, [\ion{S}{ii}]$\lambda\lambda 6717, 6731$, [\ion{N}{ii}]$\lambda 6584$,  [\ion{O}{i}]$\lambda 6300$, [\ion{O}{iii}]$\lambda 5007$, and [\ion{O}{ii}]$\lambda\lambda 3727, 3729$. 
In the following analysis, we then aim to answer five key questions:

\begin{enumerate}
\setlength\itemsep{0.3em}
\item \textit{How are galaxy populations with line emission above the defined flux thresholds in the \Euclid Wide and Deep Surveys biased with respect to stellar mass, standard scaling relations, and dominant ionising source?}
\item \textit{Are optical BPT diagrams able to distinguish between dominant ionising sources in EDS-observable galaxies?}
\item \textit{Can locally-used relations between emission-line intensities and ionising properties (i.e. SFR and AGN luminosity) be extended to intermediate redshifts? }
\item \textit{Which emission-line ratios directly trace the BHAR/SFR ratio?}
\item \textit{Do optical line-ratio calibrations for the interstellar metallicity already show a significant redshift evolution at intermediate redshifts?}
\end{enumerate}

The paper is structured as follows. The theoretical framework is described in detail in Sect. \ref{sec:theoretical FW}. In Sect. \ref{sec:validation}, we validate our approach by testing its predictions against robust theoretical and observational findings. In Sect. \ref{sec:selection bias}, we explore how observing line emitters in the EWS and EDS imposes selection bias effects on the emission-line flux versus stellar mass plane and various standard scaling relations. Additionally, we provide estimates for EDS-observable fractions of line-emitting galaxies, divided into SF and active. In Sect. \ref{sec:BPT}, we verify the use of the standard [\ion{O}{iii}]$\lambda 5700$/H$\beta$ versus [\ion{N}{ii}]$\lambda 6584$/H$\alpha$ BPT diagram to determine the dominant ionising sources for the EDS-observable sample. Section \ref{sec: galaxy prop} demonstrates that, according to our framework, locally defined calibrations between emission-line luminosities and ionising properties continue to perform well at intermediate redshifts. We further establish a strong [\ion{N}{ii}]$\lambda 6584$ emission-line dependence of the BHAR/SFR ratio and provide three novel calibrations to [\ion{N}{ii}]$\lambda 6584$-based emission-line ratios. In Sect. \ref{sec:deriving Z}, we predict that the relationship between various line-ratios and the interstellar metallicity undergo a significant evolution between redshift 0--2.5. These changes manifest as shifts in normalisation.
We discuss potential caveats of our approach in Sect. \ref{sec:discussion} and summarise our results in Sect. \ref{sec:conclusion}.

\section{Theoretical Framework}
\label{sec:theoretical FW}
\subsection{Mock light cones from the \textsc{Gaea} semi-analytic model}
\label{sec:methods_GAEA}
The GAlaxy Evolution and Assembly semi-analytic model \citep[\textsc{Gaea}\footnote{https://sites.google.com/inaf.it/gaea/},][]{DeLucia2014ElementalModel,Hirschmann2016GalaxyModel} is a successor to a model first published in \citet{DeLucia2007TheGalaxies}. Constructed upon dark matter merger trees, it traces the evolution of four baryonic components: stars in galaxies, hot gas in dark matter haloes, cold gas in galactic disks, and the gas component ejected by stellar and AGN-driven winds. \textsc{Gaea}'s  physical processes have been updated in multiple versions over the years. In this study, we make use of the most recent realisation described in \citet{Fontanot2020TheModel}, which added improved black hole (BH) accretion and AGN feedback modelling to the prescriptions for gas cooling, star formation, gas recycling, environmental processes \citep[all from original model in][]{DeLucia2007TheGalaxies}, chemical enrichment \citep[updated in][]{DeLucia2014ElementalModel}, and stellar feedback \citep[updated in][]{Hirschmann2016GalaxyModel}.

This version of \textsc{Gaea} was run on merger trees extracted from the N-body cosmological Millennium Simulation \citep{Springel2005SimulatingDistribution}, which adopted a box size of 500\,\hMpc\: and WMAP1 $\mathrm{\Lambda CDM}$ concordance cosmology ($\Omega_{\Lambda}=0.75$, $\Omega_{\mathrm{m}}=0.25$, $\Omega_{\mathrm{b}}=0.045$, $n_{\rm s}=1$, $\sigma_8=0.9$, and $\mathrm{H}_0=73\,\kmsMpc$). This large box size is crucial to make predictions which are statistically representative, given \Euclid's large areal coverage. The stellar quantities were calculated using the stellar population synthesis model from \citet{Bruzual2003Stellar2003} assuming a Chabrier IMF \citep{Chabrier2003GalacticFunction}, and the resulting physical quantities were stored in galaxy catalogues corresponding to each simulation snapshot taken at finite redshifts. 

To ensure that our \textsc{Gaea} predictions closely match the upcoming \Euclid observations, we used those catalogues to construct a mock light cone according to the algorithm described in \citet{Blaizot2005MoMaF:Facility} and \citet{Zoldan2016HI-selectedEvolution}. To avoid replications, the \textsc{Gaea} boxes at different redshift snapshots are first randomly rotated, shifted, or inverted before placing the model galaxies into an empty light cone with an aperture of 5.27 degrees, which is the largest possible diameter without exceeding the Millennium Simulation box size. Redshift varies continuously between 0 and 3.9 along the light cone and thus galaxies are extracted from the snapshot closest in redshift to the corresponding light cone region.

The resulting \textsc{Gaea} light cone catalogues (\textsc{Gaea-lc} hereafter) include all galaxies with an estimated $H$-band AB magnitude $m_{H}$ brighter than 25. We note that the EDS is expected to reach magnitude 26. However, with an apparent magnitude limit of 25, the typical mass-to-light ratios for galaxies below redshift 0.5 translates to a mass below the mass resolution of the original simulation.
We assume a conservative resolution cut for the Millennium Simulation of $10^{11} \si{\solarmass}$ for dark matter halos, which translates to an approximate resolution limit in stellar mass of $10^{9}\,\si{\solarmass}$ in \textsc{Gaea}. In applying the EWS and EDS flux limits of the NISP spectrometer to individual emission lines predicted by our model, we will demonstrate that the majority of galaxies with masses less than $ 10^9\,\si{\solarmass}$ will not be observable with \Euclid.

\subsection{Modelling of emission lines for \textsc{Gaea-lc} galaxies}
The \textsc{Gaea-lc} galaxies were post-processed with photoionisation models based on the \texttt{Cloudy} code \cite[][version c13.03]{Ferland2013THECLOUDY} to obtain nebular emission from \ion{H}{ii} regions around young stars \citep{Gutkin2016ModellingGalaxies}, AGN narrow-line regions \citep{Feltre2016NuclearWavelengths}, and post-AGB stellar populations \citep{Hirschmann2017SyntheticRatios}. 
We used the same grids of emission-line models as those described in \citet{Hirschmann2019SyntheticSources,Hirschmann2023Emission-lineJWST}, which represent updated versions of the ones detailed in \citet{Hirschmann2017SyntheticRatios}. The general modelling approach remained the same.

\subsubsection{Emission-line models for young stars, AGN narrow-line regions
and post-AGB stellar populations}
\label{sec:EL models}
For each galaxy, the \citet{Gutkin2016ModellingGalaxies} emission-line model grids for young star clusters (hereafter SF models) describe an ensemble of typical, ionisation-bounded \ion{H}{ii} regions illuminated by $10\,\mathrm{Myr}$-old stellar populations with constant star formation history. The \ion{H}{ii} regions are characterised by various model parameters, such as the \ion{H}{ii} gas density, interstellar metallicity, ionisation parameter, dust-to-metal mass ratio, and C/O abundance (see Table 1 in \citealt{Hirschmann2017SyntheticRatios}). To model the stellar component, we used the most recent version of the \citet{Bruzual2003Stellar2003} stellar population synthesis model (Charlot $\&$ Bruzual in prep.) with a standard \citet{Chabrier2003GalacticFunction} initial mass function (IMF) truncated at 0.1 and 300\,$\si{\solarmass}$. This version contains updated spectra of Wolf–Rayet stars and newer evolutionary tracks for post-AGB stars from Miller Bertolami (2016). 

The photoionisation model grids for AGN narrow-line regions (hereafter AGN models) from \citet{Feltre2016NuclearWavelengths} assume an emitted spectrum, following a broken power law, incident on gas clouds with uniform properties. Models on the grid are described by the interstellar metallicity, carbon-to-oxygen abundance, and dust-to-metal mass ratio in the narrow-line region, as well as the ionised gas density in the clouds and the ionisation parameter (see Table 1 in \citealt{Hirschmann2017SyntheticRatios}). We do not model broad-line regions, meaning we implicitly assume that all AGN are of Type 2 (see Sect. \ref{sec:discussion_coupling} for more details). 

Lastly, the model grids for evolved post-AGB stellar populations (hereafter PAGB models) from \citet{Hirschmann2017SyntheticRatios} again use the updated version of the \citet{Bruzual2003Stellar2003} stellar population synthesis code as input for \texttt{Cloudy}, this time for evolved, single-age stellar populations between 3 and $9\,\mathrm{Gyr}$ at a range of stellar metallicities. The chosen ages represent the time span in which a population of post-AGB stars has built up and produces a significant amount of ionising photons. The models are largely parameterised the same way as the \citet{Gutkin2016ModellingGalaxies} models, except for allowing the interstellar metallicity to differ from the stellar metallicity (see Table 1 in \citealt{Hirschmann2017SyntheticRatios}).

\subsubsection{Coupling the photoionisation models to \textsc{Gaea-lc}}
\label{sec:coupling}
Connecting the SF, AGN, and PAGB models described in Sect. \ref{sec:EL models} to the \textsc{Gaea-lc} catalogues of Sect. \ref{sec:methods_GAEA} was done in a self-consistent way following the methodology from previous works using these models \citep{Hirschmann2017SyntheticRatios,Hirschmann2019SyntheticSources,Hirschmann2023Emission-lineJWST,Hirschmann2023High-redshiftSimulations} but with slight modifications. In particular, our adjustments account for coupling the emission-line models to a semi-analytic model like \textsc{Gaea} rather than, as in the proceeding works, to hydrodynamical simulations explicitly containing baryonic components.

For each galaxy in the light cone, a SF, AGN, and PAGB emission-line model was chosen according to which relevant model parameters match most closely the simulated galaxy properties available from \textsc{Gaea}. Model parameters, for which no equivalent property can be recovered from \textsc{Gaea}, are fixed to standard values. The dust-to-metal mass ratio $\xi_{\mathrm{d}}$, for instance, was set to $0.3$ for all galaxies.

The most suitable SF model was selected according to the parameters closest to the simulated \textsc{Gaea} values for the global interstellar metallicity and C/O abundance, as well as the ionisation parameter. The ionisation parameter is a measure for the degree of ionisation of the ISM and, thus, depends both on the hardness and intensity of the ionising radiation coming from the source, as well as the distribution and density of the gas. For \ion{H}{ii} regions ionised by young stars in \textsc{Gaea-lc} galaxies, we follow the computation of the SF ionisation parameter $U_{\mathrm{sim,\star}}$ according to equations (1) and (2) in \citet{Hirschmann2017SyntheticRatios}. The simulated SFR of the stellar population provides the rate of ionising photons $Q_{\mathrm{sim,\star}}$, which are incident on hydrogen gas, characterised by the filling factor $\epsilon$ and the hydrogen gas density in ionised regions ($n_{\mathrm{H}, \star}$, set to $10^2\, \mathrm{cm}^{-3}$). The filling factor is calibrated such that at redshift 0 galaxies in \textsc{Gaea} reproduce the \citet{Carton2017InferringApproach} relation between $U_{\mathrm{sim,\star}}$ and interstellar metallicity (i.e. $\logten U \approx -0.8 \logten(Z_\mathrm{ISM}/Z_\odot) - 3.58$). At higher redshift, the filling factor then evolves according to the global average gas density in galaxies from the cosmological simulation IllustrisTNG (same method as in \citealt{Hirschmann2023Emission-lineJWST}).

By analogy, AGN models for nebular emission from the narrow-line region are coupled to \textsc{Gaea-lc} galaxies by matching the central (as opposed to the global) interstellar metallicity and C/O abundance, as well as the simulated ionisation parameter $U_{\mathrm{sim,\bullet}}$. \textsc{Gaea} does not trace the central metallicity directly, thus we assume it to be twice the global value. Testing other values showed that our results are insensitive to this assumption. The rate of ionising photons is now set by the AGN luminosity of the simulated galaxy. Its spectrum is assumed to follow a broken power law with adjustable index $\alpha = -1.7$ between wavelengths 0.001\,\si{\micron} and 0.25\,\si{\micron} (see Eq. 5 in \citealt{Feltre2016NuclearWavelengths}). We set the density of ionised gas clumps in the narrow-line region regions $n_{\mathrm{H}, \bullet}$ to $10^3 \mathrm{cm}^{-3}$ and modeled the volume-filling factor according to the \citet{Carton2017InferringApproach} relation, now scaling it with the central average density in IllustrisTNG galaxies for increasing redshift.  

For PAGB emission-line models, we computed the average age and metallicity of the stellar population provided by \citet{Bruzual2003Stellar2003} synthesis models, including only evolved stars older than 3\,Gyr. We matched these values to available PAGB model grid ages and metallicities from \citet{Hirschmann2017SyntheticRatios} and found the rate of ionising photons based on the mass contained in the evolved stars. Then, as before, we selected the PAGB model with the closest global interstellar metallicity, global C/O ratio and ionisation parameter  $U_{\mathrm{sim,\diamond}}$.

\subsection{Total emission-line luminosities and observer-like fluxes}
\label{sec:totlum}
After coupling the emission-line models to the \textsc{Gaea-lc} catalogues, we recovered the total emission-line luminosities for each galaxy by summing over the contributions from the matched SF, AGN, and PAGB models. 
In this study, we focus on spectroscopic diagnostics based on seven optical emission lines:
H$\alpha$, H$\beta$, [\ion{S}{ii}]$\lambda\lambda 6717, 6731$ (hereafter simply [\ion{S}{ii}]), [\ion{N}{ii}]$\lambda 6584$ ([\ion{N}{ii}]),  [\ion{O}{i}]$\lambda 6300$ ([\ion{O}{i}]), [\ion{O}{iii}]$\lambda 5007$ ([\ion{O}{iii}]), and [\ion{O}{ii}]$\lambda\lambda 3727, 3729$ ([\ion{O}{ii}]). 
For simplicity, we adopt the notation $L_{[\ion{O}{iii}]}/L_\mathrm{H \beta} = [\ion{O}{iii}]/\mathrm{H \beta}$ for luminosity ratios. In order to make targeted predictions for the observability of line-emitting galaxies with \Euclid, we compute observer-like fluxes based on the location and redshift of each galaxy in the light cone and then apply the EWS and EDS specific detection limits ($f \geq 2 \times 10^{-16}$ and $6 \times 10^{-17}\,\mathrm{erg} \, \mathrm{s}^{-1}\, \mathrm{cm}^{-2}$, respectively) to our model fluxes.

\subsection{Dust attenuation}
\label{sec:dust}
Observed galaxies will be subject to non-negligible attenuation due to their dust content, which is usually estimated with the Balmer decrement \citep{Kennicutt1992TheSpectra}. While \texttt{Cloudy} treats dust processes, including attenuation, self-consistently within \ion{H}{ii} regions, we have, thus far, not accounted for interstellar dust. Estimating the exact contribution is challenging, particularly for higher-redshift galaxies, due to the limited observational data and redshift effects, which can obscure the signs of dust. Additionally, the complexity of dust properties, intrinsic variability among galaxies, and confounding factors like star formation and AGN activity further complicate accurate estimations.

Locally found empirical relations \citep[e.g.][]{Garn2010PredictingGalaxy,Zahid2013THEGALAXIES} have established that the overall dust attenuation broadly scales with the stellar mass, with sub-dominant effects from the SFR and the metallicity. For higher redshifts, results from a series of studies using various dust indicators \citep[e.g.][]{Sobral2013AHiZELS,Dominguez2012DUSTSURVEY,Kashino2014THEEXTINCTION,Price2014DIRECTRATES,Mclure2017DustField,Cullen2017The5,Maheson2024UnravellingEvolution,Shapley2022The} have shown that, until at least redshift 3, there is no significant evolution of the relationship between dust attenuation and stellar mass. Thus, we use the local \citet{Garn2010PredictingGalaxy} relation to compute the $V$-band magnitude $A_{V}$ for each galaxy and apply a line-of-sight \citet{Calzetti2000TheGalaxies} attenuation to the predicted line fluxes. 

In order to illustrate the potential impact of dust attenuation on the observability of various line-emitting galaxy populations in the upcoming \Euclid surveys, we will mainly distinguish between three different types of samples: 
\bi 
\setlength\itemsep{0.3em}
\item The \textit{intrinsic} sample of different line-emitters as predicted by our \textsc{Gaea-lc} framework;
\item \textit{Flux-limited} galaxy populations, which for a chosen strong emission line contain only galaxies with fluxes exceeding the EWS or EDS flux limits;
\item \textit{Dust-attenuated} populations, which have the \citet{Calzetti2000TheGalaxies} curve applied before enforcing the respective flux cuts.
\ei
In Sect. \ref{sec:discussion_dust}, we elaborate on the treatment of dust attenuation in the context of our results. 

\subsection{Instrumental and environmental effects}
At this stage we do not account for additional instrumental and environmental effects which might limit the observation of emission lines. As a result, we exclude considerations of the astrophysical background, read-out and detector noise, as well as spectral resolution and recovery of blended H$\alpha$ and [\ion{N}{ii}] emission lines. We elaborate on these points in Sect.  \ref{sec:discussion_inst}. The strength of our framework is its self-consistent modelling of emission lines due to both young stars and AGN allowing us to assess the intermediate redshift validity of locally calibrated spectroscopic diagnostics from a \textit{physical} perspective. Estimates on the biasing of various scaling relations and the observability of different line-emitting galaxies should be understood as upper limits. 

\subsection{Distinguishing between dominant ionising sources in \textsc{Gaea}}
As in \citet{Hirschmann2017SyntheticRatios,Hirschmann2019SyntheticSources,Hirschmann2023Emission-lineJWST,Hirschmann2023High-redshiftSimulations}, we use the theoretical BHAR/SFR criterion to divide the \textsc{Gaea-lc} sample according to their dominant ionising source, meaning SF-dominated,  AGN-dominated, and composite galaxies, which contain significant SF and AGN contribution. For this study, we have adjusted the BHAR/SFR boundaries in order to ensure that the populations are reasonably separated in all diagrams:
\bi
\setlength\itemsep{0.3em}
 \item SF-dominated galaxies: $\mathrm{BHAR/SFR} < 10^{-3}$
 \item Composite galaxies: $10^{-3} < \mathrm{BHAR/SFR} < 10^{-2.2}$
\item AGN-dominated galaxies: $\mathrm{BHAR/SFR} > 10^{-2.2}$
\ei

\citet{Hirschmann2017SyntheticRatios,Hirschmann2019SyntheticSources,Hirschmann2023Emission-lineJWST,Hirschmann2023High-redshiftSimulations} further define galaxies to be dominated by aged PAGB stars if their H$\beta$ emission exceeds the contribution from both young stars and AGN. They generally have low star-formation rates and form a sub-category of galaxies with low-ionisation (nuclear) emission-line regions \citep[LIER/LINER, see][]{Heckman1980AnNuclei,Kauffmann2003TheAGN,Singh2013TheHoles,Belfiore2016SDSSLIERs}. 
While these galaxies do exist in our \textsc{Gaea-lc} sample, they become increasingly rare beyond the local Universe where galaxies exhibit younger stellar populations and high SFRs. This agrees with \citet{Hirschmann2023Emission-lineJWST}, who found that the number of PAGB-dominated galaxies in their sample of post-processed IllustrisTNG galaxies rapidly decreases from a few per cent below redshift 1 to a negligible fraction above redshift 1. Additionally, with luminosities of order $\mathrm{10^{39} \,erg \,s^{-1}}$, their emission is relatively faint, meaning they lie below the EDS flux limit already at around redshift 0.1.
Thus, we conclude that \Euclid will likely not observe any PAGB-dominated galaxies and exclude them from further analysis.

\begin{figure}[htbp!]
\centering
	\includegraphics[width=\columnwidth]{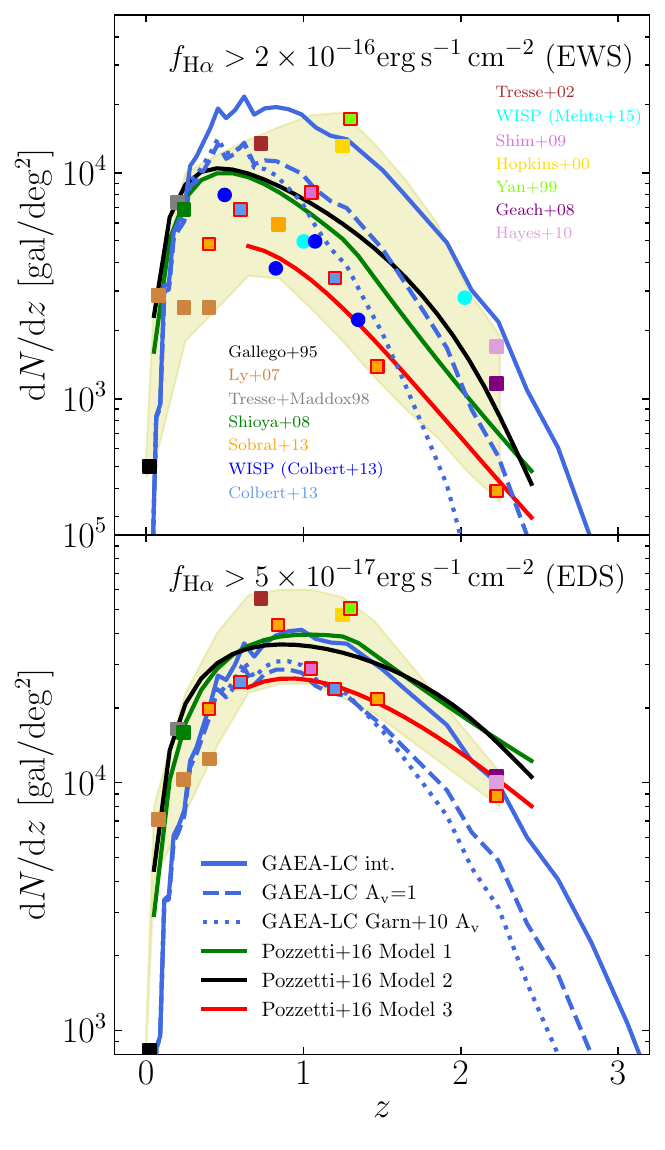}
    \caption{Redshift distribution of number density of $\mathrm{H} \alpha$ emitters above ﬂux thresholds $2\times10^{-16} \operatorname{erg} \mathrm{s}^{-1}\, \mathrm{cm}^{-2}$ (top panel) and $5\times10^{-17} \operatorname{erg} \mathrm{s}^{-1}\, \mathrm{cm}^{-2}$ (bottom panel), corresponding to the respective EWS and EDS flux limits. Predictions from the \textsc{Gaea-lc} framework (blue lines) use the intrinsic population of emitters (solid) and dust-attenuated versions with a flat $A_v$ (dashed) and a mass-dependent $A_v$ scaling \citep[dotted][]{Garn2010PredictingGalaxy}.  Models 1 (green), 2 (black), and 3 (red) from \citet{Pozzetti2016ModellingMissions} represent various model fits to collections of uncorrected $\mathrm{H} \alpha$ survey results (data points, covering yellow-shaded area) across different redshift ranges. Model 3 has been fit to data points outlined in red.}
    \label{fig:H alpha no}
\end{figure}

\begin{figure*}[htbp!]
\centering
	\includegraphics[width=\textwidth]{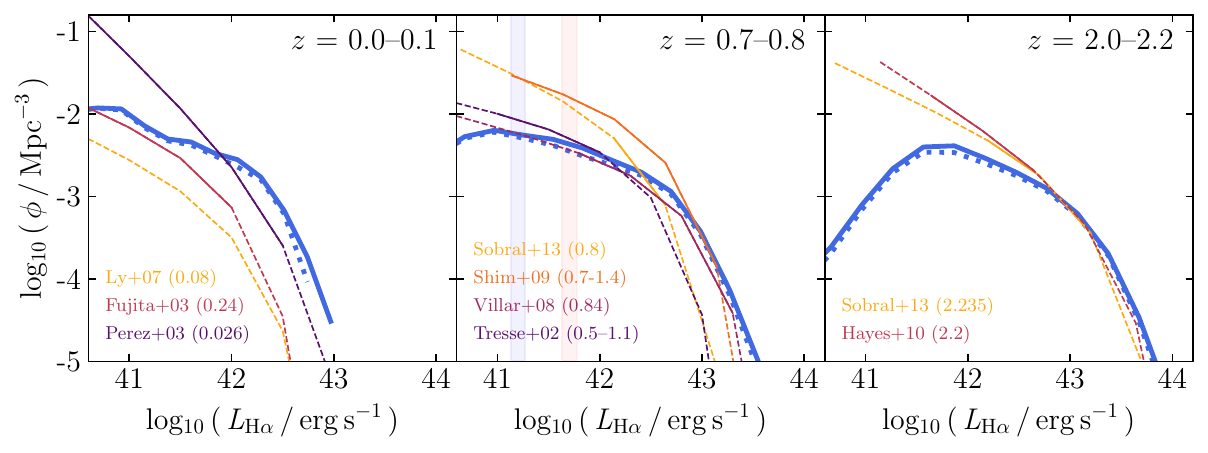}
    \caption{Redshift evolution of the $\mathrm{H} \alpha$ luminosity function for all (solid blue lines) and only SF-dominated (dotted blue lines) \textsc{Gaea-lc} galaxy populations from redshift 0 to 2.2. Overplotted are various fits to dust-corrected observational results (thin lines) from narrow-band and spectroscopic surveys within the redshift range indicated by the legend in each panel \citep{Tresse2002TheZ1,Fujita2003TheData,Perez-Gonzalez2003SpatialGalaxies,Ly2007TheField,Villar2007TheZ=0.84,Shim2009Global0.7z1.9,Hayes2010TheVLT/HAWK-I,Sobral2013AHiZELS}. Due to selection effects, the resulting Schechter fits rely on measurements covering only part of the luminosity range (solid), which are then extrapolated to low and high luminosities (dashed). For reference, the EWS (red) and EDS (blue) detection limits across the redshift range between 0.7 and 0.8 are shown in shaded areas.}
    \label{fig:H alpha LF}
\end{figure*}

\section{Validation of the method}
\label{sec:validation}
While the photoionisation models discussed in Sect. \ref{sec:EL models} have been successfully applied to galaxies formed in numerical simulations like SPHGal and the IllustrisTNG suite \citep{Hirschmann2017SyntheticRatios,Hirschmann2019SyntheticSources,Hirschmann2023Emission-lineJWST,Hirschmann2023High-redshiftSimulations}, this work represents the first instance of applying them to a semi-analytic model, like \textsc{Gaea}. As semi-analytic models do not explicitly treat gas dynamics, we had to adapt the approach and underlying assumptions in order construct the mock-emission lines for \textsc{Gaea-lc}. 
Thus, we validate our method by comparing our emission-line predictions to observational data and theoretical predictions for low redshifts. Our self-consistent modelling will then allow us to extend our predictions to higher redshifts. 

\subsection{\texorpdfstring{H$\alpha$ number counts and luminosity function}{H-alpha Luminosity function}}
As a first validation, we compare in Fig. \ref{fig:H alpha no} predicted redshift distributions of the H$\alpha$ emitter number density (blue lines) against models from \citet[][]{Pozzetti2016ModellingMissions}, which have been derived from fits to a collection of H$\alpha$ luminosity functions \citep[][]{Gallego1995TheUniverse,Ly2007TheField,Tresse19980.2,Shioya2008Field,Sobral2013AHiZELS,Colbert2013PredictingSurvey,Tresse2002TheZ1,Shim2009Global0.7z1.9,Hopkins2000STAR1.8,Yan1999TheTwo,Geach2008HiZELS:Z=2.23,Hayes2010TheVLT/HAWK-I}. Model 1 (green) and 2 (black) use the same collection of survey data and underlying Schechter function, but differ in their implementation of the redshift evolution. Model 3 (red) was determined using only surveys covering higher redshifts from 0.7--2.23 (points outlined in red) and is based on a broken power law. While Model 1 and 2 produce similar number densities, Model 3’s prediction is generally lower by a factor of 1.5–2.5. This difference can be attributed to the large underlying scatter in the observed luminosity functions, as well as the different functional form. We include the scatter of number count predictions from the collection of integrated luminosity functions, as well as direct estimates of cumulative number counts derived from the WFC3 Infrared Spectroscopic Parallels (WISP) survey \citep[circular data points]{Colbert2013PredictingSurvey,Mehta2015PREDICTINGGALAXIES}. The significant spread of data points (highlighted in yellow shaded area) is a result of varying survey set-ups, such as using different instruments with different selection functions, collecting either narrow-band or spectroscopic measurements, as well as varying areal coverage and treatments of cosmic variance.

Since the \citet{Pozzetti2016ModellingMissions} models were constructed to explore observational yields, survey results have not been corrected for extinction. For a fair comparison, they should thus be contrasted with a dust-attenuated sample of H$\alpha$-emitters predicted by our \textsc{Gaea-lc} framework. However, the prevalence and nature of interstellar dust beyond redshift 1 is largely uncertain. In order to visualise the impact different dust distributions might have, we thus present three different predictions: the intrinsic \textsc{Gaea-lc} version (solid blue lines), a \citet{Calzetti2000TheGalaxies} attenuation with a flat $A_{v}$ scaling of 1 (dashed lines) and a mass-dependent $A_v$ scaling of \citet[dotted lines]{Garn2010PredictingGalaxy}, which is adopted in the remaining paper. 

We compare the predicted number counts per $\rm deg^{2}$ above two flux thresholds. For an EWS-like threshold of $2\times10^{-16} \operatorname{erg} \mathrm{s}^{-1}\, \mathrm{cm}^{-2}$ (top panel), the intrinsic \textsc{Gaea-lc} prediction differs by a factor of 2 from Models 1 and 2, while the prediction using a flat $A_{v}$ scaling agrees almost exactly with Models 1 and 2. The curve using a \citet{Garn2010PredictingGalaxy} $A_{v}$ scaling gives the lowest prediction of the number count density, which is between Model 2 and 3 until it diverges from the observational scatter around redshift 1.8, when H$\alpha$ stops being observable with \Euclid. This steeper decrease with redshift compared to other predictions is because, at high redshift, H$\alpha$ emitters above the flux cut are generally massive and thus more strongly attenuated according to the \citet{Garn2010PredictingGalaxy} scaling. Under the EDS-like flux cut ($5\times10^{-17} \operatorname{erg} \mathrm{s}^{-1}\, \mathrm{cm}^{-2}$), number densities from the intrinsic model are in good agreement with Models 1 and 2. Predictions from the dust-attenuated \textsc{Gaea-lc} samples are in better agreement with Model 3, but decrease below the observational scatter at redshift 1.5. 

Overall, predictions for the redshift distribution of the H$\alpha$ emitter number density from our \textsc{Gaea-lc} models fall within the significant scatter of observational estimates, and broadly agree with the models from \citet{Pozzetti2016ModellingMissions} for both the EWS- and EDS-like thresholds. Varying the applied dust attenuation significantly changes the prediction beyond redshift 0.5. Which dust attenuation model agrees best with which \citet{Pozzetti2016ModellingMissions} model and the integrated observational estimates varies if considering the EWS or the EDS threshold. Our line-of-sight \citet{Calzetti2000TheGalaxies} extinction likely does not capture the full complexity of the nature of dust in this regime. Thus, exact comparisons should be approached with caution. While the prediction using the $A_{v}$ scaling from \citet{Garn2010PredictingGalaxy} drops below the observational scatter at 1.5 for the EDS and 1.8 for the EWS, this represents a range with relatively few data points, which have mostly been determined from narrow-band measurements \citep{Sobral2013AHiZELS,Hayes2009TheVLT/HAWK-I,Geach2008HiZELS:Z=2.23} that often suffer from contamination due to other emission lines. Moreover, this is at the edge of the detectable range with \Euclid, meaning this divergence from the scatter will not affect our predictions significantly. Since the \citet{Garn2010PredictingGalaxy} scaling is empirically motivated and has not shown significant evolution across the relevant range (see Sect. \ref{sec:dust} and \ref{sec:discussion_dust}), we thus continue to adopt it for the remaining paper. For this sample, the average number density from redshift 0.9--1.8 above an EWS-like threshold is 7\,800 H$\alpha$ emitters/$\rm deg^{2}$. \Euclid's mission requirements specify the redshift measurement of 1\,700 H$\alpha$ emitters $\rm deg^{-2}$. Thus, according to this estimate, \Euclid would only have to recover redshifts for around 21\% of these emitters.

We further compare our model predictions against observed H$\alpha$ luminosity functions. Figure \ref{fig:H alpha LF} shows the evolution of the intrinsic H$\alpha$ luminosity function $\phi$ between redshifts 0 and 2.2 for the full population of \textsc{Gaea-lc} galaxies (solid blue lines) and the sub-group of SF-dominated galaxies (dotted blue lines). As established in Sect. \ref{sec:intro}, the H$\alpha$ luminosity is often used as a proxy for star formation and thus, on cosmological scales, the evolution of the H$\alpha$ luminosity function traces the cosmic SFR density. The full \textsc{Gaea-lc} and the SF-dominated sample appear closely spaced in the figure, indicating that the $\mathrm{H} \alpha$ luminosity is indeed shaped by the emission from \ion{H}{ii} regions around young stars. 

Overall, cosmic star formation has sharply declined from redshift 0 to 2, and as a result, the luminosity function exhibits a similarly strong evolution. The maximum at the luminous end decreases by around $0.8\,\mathrm{dex}$ from $L_{\mathrm{H} \alpha} \sim 10^{43.8}\, \mathrm{erg} \, \mathrm{s}^{-1}$ to $L_{\mathrm{H} \alpha} \sim 10^{43}\, \mathrm{erg} \, \mathrm{s}^{-1}$. This is consistent with dust-corrected observational fits from both narrow-band and spectroscopic surveys \citep[e.g.][shown in thin lines, redshifts indicated in parentheses]{Tresse2002TheZ1,Fujita2003TheData,Perez-Gonzalez2003SpatialGalaxies,Ly2007TheField,Villar2007TheZ=0.84,Shim2009Global0.7z1.9,Hayes2010TheVLT/HAWK-I,Sobral2013AHiZELS}. We distinguish between the luminosity ranges constrained by measurements (solid) and the extrapolated fits to low and high luminosities (dashed).

At high redshift, between 2--2.2, our predictions for the luminous end of the luminosity function are in excellent agreement with observational results. Below $L_{\mathrm{H} \alpha} \sim 10^{43}\, \mathrm{erg} \,\mathrm{s}^{-1}$, our prediction slowly starts diverging towards lower values, until it reaches a turnover around $L_{\mathrm{H} \alpha} \sim 10^{41.8}\, \mathrm{erg} \,\mathrm{s}^{-1}$. This feature is an effect resulting from \textsc{Gaea}'s resolution limit at $M_\star \sim 10^{9}\,\si{\solarmass}$, as well as the applied $H$-band magnitude cut, which particularly affects low-luminosity galaxies at higher redshift. As a result, the faint end of the predicted H$\alpha$ luminosity function is underestimated. In addition, we note that observational surveys either undersample the faint end, in which case they apply estimated completeness corrections, or do not measure any low-luminosity H$\alpha$ emitters and only extrapolate the Schechter fit. Thus, at the low-luminosity end, our predictions should be compared to observational results with caution.

Across the 0.7--0.8 redshift range, the \textsc{Gaea-lc} prediction lies among the spread of the different survey results. These exhibit varying shapes and a large scatter in $\phi$ of around 0.8--1\,dex. In the case of \citet{Shim2009Global0.7z1.9} and \citet{Tresse2002TheZ1}, the resulting luminosity functions have been fit to data covering relatively large redshift ranges, averaging out any potential evolution across them. However, even \citet{Sobral2013AHiZELS} and \citet{Villar2007TheZ=0.84}, which both cover redshifts around 0.8, exhibit very different shapes. In general, we conclude that $\phi$ is poorly constrained in this redshift regime, also illustrating the need for more extensive spectroscopic surveys.

In the 0--0.1 redshift range, we slightly overpredict the luminous end with respect to observational determinations. At luminosities below $L_{\mathrm{H} \alpha} \sim 10^{42}\, \mathrm{erg} \, \mathrm{s}^{-1}$, our result lies within the large scatter among them. As for redshifts in 0.7--0.8, the surveys have targeted different redshift ranges, which could partially explain this scatter. However, the luminosity function at redshift 0.24 determined by \citet{Fujita2003TheData} lies between the results at redshift 0.08 and 0.026 from \citet{Ly2007TheField} and \citet{Perez-Gonzalez2003SpatialGalaxies}, which differ by more than 1\,dex at the faint end. This large discrepancy suggests that $\phi$ is not well-constrained in the low-redshift regime either.
Lastly, due to the geometry of the \textsc{Gaea} light cone, our sample contains limited number counts at low redshift, which makes our estimate susceptible to low number statistics.

However, we overall conclude that our framework predicts the evolution of the H$\alpha$ luminosity function in broad agreement with empirical results and any discrepancies lie within observational and modelling uncertainties.

\begin{figure*}[htbp!]
\centering
	\includegraphics[width=\textwidth]{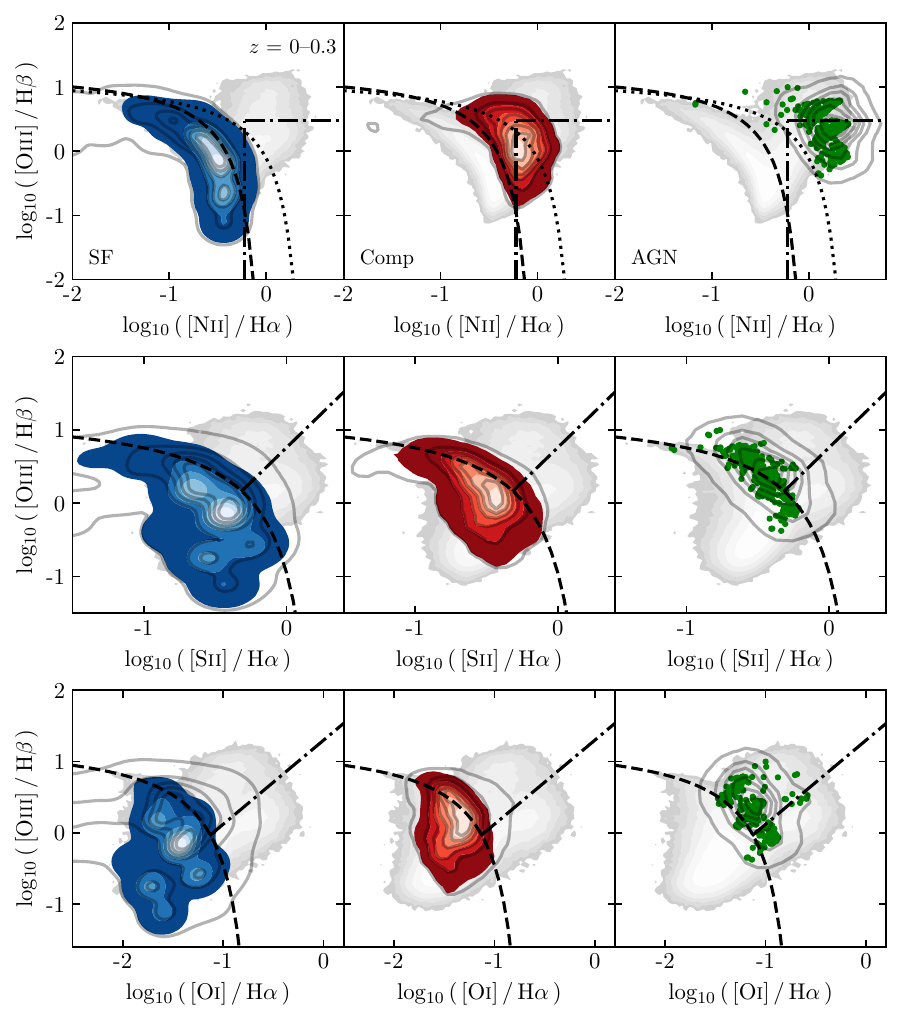}
    \caption{Location of \textsc{Gaea-lc} galaxy populations at redshift 0--0.3 in the classical BPT diagrams, [\ion{O}{iii}]/H$\beta$ versus [\ion{N}{ii}]/H$\alpha$ (top row), [\ion{O}{iii}]/H$\beta$ versus [\ion{S}{ii}]/H$\alpha$ (middle row), and [\ion{O}{iii}]/H$\beta$ versus [\ion{O}{i}]/H$\alpha$ (bottom row). Shown are simulated SDSS-like galaxies (limited to fluxes above $5 \times 10^{-17} \operatorname{erg} \mathrm{s}^{-1}\,\mathrm{cm}^{-2}$, coloured contours and data points), alongside the intrinsic \textsc{Gaea-lc} sample (grey contour lines). For comparison, SDSS-observed galaxies are plotted in the background (filled grey contours). 
    Galaxy populations are divided according to dominant ionising source, meaning SF-dominated galaxies (blue, left column), composite galaxies (red, middle column), and AGN-dominated galaxies (green, right column). Overplotted are empirical selection criteria meant to broadly distinguish SF galaxies (below dashed lines, \citealt{Kewley2001OpticalGalaxies} in top row, \citealt{Kauffmann2003TheAGN} in middle and bottom row) and active galaxies (above dashed lines). An additional criterion separates composite galaxies (above dashed line and below dotted line, \citealt{Kewley2001OpticalGalaxies}) from purely AGN-dominated galaxies in the [\ion{O}{iii}]/H$\beta$ versus [\ion{N}{ii}]/H$\alpha$ diagram. In all diagrams, LI(N)ER are expected to fall in the bottom right corner (rectangle defined by dash-dotted lines in top row, \citealt{Kauffmann2003TheAGN} and area below dash-dotted lines in middle and bottom row \citealt{Kewley2006TheNuclei}).}
    \label{fig:BPT lowz}
\end{figure*}
\subsection{Distinguishing between ionising sources using BPT diagnostic diagrams}
\label{sec:BPT validation}
In Fig. \ref{fig:BPT lowz}, we show the locations of the predicted SF-dominated (left column), composite (middle column), and AGN-dominated (right column) galaxy populations at redshift less than 0.3 in the standard BPT diagrams: [\ion{O}{iii}]/H$\beta$ against [\ion{N}{ii}]/H$\alpha$ (top row, \citealt{Baldwin1981ClassificationObjects.}), [\ion{O}{iii}]/H$\beta$ against [\ion{S}{ii}]/H$\alpha$ (middle row), and [\ion{O}{iii}]/H$\beta$ against [\ion{O}{i}]/H$\alpha$ (bottom row, both \citealt{Veilleux1987SpectralGalaxies}). In addition to the intrinsic \textsc{Gaea-lc} samples (grey contour lines), we contrast the observed SDSS sample (filled grey contours in background) with simulated SDSS-like galaxy populations (blue contours for SF-dominated, red contours for composites, green data points for AGN-dominated). We note that the observed SDSS galaxies shown here are not separated according to type and instead represent the combined sample. As in \citet{Hirschmann2017SyntheticRatios,Hirschmann2019SyntheticSources,Hirschmann2023Emission-lineJWST,Hirschmann2023High-redshiftSimulations}, SDSS-like galaxies were selected by applying a flux limit of $5 \times 10^{-17} \operatorname{erg} \mathrm{s}^{-1}\, \mathrm{cm}^{-2}$ \citep[see Table 1 in][]{Juneau2014ACTIVEEVOLUTION} to all four lines. We note that, in some instances, the grid parameterisation of the photoionisation models results in visible accumulations of galaxies at discrete points in the diagrams, such as for SF-dominated galaxies in [\ion{S}{ii}]/H$\alpha$ and [\ion{O}{i}]/H$\alpha$ plots and AGN-dominated galaxies in all plots. For the latter populations, only a few hundreds of galaxies exceed the SDSS flux cut and thus, we show the scatter of the individual data points. In general, simulated SDSS galaxies occupy the same region as the observed SDSS galaxies. The flux limit mostly cuts out galaxies with particularly low [\ion{N}{ii}]/H$\alpha$, [\ion{S}{ii}]/H$\alpha$ and [\ion{O}{i}]/H$\alpha$. In the photoionisation models by \citet{Gutkin2016ModellingGalaxies} these represent the lowest metallicity galaxies, in line with our expectation that these tend to be low-luminosity \citep[see][]{Tremonti2004TheSurvey}.  

In order to test our framework, we then compare our \textsc{Gaea-lc} samples to optical criteria used to distinguish between dominant ionising sources in local galaxies. By combining photoionisation and stellar population synthesis models, \citet{Kewley2001OpticalGalaxies} set a theoretical upper limit to the location of star-forming galaxies above which the emission from galaxies would not be explainable without a strong AGN component (dotted line in top row, dashed lines in middle and bottom row). Based on SDSS observations of nearby AGN, \citet{Kauffmann2003TheAGN} found that in their sample, galaxies containing an AGN are confined above the dashed line in the [\ion{N}{ii}]/H$\alpha$ diagram. Thus, the area between the dotted and dashed lines can be understood as a transition region where we expect composite galaxies to be located.

As expected, at  redshift less than 0.3, the optical selection criteria for BPT diagrams are successful at separating the sample according to ionising source. The majority of \textsc{Gaea-lc} galaxies selected according to the theoretical BHAR/SFR criteria are confined to the predicted locations for SF-dominated, composite, and AGN-dominated populations. Composite galaxies partially extend into the AGN-dominated regime in the [\ion{N}{ii}]/H$\alpha$ diagram, but, nevertheless, occupy a distinct region. There are no selection criteria to identify composite galaxies in the [\ion{S}{ii}]/H$\alpha$ diagram and [\ion{O}{i}]/H$\alpha$ diagram and they mostly overlap with the SF-dominated population. In general, these diagrams also provide a less clear distinction of SF- and AGN-dominated galaxies compared to the [\ion{N}{ii}]/H$\alpha$ diagram. As a result, we will focus on this diagram when we extend our predictions to intermediate redshifts in Sect. \ref{sec:BPT}.

\begin{figure}
\centering
	\includegraphics[width=\columnwidth]{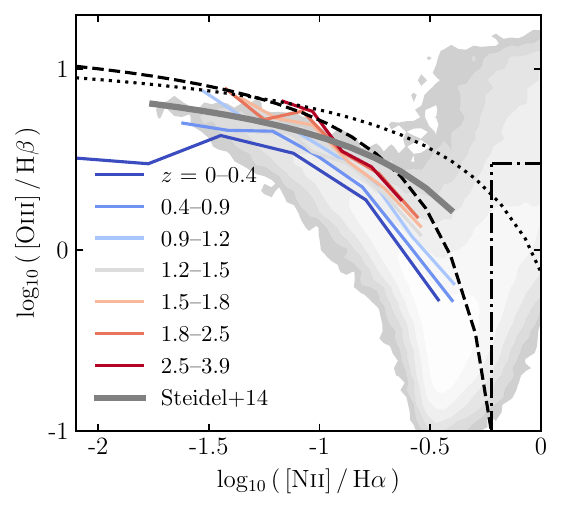}
    \caption{Evolution of average [\ion{O}{iii}]$ / \mathrm{H \beta}$ versus [\ion{N}{ii}]$ / \mathrm{H \alpha}$ of SF-dominated \textsc{Gaea-lc} galaxies at different redshift intervals, as indicated by the legend (coloured lines). Shown are simulated SDSS-like galaxies with masses greater than $10^9 \si{\solarmass}$. Overplotted is the mean relation found by \citet{Steidel2014mossfire} in their sample of star-forming galaxies at redshift 2.3 (thick grey line) and, as in Fig. \ref{fig:BPT lowz}, the SDSS sample (grey contours) and empirical criteria distinguishing between ionising sources \citep[black lines]{Kewley2001OpticalGalaxies,Kauffmann2003TheAGN}.}
    \label{fig:BPT SF branch}
\end{figure}

\subsection{\texorpdfstring{Evolution of the star-forming branch in the $[\ion{O}{iii}] / H\beta$-versus-$ [\ion{N}{ii}] / H\alpha$ diagram}{Evolution of the star-forming branch}}
In this Section, we verify the observed increase from low to high redshift of the [\ion{O}{iii}]/H$\beta$ ratio at fixed [\ion{N}{ii}]/H$\alpha$ ratio for SF galaxies. This was initially observed around redshift 2 \citep[e.g.][]{Shapley2004EvidenceZ2,Liu20081.01.51,Hainline2009Rest-frame2,Bian2010Lbt/luciferJ0900+2234,Lehnert2009PhysicalRedshift2,Yabe2012NIRRelation,Masters2014PhysicalFire,Steidel2014mossfire, Shapley2015TheGalaxies,Kashino2017TheMedium,Strom2017Ratio}, while recent JWST/NIRSpec data showed a continuation of this trend to redshift 5 \citep[e.g.][]{Cameron2023ARelations,Sanders2023ExcitationJWST/NIRSpec}.

From our \textsc{Gaea-lc} sample, we selected SF-dominated galaxies with resolved stellar masses ($M_{\star} \geq 10^9 \,\si{\solarmass}$) and SDSS-like fluxes ($\geq 5 \times 10^{-17}\, \operatorname{erg} \mathrm{s}^{-1}\, \mathrm{cm}^{-2}$), for which we then computed the average [\ion{O}{iii}]/H$\beta$ at fixed [\ion{N}{ii}]$ / \mathrm{H}\alpha$. 
The result is shown in Fig. \ref{fig:BPT SF branch} for different redshift bins between redshift 0 and 3.9 (thin coloured lines, indicated in legend), alongside the optical selection criteria \citep[black lines, as in Fig. \ref{fig:BPT lowz}]{Kewley2001OpticalGalaxies,Kauffmann2003TheAGN,Kewley2006TheNuclei} and SDSS-observed galaxies (grey contours in background).
At fixed [\ion{N}{ii}]$ / \mathrm{H} \alpha$, the mean [\ion{O}{iii}]$ / \mathrm{H} \beta$ ratio increases with redshift. 
According to detailed investigations by \citet{Hirschmann2017SyntheticRatios,Hirschmann2023Emission-lineJWST}, the redshift evolution of the [\ion{O}{iii}]$ / \mathrm{H} \beta$ at fixed [\ion{N}{ii}]$ / \mathrm{H} \alpha$ is driven by the elevated SFR and global gas density at higher redshift, increasing the ionisation parameter $U_{\star}$ and, as a result, the probability for doubly-ionised oxygen.
Comparing the predicted average relation at redshift 1.8--2.5 with the fit of 219 observed SF galaxies at redshift 2.3 from \citet[thick grey line]{Steidel2014mossfire}, we find good agreement. We note that the average relations from our simulated galaxies appear slightly steeper than the \citet{Steidel2014mossfire} determination, which can be partly explained by the difficultly of correctly identifying composites observationally and the slight dependence on the choice of BHAR/SFR boundaries used in our theoretical definition.

\begin{figure}[htbp!]
\centering
	\includegraphics[width=0.9\columnwidth]{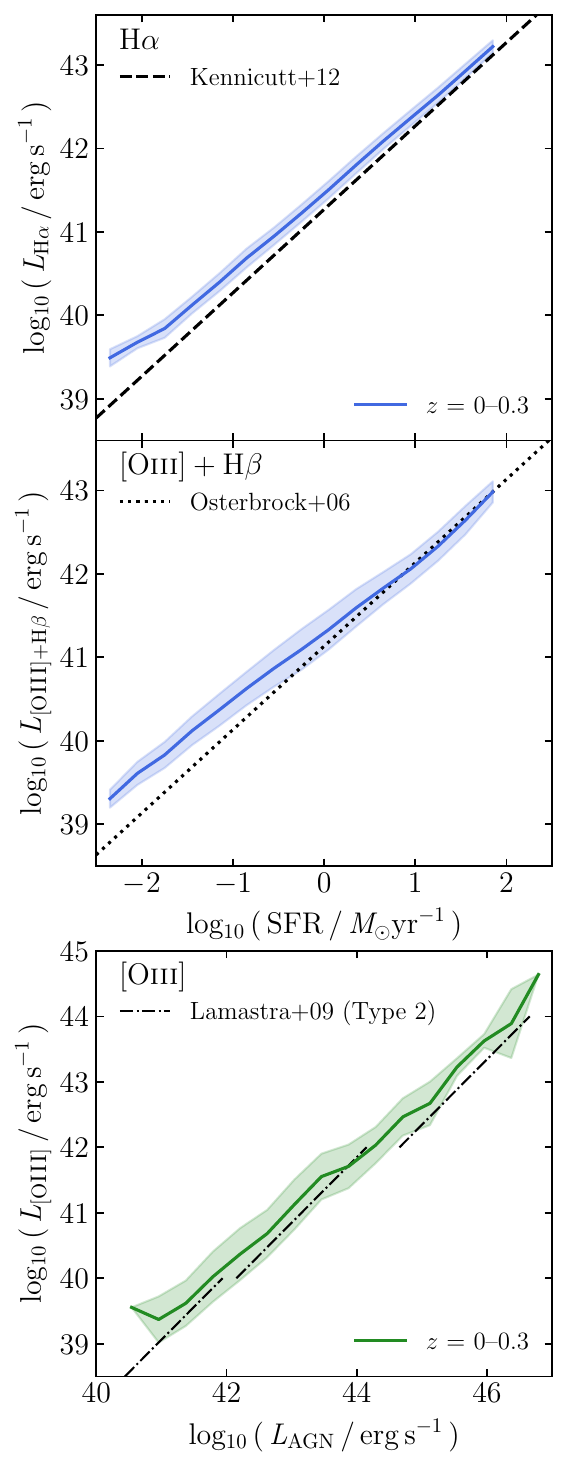}
    \caption{Average H$\alpha$ (first panel, blue line) and [\ion{O}{iii}] + H$\beta$ (second panel, blue line) line luminosities versus SFR for \textsc{Gaea-lc} galaxies with specific SFR $> 10^{-10.5} \,\mathrm{ yr^{-1}}$ in the redshift range 0--0.3, alongside the 1 standard deviation scatter (shaded area). Overplotted, for comparison, are widely used relations from \citet[for H$\alpha$, dashed line]{Kennicutt2012} and \citet[$\mathrm{for \ [\ion{O}{iii}] +H\beta, \ dotted \ line}$]{Osterbrock2006AstrophysicsNuclei}. The bottom panel shows the average [\ion{O}{iii}] line luminosity versus AGN luminosity for active \textsc{Gaea-lc} galaxies with a 1$\sigma$ scatter (green line and shaded area), plotted alongside the relation found by \citet[dash-dotted line]{Lamastra2009TheSources} from a sample of 61 type-2 AGN with $z < 0.83$.}
    \label{fig:SFR+LAGN lowz}
\end{figure}

\subsection{Relation of strong line luminosities to SFRs and AGN luminosities at low redshift}
As a last test of our methodology, we demonstrate that its low-redshift predictions reproduce local calibrations between strong line luminosities and galaxy properties, such as the SFR and the AGN luminosity.
In the first two panels of Fig. \ref{fig:SFR+LAGN lowz}, we show the average H$\alpha$ ($L_\mathrm{{H\alpha}}$, top panel) and [\ion{O}{iii}]+H$\beta$ ($L_\mathrm{[\ion{O}{iii}]+H\beta}$, middle panel) luminosity at fixed SFR for \textsc{Gaea-lc} galaxies between redshift 0--0.3 (blue line with shaded area indicating 1 standard deviation). These proxies are only expected to be robust for SF galaxies (also see Fig. 10 in \citealt{Hirschmann2023Emission-lineJWST}). 
We note the distinction between SF galaxies, which are commonly defined as having high specific SFRs $(\mathrm{sSFR} \, \equiv \mathrm{SFR}/M_{\star})$, and SF-dominated galaxies, in which the ionisation budget due to young star clusters is greater than the contribution from other sources. While most SF galaxies are SF-dominated, SF-dominated galaxies are not necessarily highly star-forming in our simulations. However, we can generally consider our predictions for SF-dominated galaxies to be a good proxy for SF galaxies. 

As here we directly compare with observationally used relations, we select SF \textsc{Gaea-lc} galaxies by applying a sSFR cut of $10^{-10.5} \,\mathrm{ yr^{-1}}$, with no additional flux or mass cuts. Selecting SF-dominated galaxies produced an identical relation. Alongside our predictions, we plot local calibrations for the respective relationships. The \citet[top panel, dashed line]{Kennicutt2012} relation, originally published in \citet{Murphy2011CalibratingNGC6946}, was derived from evolutionary synthesis models for SF galaxies based on a \citet{Kroupa2003Galactic-FieldStars} IMF, which yields nearly identical results to a \citet{Chabrier2003GalacticFunction} IMF \citep[see][]{Chomiuk2011TowardGalaxies}. However, our models cover a range of metallicities from $10^{-3}\,Z_{\odot}$ to $1\,Z_{\odot}$, while \citet{Murphy2011CalibratingNGC6946} assume solar metallicity $Z_{\odot}$. Thus, we do not expect our predictions to match these relations exactly. Across all SFR, our H$\alpha$ luminosity-SFR relation exhibits the same slope as the \citet{Kennicutt2012} calibration, but is slightly offset toward higher $L_\mathrm{{H\alpha}}$ values, which can be explained by the slightly different modelling assumptions. This difference increases at low SFR, which is where the magnitude cut introduces a bias toward an increased $L_\mathrm{{H\alpha}}$ in the remaining galaxies. Our $L_\mathrm{[\ion{O}{iii}]+H\beta}$-SFR prediction agrees well with \citet[middle panel, dotted line]{Osterbrock2006AstrophysicsNuclei}, especially at $\logten \left(\mathrm{SFR / \si{\solarmass} yr^{-1}}\right)>-0.5$, below which it diverges for similar reasons as detailed above.

In the bottom panel of Fig. \ref{fig:SFR+LAGN lowz}, we explore the relationship between the [\ion{O}{iii}] luminosity $L_{\mathrm{[\ion{O}{iii}]}}$ and the bolometric AGN luminosity $L_{\mathrm{AGN}}$. For active (meaning composite and AGN-dominated) galaxies at redshift 0--0.3, we show the predicted mean [\ion{O}{iii}] luminosity at fixed AGN luminosity (green line with shaded 1 standard deviation). We compare to an empirical relation by \citet{Lamastra2009TheSources}, who, based on a sample of 61 type-2 AGN with redshift less than 0.83, found a luminosity-dependent [\ion{O}{iii}]-bolometric correction factor in the ranges $\logten (L_{[\ion{O}{iii}]}) = 38$--40, 40--42 and 42--44 (dash-dotted lines). In general, we note excellent agreement with our \textsc{Gaea-lc} relation across the entire $L_{\mathrm{AGN}}$ range.

\section{Exploring Euclid's selection bias}
\label{sec:selection bias}
Section \ref{sec:validation} has demonstrated that our model framework, which connects emission-line models to simulated galaxies from the \textsc{Gaea} semi-analytic model, successfully reproduces a wide range of locally observed emission-line properties. 
As a result, we have established a self-consistent, physically validated sample of line-emitting galaxies between redshift 0--3.9, which we can now use to put \Euclid forecasts on solid ground. In this Section, we explore the selection bias resulting from the observation of the BPT emission lines (H$\alpha$, H$\beta$, [\ion{O}{iii}], and [\ion{N}{ii}]) by considering the EWS and EDS specific flux limits at the relevant redshifts. 
Specifically, we will assess the effects on the line flux-stellar mass planes, standard scaling relations, the prevalence of luminous AGN, and the observability of line-emitting populations. We use `detectability' to describe the redshift range in which an emission line of a given rest wavelength falls into the wavelength sensitivity range of \Euclid's grisms. For a given redshift, we then define `observability' as the number of galaxies emitting line intensities within the detectable range and above the flux limits, respectively for the EWS and EDS.  
We note that this definition is distinct from the \Euclid mission requirement of `completeness' \citep[see][]{Racca2016TheDesign}, which is defined as the fraction of galaxies at redshift 0.9--1.8 emitting fluxes above the limit $2\times10^{-16}\, \operatorname{erg} \mathrm{s}^{-1}\, \mathrm{cm}^{-2}$, for which a redshift measurement can be recovered. 

\subsection{Evolution of line detectability with the blue and red grisms}
\begin{figure}[htbp!]
\centering
	\includegraphics[width=\columnwidth]{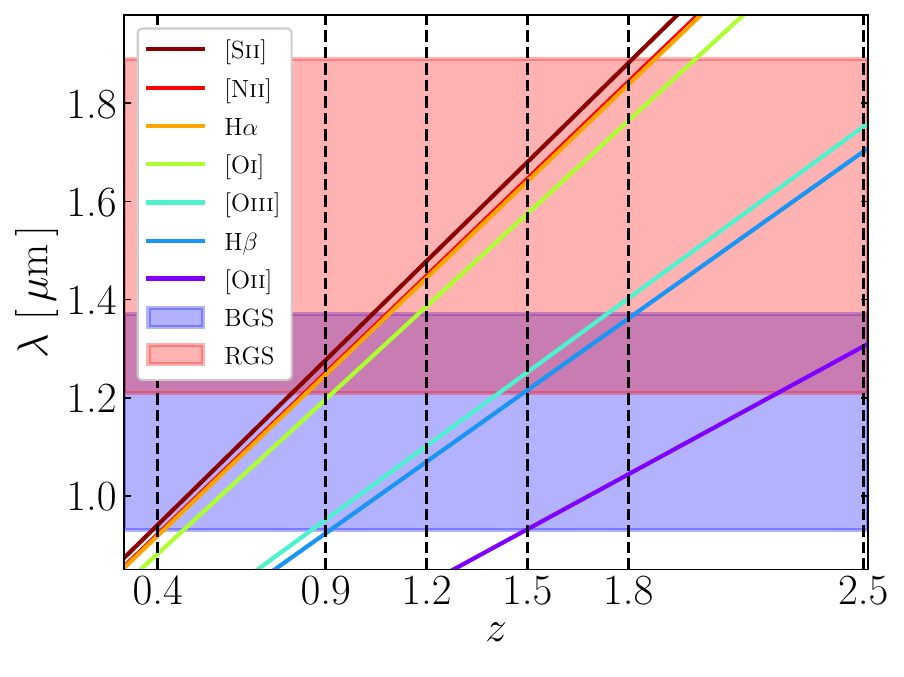}
    \caption{Redshifted wavelengths of strong emission lines in the rest-frame optical (crimson: [\ion{S}{ii}]$\lambda 6724$, red: [\ion{N}{ii}]$\lambda 6584$, yellow: H$\alpha$, green: [\ion{O}{i}]$\lambda 6300$, cyan: [\ion{O}{iii}]$\lambda 5007$, blue: H$\beta$, and purple: [\ion{O}{ii}]$\lambda 3727$)  and their resulting detectability using Euclid's blue grism (BGS, blue shaded area) and red grism (RGS, red shaded area). The redshift bins used for the following analysis (vertical dashed lines) were chosen according to the overlap of observed wavelengths with the BGS and RGS sensitivity ranges, 0.93--1.37\,\si{\micron} and 1.21--1.89\,\si{\micron}, respectively.}
    \label{fig:specgrisms}
\end{figure}

Before imparting on our analysis, we first illustrate at which redshifts key optical emission lines will be detectable with Euclid's red and blue grisms, given their respective sensitivity ranges. Figure \ref{fig:specgrisms} shows the spectral coverage of strong emission lines [\ion{S}{ii}] (crimson), [\ion{N}{ii}] (red), H$\alpha$ (yellow), [\ion{O}{i}] (green), [\ion{O}{iii}] (cyan), H$\beta$ (blue), and [\ion{O}{ii}] (purple) as function of redshift and where they overlap with the RGS (red area) and BGS (blue area) rest frame sensitivity ranges.

[\ion{S}{ii}], [\ion{N}{ii}], H$\alpha$, and [\ion{O}{i}] are all roughly detectable from redshift 0.4--0.9 in the BGS and then fall into the sensitivity range of the RGS until roughly redshift 1.8. Above redshift 1.8, they will not be detected by \Euclid. [\ion{O}{iii}] and H$\beta$, on the other hand, are detectable with the BGS at redshift 0.9--1.5, then with the RGS from 1.5--2.5. [\ion{O}{ii}] is detectable in the BGS from 1.5--2.5, while the RGS dectectability extends from 2.3--4.1, largely outside the range considered here.
Based on the overlap of these regimes, we chose the redshift bins 0.4--0.9, 0.9--1.2, 1.2--1.5, 1.5--1.8, and 1.8--2.5 (dashed vertical lines) as a basis for the following sections. Depending on the figure, we either focus on bins in which the chosen lines are detectable with \Euclid, sometimes combining multiple bins into one, or show predictions for all five bins across redshift 0.4--2.5, in order to demonstrate the physical evolution of the underlying emission-line properties, regardless of detectability with \Euclid.

\subsection{Fluxes and observability of line-emitting galaxies according to their redshift and stellar mass}
\label{sec:flux mstar}
\begin{figure*}[htbp!]
\centering
	\includegraphics[width=0.9\textwidth]{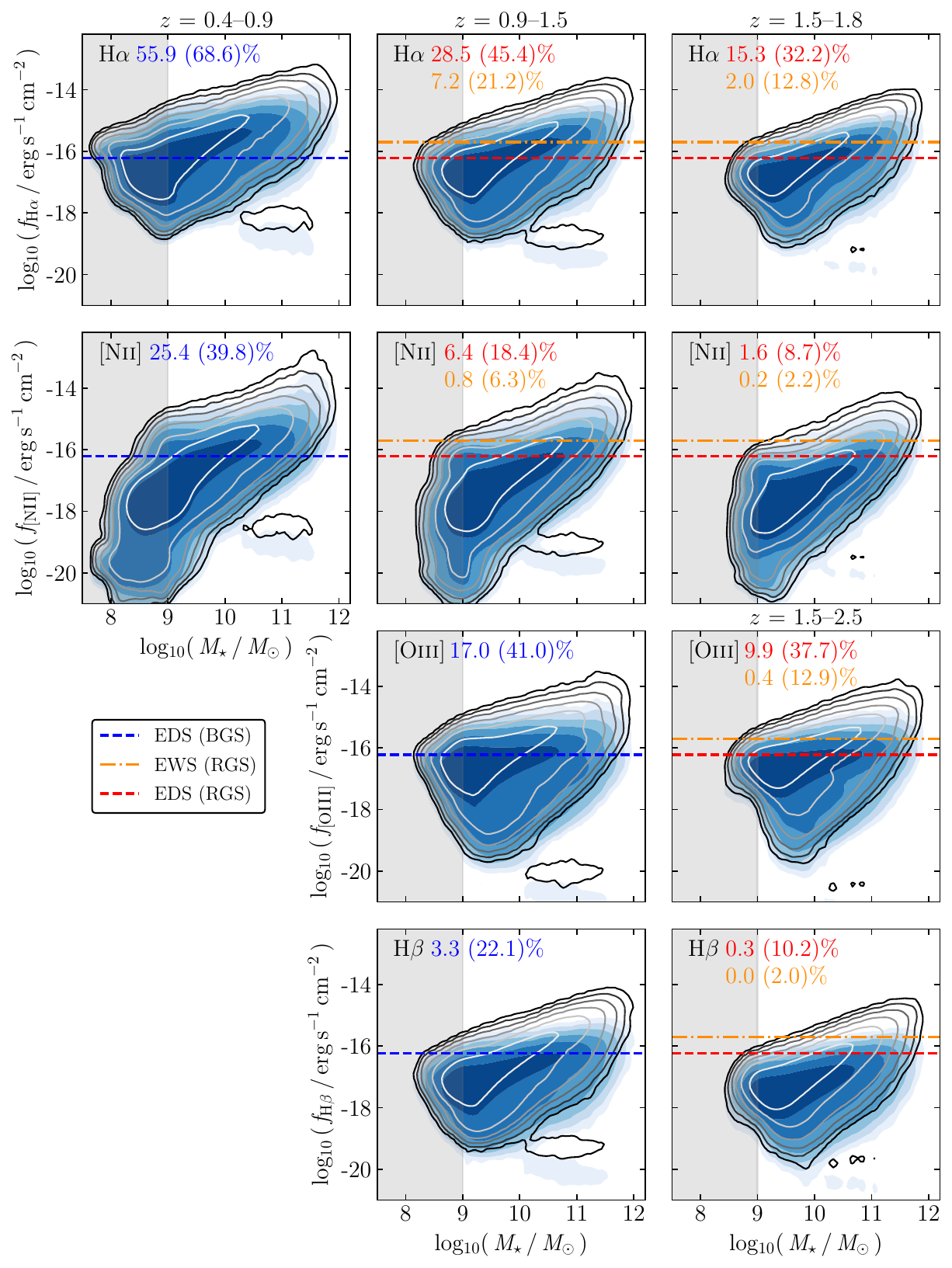}
    \caption{Location of intrinsic (grey contour lines) and dust-attenuated, but not flux-limited (blue contours) \textsc{Gaea-lc} galaxy populations in the H$\alpha$ (top row), [\ion{N}{ii}] (second row), H$\beta$ (third row), and [\ion{O}{iii}] (bottom row) line flux-stellar mass plane at different redshift ranges, following their observability with \Euclid's grisms given the line's wavelength (different columns as indicated by the legend). Overplotted are the flux limits of the blue and red grisms in the Deep Survey mode (EDS, blue and red dashed lines), and the red grism in the Wide Survey mode (EWS, yellow dash-dotted lines). Grey panels mark \textsc{Gaea}'s resolution limit of stellar masses below $10^{9}\,\si{\solarmass}$. Percentages indicate the observable fractions of the resolved sample above $10^{9}\,\si{\solarmass}$ when applying the EWS or EDS limit to the dust-attenuated emission-line fluxes (unattenuated fluxes in parentheses, colours reflecting survey mode and grism).}
    \label{fig:flux Mstar}
\end{figure*}

In Fig. \ref{fig:flux Mstar}, we visualise the location of \textsc{Gaea-lc} galaxies in the H$\alpha$ (first row), [\ion{N}{ii}] (second row), H$\beta$ (third row), and [\ion{O}{iii}] (fourth row) flux versus stellar mass plane. We note that the H$\alpha$ and [\ion{N}{ii}] lines will be blended in \Euclid data, however the line fluxes can be recovered using a simultaneous 3-Gaussian fit, as described in Sect. \ref{sec:discussion_inst}.

We contrast the intrinsic \textsc{Gaea-lc} sample (grey contour lines) with the dust-attenuated, but not flux-limited sample (blue contours). We show the line-emitting populations across the redshift bins for which the line is detectable (as indicated at the top of each column and the middle of the right column). The respective grism flux limits in the survey configurations are indicated, noting again that the blue grism (dashed blue lines) is only deployed in the EDS, while the red grism is active in both the EWS (yellow dash-dotted lines) and the EDS (red dashed lines). In the upper left corners, we indicate the observable percentage of dust-attenuated galaxies, according to the survey's grism flux threshold (colours corresponding to the configuration in the legend). Percentages for only flux-limited and unattenuated populations are shown in parentheses. We excluded galaxies below the conservative \textsc{Gaea} mass resolution limit of $10^{9}\,\si{\solarmass}$ (grey shaded area in plot) from this calculation.

For all emission lines, we note a clear stellar-mass dependence of the flux, with the most massive galaxies emitting the highest fluxes. This is due to higher SFRs in more massive galaxies, which increase the number of ionising photons. As a result, the observable galaxy populations will be dominated by massive galaxies. Our estimate of dust attenuation shifts the \textsc{Gaea-lc} sample toward lower fluxes, reducing the number of observable galaxies. Due to the stellar mass-dependence of the scaling law, this effect is more pronounced at higher masses, reducing fluxes by up to 0.6\,dex. Overall, this reduces the percentage of observable fluxes by 15-20\%. In the following, we refer to the observable percentages of the dust-attenuated line-emitting populations only.

Due to the distance dependence of line fluxes, observable percentages decrease significantly with increasing redshifts. While at redshift 0.4--0.9 we predict roughly 56\% of the dust-attenuated H$\alpha$-emitting population to lie above the EDS limit, this is reduced to 29\% at redshift 0.9--1.5 and 15\% at redshift 1.5--1.8. In the EWS, where H$\alpha$ is only detectable at redshift 0.9--1.8, these numbers are even further decreased to 7\% and 2\%, respectively. H$\alpha$ is the brightest line with the highest observable percentages, but the other three lines generally follow a similar evolution. [\ion{N}{ii}] is already detectable in the EDS configuration at redshift 0.4--0.9 with a predicted observability of around 25\%. It falls steeply to 6\% and then 1.6\% at redshift 0.9--1.5 and 1.5--1.8. In the EWS mode, observability is even lower in these regimes, with 0.8\% at  redshift 0.9--1.5 and 0.2\% at 1.5--1.8. This steep redshift decline of the observability of [\ion{N}{ii}]-emitters is likely not only caused by the physical reduction in flux due to the luminosity distance, but also by the intrinsically lower prevalence of strong [\ion{N}{ii}]-emission at higher redshift compared to low redshift. \citet{Hirschmann2017SyntheticRatios} explained the increased [\ion{N}{ii}]/H$\alpha$ ratio with decreasing redshift by inferring a higher gas-phase and stellar metallicity, as well as a decreased (s)SFR. High gas-phase metallicity boosts secondary production of nitrogen, while high stellar metallicity results in softer ionising radiation, which makes singly-ionised nitrogen ($\rm N^{+}$) more likely than multiply-ionised nitrogen. Similarly, a lower (s)SFR implies a lower ionisation parameter $U_{\star}$, which further favours $\rm N^{+}$ over higher ionisation states.   

In contrast, the observability of [\ion{O}{iii}] declines less steeply with redshift, from 17\% to 10\% between redshift 0.9--1.5 and 1.5--2.5, in agreement with the expectation of a greater number of bright [\ion{O}{iii}]-emitters at high redshift. Due to the rising metallicity toward low redshift, the electron temperature in the gas decreases, which disfavours collisional excitations into the [\ion{O}{iii}] state \citep{Gutkin2016ModellingGalaxies,Hirschmann2017SyntheticRatios}. In analogy with [\ion{N}{ii}], the softer ionising radiation of metal-rich stars and lower SFR in the recent Universe favours [\ion{O}{ii}] at the expense of [\ion{O}{iii}].
In the given ranges for H$\beta$, we predict 0.3--3.3\% above the flux limit in the EDS and basically none, except a few extreme objects, in the EWS. For all lines, we expect the full intrinsic populations to only be observable for massive galaxies with $\logten(M_{\star}/\si{\solarmass})$ above 11.5--11.7 in the EDS and above 11.7--11.9 in the EWS, depending on the line and redshift range.

At redshift 0.4--0.9, a significant portion of galaxies lie in the unresolved mass regime, while, as a result of the magnitude cut, the light cone contains fewer unresolved low-mass galaxies at high redshift. However, for all lines and at all redshifts, galaxies above the EDS flux boundaries are at most 1.5\% unresolved, such that we do not expect \textsc{Gaea}'s resolution limit to affect our predictions. 

Lastly, we point toward the population in the lower right corner of each panel, which is separate from the main population and appears to grow between redshift 1.8 and 0.9. This region contains the PAGB-dominated galaxies, which, with fluxes of $10^{-18}$--$10^{-20} \, \mathrm{erg} \, \mathrm{s}^{-1}\, \mathrm{cm}^{-2}$, lie far below both the EWS and EDS limits.

In conclusion, we predict that the observable populations of H$\alpha$, [\ion{N}{ii}], [\ion{O}{iii}] and H$\beta$-emitters are perhaps unsurprisingly biased towards the brightest and most massive galaxies, a trend which becomes more pronounced with increasing redshift. We expect \Euclid to observe around 56\% of H$\alpha$-emitters at redshift 0.4--0.9, and fewer than 30\% for the other three lines and covered redshift ranges, given our underlying galaxy sample with a mass resolution limit of $10^{9} \, \si{\solarmass}$ and magnitude cut of $m_H = 25$.

\begin{figure*}[htbp!]
\centering
	\includegraphics[width=\textwidth]{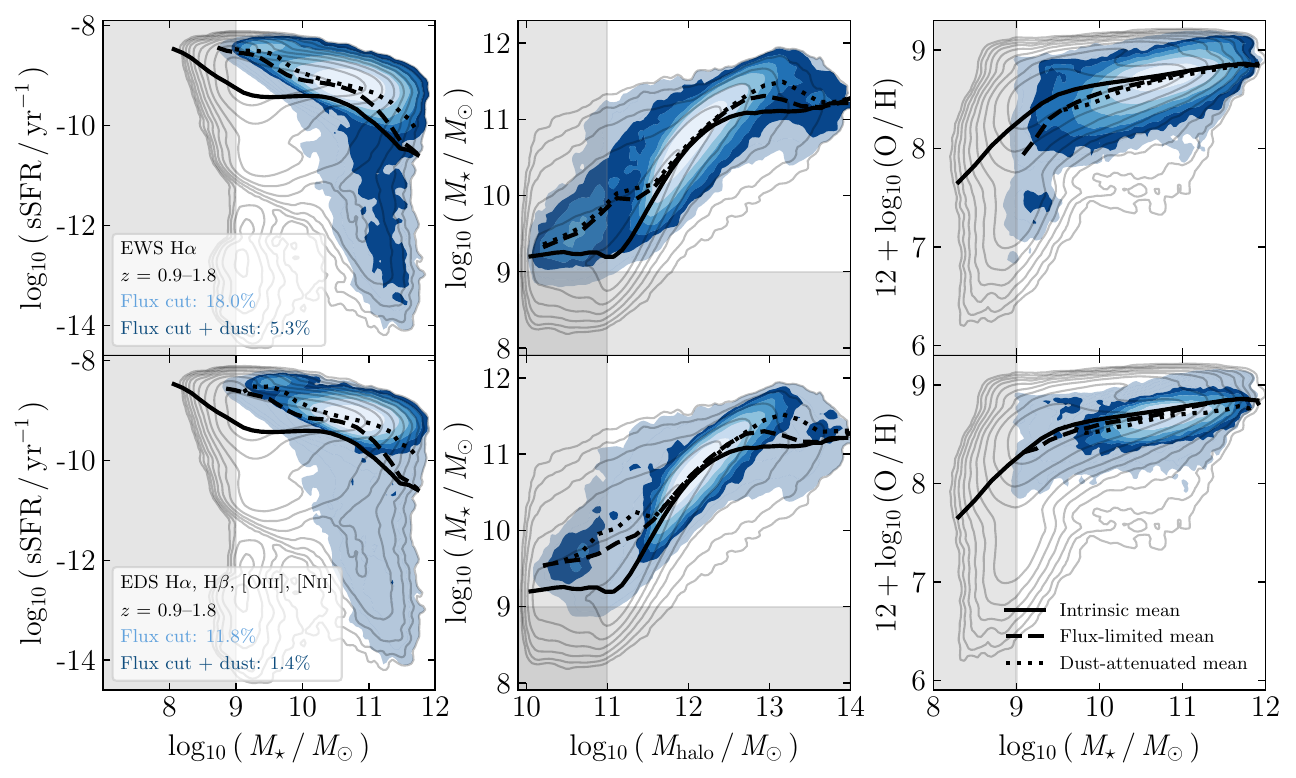}
    \caption{Scaling relations between sSFR and $\rm M_{\star}$ (left column), $\rm M_{\star}$ and $\mathrm{M_{halo}}$ (middle column), and O/H abundance versus $\rm M_{\star}$ (right column) for intrinsic (grey contour lines), flux-limited (light blue area), and dust-attenuated (blue contours) \textsc{Gaea-lc} populations in the redshift range 0.9--1.8. Shown are galaxies emitting H$\alpha$ fluxes above the EWS cut (top row) and H$\alpha$, H$\beta$, [\ion{O}{iii}], and [\ion{N}{ii}] fluxes above the EDS cut (bottom row). The redshift range reflects the entire detectable range for the chosen lines in the given survey configuration. Percentages in the left panels indicate the observable percentage of the flux-limited sample (light blue) and dust-attenuated sample (dark blue). Each panel also contains the mean of the intrinsic (solid lines), flux-limited (dashed), and dust-attenuated (dotted) \textsc{Gaea-lc} populations. Grey panels indicate \textsc{Gaea}'s resolution limits in stellar mass ($M_{\star} < 10^{9}\,\si{\solarmass}$) and halo mass ($M_\mathrm{{halo}} < 10^{11}\,\si{\solarmass}$). }
    \label{fig:selection}
\end{figure*}

\subsection{Biasing of standard scaling relations when observing line-emitters in the Euclid Wide and Deep Surveys}
To further explore the selection bias imposed by surveying different emission lines, we examine how they trace various standard scaling relations in Fig. \ref{fig:selection}.
We test two scenarios; surveying $\mathrm{H\alpha}$ emitters in the EWS mode (top row, hereafter EWS-H$\alpha$) and galaxies with simultaneous H$\alpha$, H$\beta$, [\ion{O}{iii}], and [\ion{N}{ii}] observability in the EDS (bottom row, hereafter EDS-BPT). The EDS-BPT configuration represents the sample for which the standard [\ion{O}{iii}]/H$\beta$ versus [\ion{N}{ii}]/H$\alpha$ BPT diagrams can potentially be used to determine the dominant ionising source. In both scenarios, the relevant lines are detectable between roughly redshift 0.9--1.8. We note that, in the EWS, the BPT lines will only be simultaneously detectable within the narrow redshift range 1.5--1.8.

For the $\mathrm{sSFR}$-$M_{\star}$ (left column), $M_{\star}$-$M_\mathrm{halo}$ (middle column), and $\mathrm{O/H\  abundance}$-$M_{\star}$ (right column) relations, we show the intrinsic (grey contour lines), flux-limited (light blue area), and the dust-attenuated and flux-limited (blue contours, hereafter just called dust-attenuated) \textsc{Gaea-lc} galaxy populations. 
For ease of comparison, we plot the mean relations for the intrinsic \textsc{Gaea-lc} sample (solid line), flux-limited (dashed line), and dust-attenuated (dotted line) populations. Additionally, we indicate the resolution limit in stellar masses, $M_{\star} < 10^{9}\,\si{\solarmass}$, and in halo masses, $M_{\rm halo} < 10^{11}\,\si{\solarmass}$ (grey shaded areas). In percentages in the left-most panels, we show the respective observability for resolved galaxies in the flux-limited (light blue) and dust-attenuated (dark blue) sample.

For the flux-limited populations, the predicted $\mathrm{sSFR}$-$M_{\star}$ relations are biased toward highly star-forming and massive galaxies. However, a small quiescent population of galaxies with $M_{\star} > 10^{11}\si{\solarmass}$ is present in both survey configurations. Applying the \citet{Calzetti2000TheGalaxies} law biases the remaining sample to even higher stellar masses and, in the EDS-BPT configuration, eliminates the quiescent population. The average $\mathrm{sSFR}$-$M_{\star}$ relation is increased by 0.5--0.8\,dex at fixed $M_{\star}$ for both configurations when compared to the average of the intrinsic population. In general, we expect the EDS-BPT configuration to recover a smaller percentage of galaxies. For the EWS-H$\alpha$ configuration the dust-attenuated (flux-limited) sample represents around 5.3\% (18\%) of intrinsic line-emitters, compared to 1.4\% (11.8\%) for the dust-attenuated (flux-limited) populations in the EDS-BPT configuration. 

For the $M_{\star}$-$M_\mathrm{halo}$ relation, we show the distribution of stellar masses for all central and satellite galaxies against the mass of their sub-halos. Enforcing only the flux limit, without accounting for dust, preferentially excludes galaxies with low halo and stellar masses resulting in similar distributions for both survey configurations. Dust-attenuated populations exhibit further restricted distributions, which, for the EDS-BPT configuration, is more closely constrained around the average relation and includes gaps around $M_{\mathrm{halo}} = 10^{11}$ and $10^{13}\,\si{\solarmass}$. The population between $M_{\mathrm{halo}} = 10^{11}$ and $10^{12}\,\si{\solarmass}$ is below the resolution limit and mostly represents satellites in subhalos. Therefore, the neighbouring gap to the main population is likely not of physical origin. The gap at $M_{\mathrm{halo}} = 10^{13}\,\si{\solarmass}$ is a product of small number statistics, as our sample contains only few halos of such high masses. For both survey configurations, the average relation in the flux-limited sample shows an increased $M_{\star}$ at low and high $M_{\mathrm{halo}}$ compared to the entire \textsc{Gaea-lc} sample. We note that the mean relations appear unbiased around $10^{12}\,\si{\solarmass}$, the turnover of the $M_{\star}$-$M_\mathrm{halo}$ relation. 

Lastly, due to the correlation of mass and metallicity, the strong bias in stellar mass introduces a bias toward high metallicties in the $\mathrm{O/H\ abundance}$-$M_{\star}$ relation. For the flux-limited sample, the EWS-H$\alpha$ configuration recovers some metal-poor galaxies below $\mathrm{\logten(O/H)+12} \sim 8$, whereas the EDS-BPT configuration is expected to only survey metal-rich galaxies. 
In the dust-attenuated samples, no metal-poor galaxies remain after the flux cut for both surveys, confining the distribution between $\mathrm{\logten(O/H)+12} \sim 8$ and 9. However, both biased mean relations are up to 0.15\,dex below the mean for the full sample, as the flux limit also excludes some of the most metal-rich galaxies. 

In conclusion, we expect the EWS-H$\alpha$ survey between redshift 0.9--1.8 to recover 5.3--18\% of the intrinsic population, depending on dust attenuation, while BPT lines should be recoverable for 1.4--11.8\% of galaxies in the EDS. In both observing scenarios, the resulting samples will be biased toward massive, highly star-forming, and metal-rich systems. 

\begin{figure}[]
\centering
	\includegraphics[width=\columnwidth]{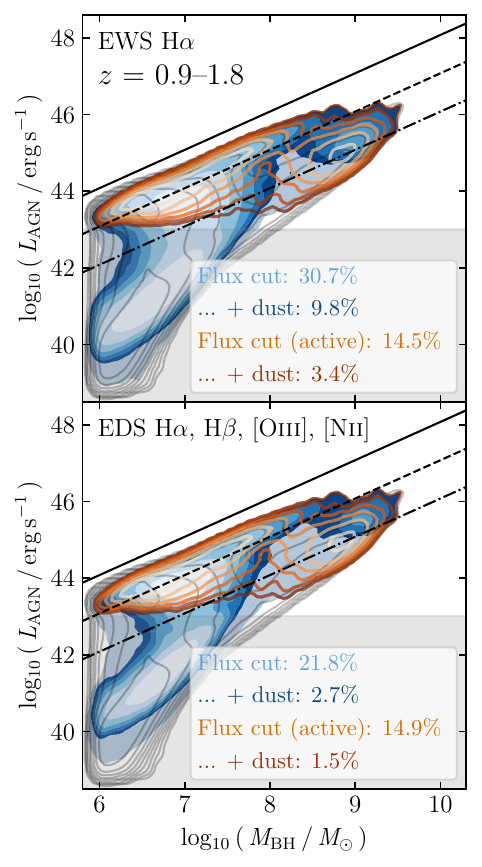}
    \caption{Scaling relation between the $L_\mathrm{AGN}$ and black hole mass $\mathrm{M_{BH}}$ for \textsc{Gaea-lc} populations in the redshift range 0.9--1.8, following the same layout as Fig. \ref{fig:selection}. Additionally shown is the active sub-sample of the dust-attenuated population (orange contours) and its observability compared to the entire active population (orange percentages). For comparison, black lines show fractions of the Eddington limit: 1 $L_{\mathrm{edd}}$ (solid), 0.1 $L_{\mathrm{edd}}$ (dashed), and 0.01 $L_{\mathrm{edd}}$ (dash-dotted).}
    \label{fig:MBH LAGN}
\end{figure}

We further explore the prevalence of luminous AGN in the two survey configurations by examining the AGN luminosity $L_\mathrm{AGN}$ versus black hole mass $M_\mathrm{BH}$ relation at redshift 0.9--1.8 in Fig. \ref{fig:MBH LAGN}. 
To ensure a physically robust sample of AGN, we limit the selection to galaxies containing a black hole with a mass greater than the seed mass ($M_\mathrm{BH}>10^{5}\, \si{\solarmass}$) and a significant luminosity output ($L_\mathrm{AGN}> \mathrm{10^{39} \,erg \,s^{-1}}$).
The general layout follows the same as in Fig. \ref{fig:selection}; we plot the intrinsic (grey contours), flux-limited (light blue area), and dust-attenuated (blue contours) populations for the EWS-H$\alpha$ (top panel) and the EDS-BPT (bottom panel) survey configurations. Percentages in the bottom right corner of both panels mark the observability in corresponding colours. As we expect the upper end of $L_\mathrm{AGN}-M_\mathrm{BH}$ to be set by the most powerful AGN, we additionally show the flux-limited population for active galaxies only (orange contour lines) and the corresponding observable percentage compared to the intrinsic active population (light orange). The distribution of the dust-attenuated active population (not explicitly indicated here) constitutes the overlap of the full dust-attenuated sample and the flux-limited active sample. In the corner, we include the predicted observable percentage of this population (dark orange).
We compare the results to relations expected for black holes accreting at fractions of the Eddington luminosity; 1 $L_{\mathrm{edd}}$ (solid black line), 0.1 $L_{\mathrm{edd}}$ (dashed line), and 0.01 $L_{\mathrm{edd}}$ (dash-dotted line). We estimate that the \textsc{Gaea} $L_\mathrm{AGN}$ convergence limit starts affecting our predictions below $10^{43} \mathrm{\,erg \,s^{-1}}$ (grey shaded area, see also Fig. 2 in \citealt{Fontanot2020TheModel}).

According to our framework, both the EWS-H$\alpha$ and the EDS-BPT survey configurations will recover galaxies containing AGN with black hole masses between $10^6$ and $10^{9.5} \,\si{\solarmass}$. The distributions of the flux-limited populations occupy almost the same regions as the intrinsic sample. Adding dust-attenuation biases the distributions slightly by excluding galaxies with $L_\mathrm{AGN}$ below $10^{39} \mathrm{\,erg \,s^{-1}}$. Additionally, a notable gap appears below the resolution limit for galaxies with $M_\mathrm{BH} \sim 10^{6}\, \si{\solarmass}$ and $L_\mathrm{AGN} \sim \mathrm{10^{42} \,erg \,s^{-1}}$. This gap is caused by the bimodal prescription for the AGN luminosity in \textsc{Gaea} distinguishing between accretion at low and high Eddington fractions (see \citealt{Fontanot2020TheModel}). As before, the distribution of the dust-attenuated EDS-BPT sample appears more restricted than the EWS-H$\alpha$ sample.

Regarding predicted observabilities, we note that the percentages for the flux-attenuated and dust-attenuated populations are larger compared to the numbers quoted in Fig. \ref{fig:selection}. By only including galaxies with a luminous black hole more massive than \textsc{Gaea}'s seed masses, we biased our initial intrinsic sample toward brighter and more massive galaxies. As in Fig. \ref{fig:selection}, the EWS-H$\alpha$ configuration is expected to produce higher observable percentages at 9.8\% (30.7\%) when considering the dust-attenuated (flux-limited) sample compared to 2.7\% (21.8\%) in the EDS-BPT populations. 

Active galaxies in the flux-limited populations produce high AGN luminosities with $L_\mathrm{AGN}$ between $10^{43}$ and $10^{46} \mathrm{\,erg \,s^{-1}}$, entirely above the resolution limit, and fall between the expected relations for accretion at Eddington fractions of 0.01--1 (see \citealt{Fontanot2020TheModel} for the implementation of black hole growth and feedback in \textsc{Gaea}). Between the two survey configurations, the distributions appear almost identical and produce similar observable percentages. At 1.5--3.4\%, we predict the observability of the dust-attenuated active sub-sample to be low in both survey configurations. This is because the H$\alpha$ and H$\beta$ lines are particularly faint for active galaxies in this regime (also see Sect. \ref{sec:no count}) and thus only the galaxies with the brightest AGN luminosities produce detectable emission-line intensities. 

Nevertheless, we predict line-emitting galaxies in the EWS-H$\alpha$ and EDS-BPT survey configurations to contain AGNs with a wide range of masses and luminosities. In order to understand the impact on their host galaxies and disentangle observational signatures, it is thus important to be able to distinguish between SF-dominated and AGN-dominated galaxies. 

\begin{figure*}[htbp!]
\centering
	\includegraphics[width=0.95\textwidth]{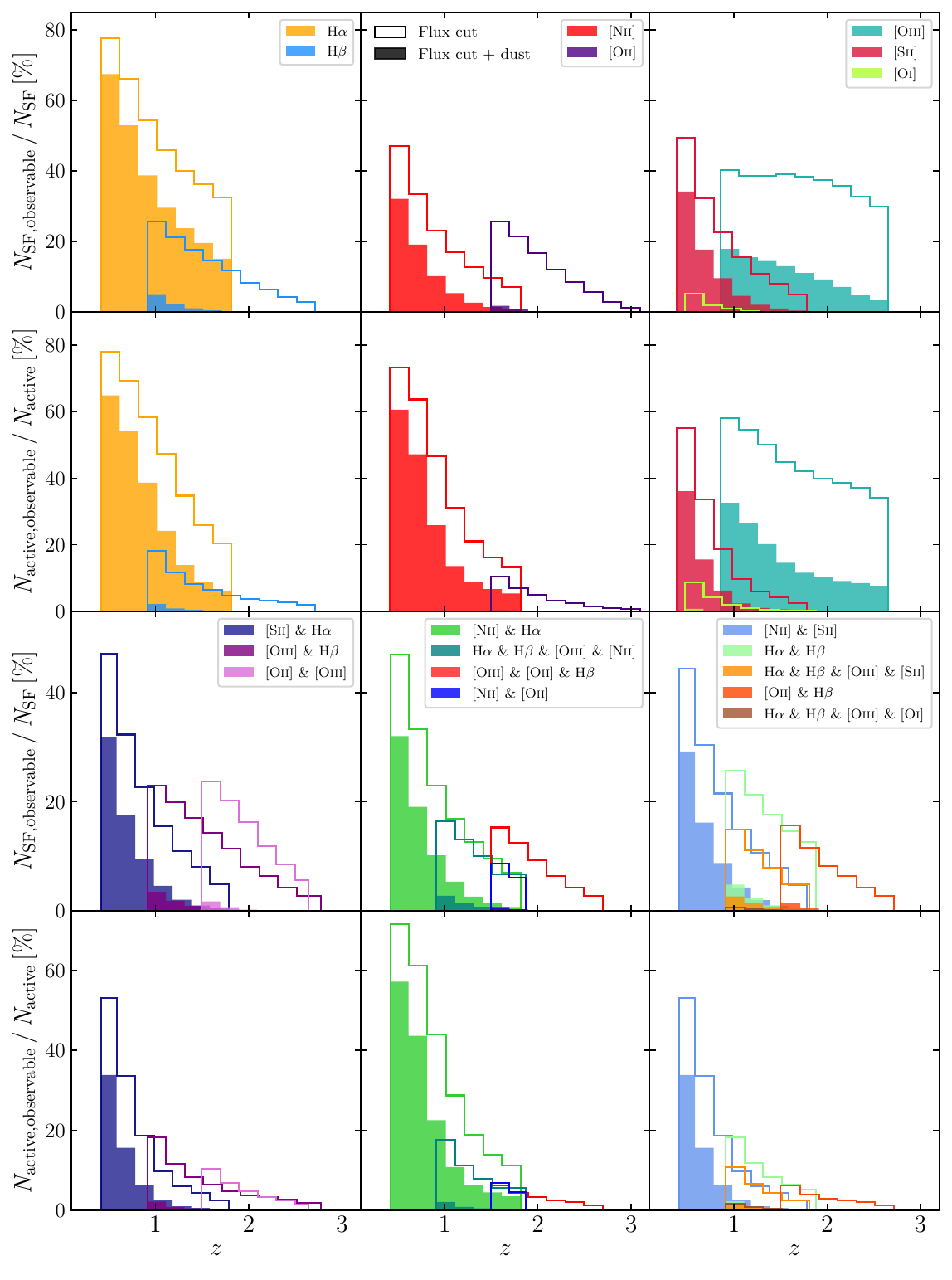}
    \caption{EDS observable fractions of SF (row 1 and 3) and active (row 2 and 4) galaxies in different redshift bins. Shown are the flux-limited (outlined histograms) and dust-attenuated (filled histograms) populations emitting H$\alpha$ (yellow),  H$\beta$ (light blue), [\ion{N}{ii}]$\lambda 6584$ (red), [\ion{O}{ii}]$\lambda 3727$ (purple), [\ion{O}{iii}]$\lambda 5007$ (cyan), [\ion{S}{ii}]$\lambda 6724$ (crimson),  [\ion{O}{i}]$\lambda 6300$ (green yellow), and their combinations (colours indicated in row 3 panels). }
    \label{fig:no count hists}
\end{figure*}

\subsection{Predicted observability of star-forming and active line-emitting galaxies in the EDS}
\label{sec:no count}
In Fig. \ref{fig:no count hists}, we further assess the redshift distribution of line-emitting galaxies observable in the EDS, divided into SF-dominated (first and third row) and active (second and fourth row) populations. 
We show the \textsc{Gaea-lc} prediction for observable percentages of the flux-limited (outlined histograms) and dust-attenuated (filled histograms) line-emitting populations in different redshift bins. The quoted percentages are with respect to our intrinsic \textsc{Gaea-lc} population, which includes a mass resolution cut of $\rm 10^{9} \si{\solarmass}$ and $H$-band magnitude cut of 25. Compared to previous figures, we chose a smaller bin spacing of $\Delta z = 0.2$ in order to sample the detectable redshift ranges of the emission lines more finely. To account for uneven redshift ranges, we allow the highest redshift bin to be larger than $\Delta z = 0.2$.
We display the redshift distributions as histograms for individual emisssion lines (row 1 and 2); H$\alpha$ (yellow),  H$\beta$ (light blue), [\ion{N}{ii}] (red), [\ion{O}{ii}] (indigo), [\ion{O}{iii}] (cyan), [\ion{S}{ii}] (crimson), and [\ion{O}{i}] (green-yellow), as well as combinations of them (row 3 and 4, colours as indicated in the legend) frequently used in spectroscopic diagnostics. We also include histograms for the combination of the H$\alpha$ and H$\beta$ lines (mint green), whose ratio, the Balmer decrement, can be used to estimate dust attenuation. 

Both SF and active line-emitting galaxies exhibit decreased observability percentages with increasing redshift due to the distance-dependence of observed fluxes. We note that, overall, observability declines more steeply within active line-emitting populations. Flux-limited histograms represent an upper limit of observable percentages of line-emitters, which will be drastically reduced by the presence of dust. Estimating the precise impact of this effect is challenging (see Sect. \ref{sec:discussion_dust}), thus exact numbers for observable percentages of dust-attenuated line-emitters should be treated with caution. Overall, we estimate that, based on the \citet{Calzetti2000TheGalaxies} law with mass-dependent scaling, dust attenuation will decrease the observable percentages by a further 20--30\% with respect to the intrinsic \textsc{Gaea-lc} populations. This could particularly be a problem when trying to recover fainter lines, for which predicted percentages are reduced to single digits for lower redshifts and below 1\% at higher redshifts. Additionally, wavelengths toward the blue end of the spectrum are more strongly attenuated than those toward the red end. In the following descriptions we refer to the dust-attenuated sample and include the flux-limited case as an upper limit in parentheses. 

We first consider the observability of individual emission lines in SF and active galaxy populations. As noted before, H$\alpha$ is by far the strongest emission line out of the ones presented here. At redshift 0.4--0.6, we expect H$\alpha$-emitters to be around 70\% (reduced from 80\%) observable for both SF and active galaxies in the dust scenario. Until redshift 1.6--1.8, this number decreases to around 15\% (30\%)  for SF and 7\% (20\%) for active galaxies, before the H$\alpha$ wavelength is redshifted out of the sensitivity range of the EDS. 

[\ion{N}{ii}] is the second strongest emission line in these populations, reaching observable percentages of up to 30\% (50\%) in the SF galaxies and 60\% (70\%) in the active sample. This difference is due to the relatively high ionisation potential of nitrogen, which is more easily ionised by the harder ionising radiation from AGN compared to young stellar populations. Furthermore, we expect the increased observability at low redshift to be partially caused by the global decrease of SFR, as well as the increase of metallicity, which strengthens [\ion{N}{ii}] emission due to secondary nitrogen production. 

In both the SF and active populations, [\ion{S}{ii}]-emitters show similar observability percentages to [\ion{N}{ii}]-emitters. They decline from 35\% (50\%) to below (10\%) 1\% between redshift 0.4--0.6 and 1.6--1.8. This is due to the dependence of [\ion{S}{ii}] on SFR and interstellar metallicity, in a similar way to [\ion{N}{ii}], which is expected to increase emissions in low-redshift galaxies. However, due to the absence of any secondary production channels for sulfur, this dependence is weaker. 

In contrast to line-emitting galaxies discussed above, the observability of [\ion{O}{iii}]-emitters evolves to a lesser degree with redshift. For SF galaxies, [\ion{O}{iii}] observability is between roughly 5--20\% for dust-attenuated populations, whereas only flux-limited percentages stay between 30--40\%. This can be explained by the increased [\ion{O}{iii}] emission for metal-poor galaxies, which are more prevalent at high redshift and thus compensate for the flux decrease due to increasing luminosity distance. The observable percentages of active [\ion{O}{iii}]-emitters also appear constant above redshift 1.6, which is likely caused by a larger ionisation parameter due to elevated central gas densities (see \citealt{Hirschmann2017SyntheticRatios}).

H$\beta$, [\ion{O}{ii}], and especially [\ion{O}{i}] emission are notably fainter for the \textsc{Gaea-lc} populations. The predicted dust attenuation makes most of the resulting histogram bins barely visible for these lines. Around redshift 1, H$\beta$ reaches about 20-30\% observability for the flux-limited sample of SF and active galaxies, while the other lines are even fainter. If our estimate is reasonable, dust would reduce the percentage of EDS-observable H$\beta$, [\ion{O}{ii}], and [\ion{O}{i}]-emitters to a few per cent at the lower redshift end and below 1\% at the higher redshift end.

We now turn our attention to the predicted observability of galaxies emitting multiple lines. The redshift bins have been adjusted according to the overlap of the relevant redshifted wavelengths with the EDS sensitivity range. As H$\alpha$, [\ion{N}{ii}], and [\ion{S}{ii}] are the strongest lines, the percentages for their simultaneous observability are also the highest. All combinations reach around 30\% (50\%) observability at redshift 0.4--0.6 in both the SF and active populations, except for [\ion{N}{ii}] with H$\alpha$, which reach close to 60\% (70\%) observability for active galaxies.

All other emission-line combinations contain at least one line that appears particularly weak in this redshift range. As a result, predicted observable percentages for the dust-attenuated sample drop below 1\% at the higher end of their redshift ranges. At the lower redshift end, we predict generally less than 5\% observability. Even though both SF and active [\ion{O}{iii}]-emitters themselves are at least 20--30\% observable around redshift 1, in combination with H$\beta$ and/or [\ion{O}{ii}], observability stays below 5\%. This represents a reduction from around 20\% observability for flux-limited populations. [\ion{O}{ii}] with H$\beta$ suffers similar losses due to dust attenuation: from 15\% to less than 5\% for SF galaxies at redshift 1.5 and 6\% to less than 1\% for active galaxies. [\ion{N}{ii}] with [\ion{O}{ii}] is also reduced from around 10\% to less than 1\%. Additionally, the wavelengths of [\ion{N}{ii}] and [\ion{O}{ii}] are spaced far apart, which significantly restricts the redshift range in which they can both be detected with \Euclid's grisms. This also affects the usefulness of emission-line calibrations based on these lines.

In the upper limit of the flux-limited sample, the combination of H$\alpha$ and H$\beta$ shows percentages between 10--25\% for SF galaxies and 5--20\% for active galaxies across redshift 0.9--1.8. The estimated dust-attenuation reduces these numbers drastically to at most 5\% for SF galaxies and 2\% for active galaxies. This is particularly due to the relatively stronger effect on the H$\beta$ line, caused by its shorter wavelength. This results in an increased Balmer decrement ($\rm H \alpha/\rm H\beta$), which can be used to estimate the presence of dust in observed galaxies, if the H$\beta$ line can still be recovered.

Galaxy populations emitting lines which make up the standard BPT diagrams, [\ion{O}{iii}], H$\beta$, and H$\alpha$ in combination with [\ion{N}{ii}], [\ion{S}{ii}], or [\ion{O}{i}], will roughly be observed between redshift 0.9--1.8. Upper limits without dust show around 5--15\% observability across this range for the [\ion{N}{ii}] and [\ion{S}{ii}]-BPT combinations. In the case with dust, we predict that for both SF and active galaxies, the [\ion{N}{ii}]-BPT combination will only be roughly $3\%$ observable at redshift 0.9--1.1 and decline to around 0.3\% observability at redshift 1.5--1.8. These numbers are in agreement with the predicted observability of 1.4\% for the EDS-BPT configuration in Fig. \ref{fig:selection}, which was averaged across all galaxies in the 0.9--1.8 redshift range. The [\ion{S}{ii}]-BPT line combination shows percentages of around 2\% at most for both SF galaxies and the active population. As the [\ion{O}{i}] line is already so weak by itself, the [\ion{O}{i}]-BPT combination is never more than 1.5\% observable in even the upper limit of the flux-limited populations.

In conclusion, we expect the presence of dust to reduce the number of galaxies with recoverable emission lines drastically, typically by an additional 20\% with respect to the intrinsic population. This is particularly significant for intrinsically fainter lines, which will be difficult to measure already at the low end of their detectable redshift range. For H$\beta$, [\ion{O}{ii}], and [\ion{O}{i}] emission we do not expect to reach more than 5\% observability of SF and active galaxies. This is the case both for these lines individually and in combination with other lines. Already in the upper limit of the no dust case, we predict percentages of 10--25\% at most. For stronger lines, like H$\alpha$, [\ion{N}{ii}], [\ion{S}{ii}], and [\ion{O}{iii}], percentages are relatively high for flux-limited populations, and, as a result, dust losses are less fatal. Without dust, we recover percentages between 40--80\% at lower redshifts and around 10--30\% at the higher redshifts. After accounting for dust, we predict the EDS to recover up to 30--70\% of both SF and active galaxies at lower redshifts, while at higher redshifts these numbers decline to below 10\%. This is similarly true for all combinations of H$\alpha$, [\ion{N}{ii}], and [\ion{S}{ii}].

While some of these numbers appear discouragingly low, we stress that with a planned sky coverage of 50\, $\mathrm{deg}^2$, the EDS will nevertheless recover thousands to hundreds of thousands of SF and active galaxies emitting these emission lines and their various combinations. Detailed number count predictions for different line-emitters will be published in Mancini et al. (in prep.), who are comparing both the \textsc{Gaea-lc} framework and the Millennium Mambo catalogues. Furthermore, we reiterate that the exact prevalence and nature of dust in this redshift regime is fundamentally uncertain and warrants an extensive study in itself. Hence, these estimates should be considered carefully. 

\begin{figure*}[htbp!]
\centering
	\includegraphics[width=\textwidth]{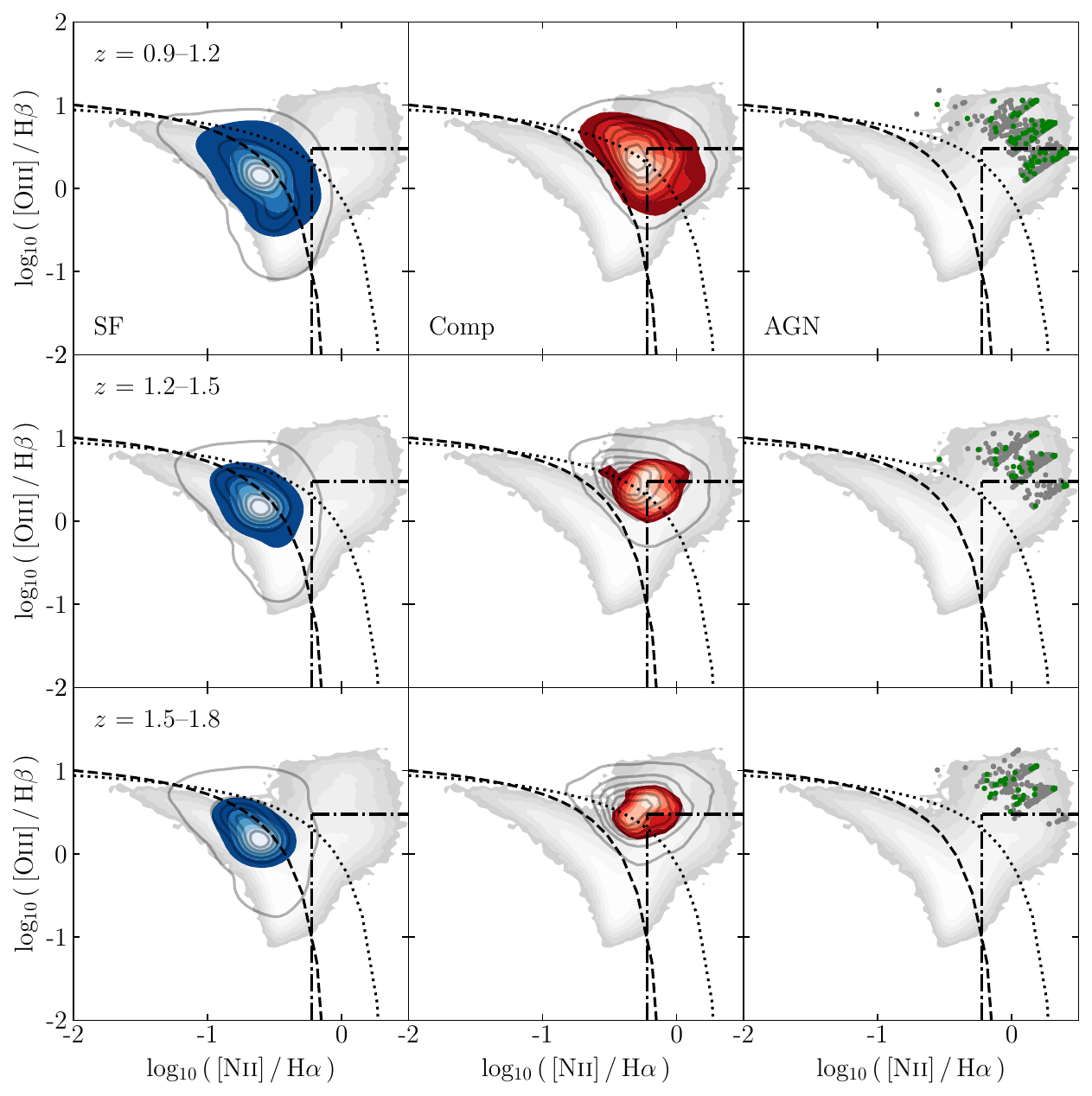}
    \caption{Location of EDS-observable \textsc{Gaea-lc} galaxy populations in the [\ion{O}{iii}]/H$\beta$ versus [\ion{N}{ii}]/H$\alpha$ BPT diagram, divided into redshift bins 0.9--1.2 (top row), 1.2--1.5 (middle row), and 1.5--1.8 (bottom row). Shown are the flux-limited (grey contour lines and data points) and dust-attenuated samples (coloured contours and data points). The columns and colour coding for different dominant ionising sources, the empirical selection criteria to distinguish between them and the SDSS-observed galaxies, shown for comparison, follow the same layout as Fig. \ref{fig:BPT lowz}.}
    \label{fig:BPT Euclid}
\end{figure*}

\section{Distinguishing between dominant ionising sources in galaxy populations observable in the Euclid Deep Survey}
\label{sec:BPT}
Measurable line intensities are particularly valuable for their potential to characterise observed sources using spectroscopic diagnostics. In these remaining Sections, we will verify which diagnostics can be applied to future \Euclid data. 

As explained in Sect. \ref{sec:intro}, it is unclear whether standard BPT diagrams are a robust diagnostic to separate SF-dominated, composite, and AGN-dominated galaxy populations beyond redshift 1 \citep{Hirschmann2019SyntheticSources,Hirschmann2023Emission-lineJWST,Kocevski2023HiddenCEERS}. A successful application to \Euclid-observable populations, in particular, also depends on the properties of these galaxies, as theoretical works \citep[e.g.][]{Groves2006Emission-lineAGN,Feltre2023OpticalGalaxies,Hirschmann2019SyntheticSources} indicate that the interstellar metallicity of galaxies, in addition to the type of ionising radiation, influences their location on the standard [\ion{O}{iii}]$ / \mathrm{H} \beta$ versus [\ion{N}{ii}]$ / \mathrm{H} \alpha$ BPT diagram.

In Fig. \ref{fig:BPT Euclid}, we thus verify if the EDS-BPT sample of \textsc{Gaea-lc} galaxies at redshift 0.9--1.8 conforms to locally calibrated optical selection criteria for the [\ion{O}{iii}]$ / \mathrm{H} \beta$-versus-[\ion{N}{ii}]$ / \mathrm{H} \alpha$ BPT diagrams. As in Sect. \ref{sec:BPT validation}, we divide the \textsc{Gaea-lc} sample into different types according to the BHAR/SFR criterion, resulting in SF-dominated (left column), composite (middle column), and AGN-dominated galaxies (right column). In order to assess a potential evolution of the location of different galaxy types in the diagram, we split the sample into the redshift bins 0.9--1.2 (top row), 1.2--1.5 (middle row), and 1.5--1.8 (bottom row).
The general layout and colour-code is the same as in the top panels in Fig. \ref{fig:BPT lowz}, except that now showing the flux-limited (grey contour lines and data points) and dust-attenuated populations (coloured contours and data points) for each galaxy type. As in Fig. \ref{fig:BPT lowz}, we show individual data points for the AGN-dominated sample to avoid artefacts caused by the discrete model grids and the small number of AGN-dominated galaxies surpassing the flux limit.

For both the flux-limited and dust-attenuated cases, the predicted galaxy populations conform well to locally calibrated criteria \citep{Kewley2001OpticalGalaxies,Kauffmann2003TheAGN}. We note that, compared to the flux-limited populations, the dust-attenuated populations exhibit reduced ranges of $[\ion{O}{iii}]/\mathrm{H}\beta$ and $[\ion{N}{ii}]/\mathrm{H}\alpha$. This trend gets more pronounced with increasing redshift. While there is some overlap of the SF-dominated galaxies into the transitional composite region, we note that the majority of galaxies is confined below the criterion by \citet{Kauffmann2003TheAGN}. For the SF-dominated populations, we also see a slight increase with redshift of $\rm [\ion{O}{iii}]/H\beta$ at fixed $\rm [\ion{N}{ii}] / \mathrm{H} \alpha$, consistent with Fig. \ref{fig:BPT lowz}. Nevertheless, they remain well-separated from AGN-dominated galaxies at all redshifts shown here. 

In contrast, \citet{Hirschmann2019SyntheticSources}, who use the same photoionisation models, found that a global increase of $\rm [\ion{O}{iii}]/H\beta$ and decrease of $\rm [\ion{N}{ii}] / \mathrm{H} \alpha$ with redshift makes SF and active galaxies indistinguishable already at redshift 1. They attribute this to a drop in interstellar metallicity (in addition to a rise in SFR, also see Sect. \ref{sec:flux mstar}), but demonstrate that metal-rich galaxy types remain separable above redshift 1. From Fig. \ref{fig:selection} we know that the EDS-BPT selected \textsc{Gaea-lc} population is biased toward high oxygen abundances, which explains why, for this sample, the galaxy types remain distinguishable.

We conclude that the [\ion{N}{ii}]-BPT diagram is a reliable diagnostic diagram to distinguish between dominant ionising sources in the EDS-observable galaxy populations at redshift 0.9--1.8. However, as seen in Sect. \ref{sec:selection bias}, the relevant sample of galaxies represents only a few per cent of the predicted intrinsic population. Additionally, we remind the reader that de-blending $[\ion{N}{ii}]$ and $\mathrm{H}\alpha$ in \Euclid spectra will be challenging (see Sect. \ref{sec:discussion_inst}), which will further reduce the sample for which diagnostics relying on separable $[\ion{N}{ii}]$ and $\mathrm{H}\alpha$ estimates can be used.

\begin{figure}[htbp!]
\centering
	\includegraphics[width=0.87\columnwidth]{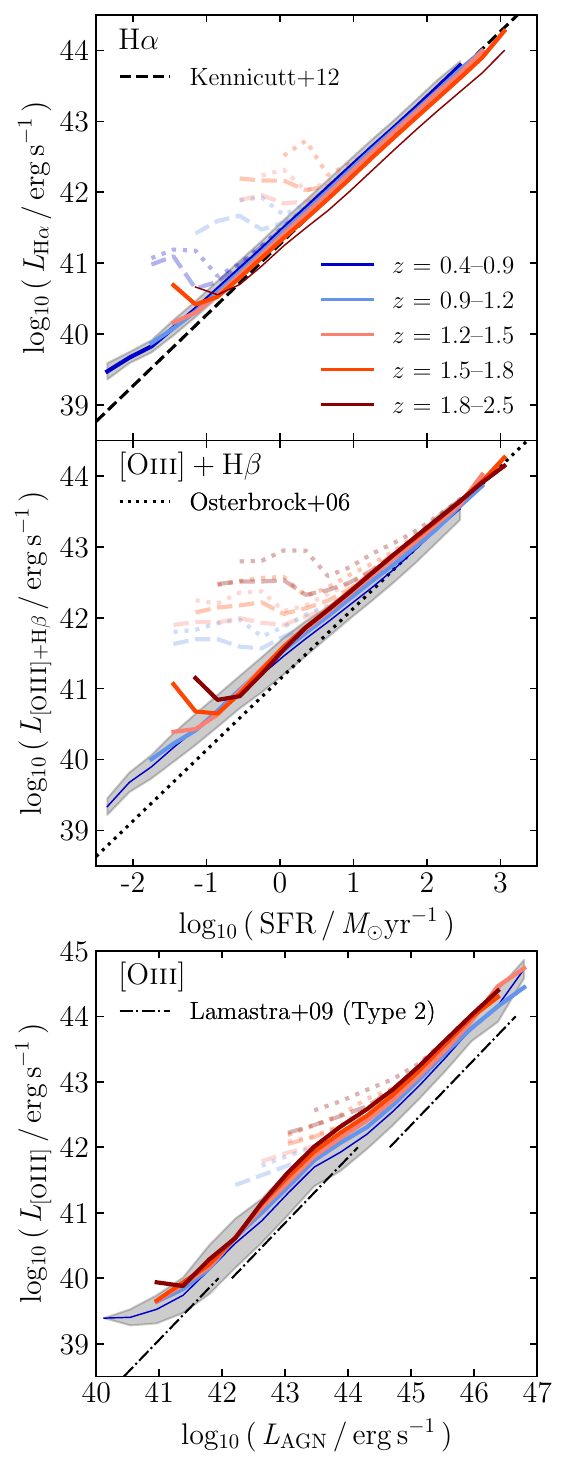}
    \caption{Average H$\alpha$ (top panel) and $\mathrm{[\ion{O}{iii}]} + \mathrm{H}\beta$ (middle panel) line luminosities versus SFR for \textsc{Gaea-lc} star-forming galaxies in different redshift intervals (shown in different coloured solid lines as indicated by the legend). For completeness, the three relations are shown for the same redshift intervals, but thick lines indicate that the emission line is within the EDS wavelength sensitivity range. For those, the flux-limited (dashed lines) and dust-attenuated relations (dotted lines) are plotted in fainter colours. As in Fig. \ref{fig:SFR+LAGN lowz}, predictions are compared to local calibrations \citep{Kennicutt2012,Osterbrock2006AstrophysicsNuclei}. 
    In the bottom panel, the average [\ion{O}{iii}] line luminosity is plotted against AGN luminosity for active \textsc{Gaea-lc} galaxies and the relation found by \citet[dash-dotted]{Lamastra2009TheSources}. The coloured lines for different redshift bins follow the same key as above.}
    \label{fig:SFR+LAGN highz}
\end{figure}

\section{Deriving ionising properties from strong line luminosities at intermediate redshifts}
\label{sec: galaxy prop}
Optical emission-line intensities, and their ratios, are also widely used as proxies to estimate key galaxy properties, as they encode signatures from both the local ISM gas, as well as the nature of the ionising radiation. 
Using observational data and photoionisation models, various emission-line calibrations have been derived for ionising properties, like the SFR and AGN luminosity. While they have been validated for the interpretation of observational spectra at low redshift, it is unclear if their use can be extended beyond the local Universe. In the more distant Universe, the ionisation and ISM conditions might differ significantly which could cause the locally assumed relationship between emission lines and galaxy properties to break down.

In this Section, we thus aim to examine the potential evolution of locally-established calibrations across the intermediate redshift range 0.4--2.5, where \Euclid will recover the majority of the relevant emission lines. This is meant as a guide for future data releases to determine which, if any, of the widely used calibrations could be used to characterise sources. Additionally, we present new calibrations which relate [\ion{N}{ii}]-based emission-line ratios to the $\rm BHAR/SFR$ ratio.

As here we focus on the intrinsic physical limitations of these relations predicted by our framework, we show results from the intrinsic \textsc{Gaea-lc} population, as well as indicate potential biasing as a result of \Euclid-like flux cuts.

\subsection{Deriving SFR and AGN Luminosity}
\label{sec:SFR+AGN}
The SFR and AGN luminosity play a crucial role in shaping the evolution of a galaxy across cosmic time. They are also the main drivers of ionising radiation, thus setting the degree of ionisation of the ISM gas and the strength of different emission lines.
In Fig. \ref{fig:SFR+LAGN highz}, we show average H$\alpha$ and $\rm [\ion{O}{iii}]+H\beta$ emission-line luminosities versus the SFR for SF galaxies (top and middle panels) and the [\ion{O}{iii}] versus AGN luminosity for active galaxies (bottom panel). We follow the layout of Fig. \ref{fig:SFR+LAGN lowz}, but now extend our predictions to higher redshift intervals between redshift 0.4 and 2.5 (solid lines, colours indicated in legend), with a scatter of 1 standard deviation for the 0.4--0.9 bin (grey shaded area). For completeness, we plot the emission-line relations in all redshift bins (0.4--0.9, 0.9--1.2, 1.2--1.5, 1.5--1.8, and 1.8--2.5) and indicate where the emission line is within \Euclid's sensitivity range with thicker lines. For these lines we further show the impact of the EDS flux limit (dashed lines in fainter colours) and dust attenuation (dotted lines in fainter colours).

We note only a negligible, if any, evolution for all relations, meaning they continue to broadly follow locally used relations up to at least redshift 1.8--2.5. In fact, the H$\alpha$ relation, which was offset at redshift 0--0.3, approaches the \citet{Kennicutt2012} relation more closely at higher redshifts. This evolution is caused by a systematically higher ionisation parameter at higher redshifts, which implies a larger \ion{H}{ii} region and an associated greater dust attenuation. The predicted $L_\mathrm{[\ion{O}{iii}]+H\beta}$-SFR relation exhibits no clear evolution with redshift and continues to agree well with the \citet{Osterbrock2006AstrophysicsNuclei} relation. In both relations, the \textsc{Gaea-lc} magnitude cut causes a luminosity increase at low SFRs for high redshift, as here the light cone only includes a few unusually bright objects. For the $L_\mathrm{[\ion{O}{iii}]}$-$L_\mathrm{AGN}$ relation, we predict an increase up to 0.5\,dex above the \citet{Lamastra2009TheSources} relation. However, the general slope stays the same and we find that, especially in the lower redshift bins, \citet{Lamastra2009TheSources} provide a good estimate of the AGN luminosity.

When applying the EDS flux limit and \citet{Calzetti2000TheGalaxies} dust attenuation, predicted relations show a systematic bias toward increasingly bright line emitters with increasing redshift. At redshift 0.4--0.9, $L_\mathrm{H\alpha}$ is a robust proxy for the SFR up to around $10^{41} \, \rm erg \, s^{-1}$, while at redshift 1.5--1.8, we can expect $L_\mathrm{H\alpha}$ below $10^{42}$--$10^{42.5} \, \rm erg \, s^{-1}$ to result in biased SFR measurements. For $L_\mathrm{[\ion{O}{iii}]+H\beta}$, we predict that at redshift 0.9--1.2 the relation can be used down to roughly $10^{42} \, \rm erg \, s^{-1}$, while at redshift 1.8--2.5 around $10^{42.5}$--$10^{43} \, \rm erg \, s^{-1}$ are needed to ensure results are unbiased. The $L_\mathrm{[\ion{O}{iii}]}$-$L_\mathrm{AGN}$ relations from flux-limited and dust-attenuated populations follow a similar pattern. However, they appear truncated compared to the intrinsic population, as no active galaxies with $L_\mathrm{AGN}$ below $10^{42} \, \rm erg \, s^{-1}$ and $L_\mathrm{[\ion{O}{iii}]}$ above $10^{41.5} \, \rm erg \, s^{-1}$ exist in our sample. If such objects are recovered in the EDS, there are likely unusual and extreme objects.

One other disadvantage of these line intensity calibrations is that they are only valid for the dominant ionising property in SF and active galaxies, respectively. Otherwise the relation is contaminated by the contribution from other sources (see also Fig. 10 in \citealt{Hirschmann2023Emission-lineJWST}). This requires a prior classification according to galaxy type. Using standard BPT diagrams for this purpose is only possible if [\ion{O}{iii}], H$\beta$, H$\alpha$, and [\ion{N}{ii}] can be simultaneously measured. However, as seen in Sect. \ref{sec:selection bias}, we expect this to only be the case for around 1.4\% of galaxies in our underlying sample, with additional reductions due to line blending. Thus, it is useful to explore new emission-line based tracers which could constrain the strength of ionising sources across all galaxy types.

\begin{figure*}[htbp!]
\centering
	\includegraphics[width=\textwidth]{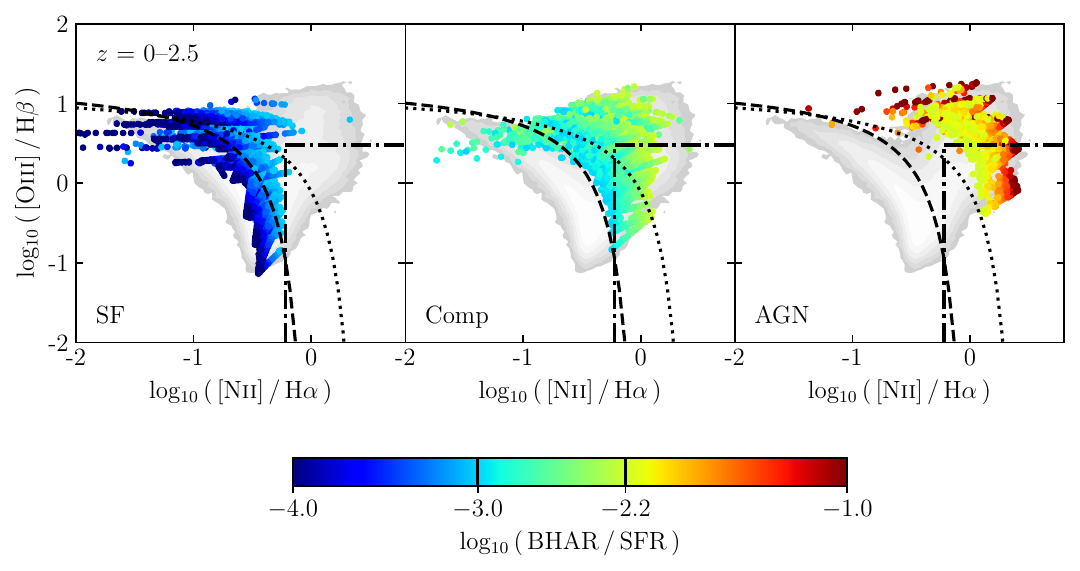}
    \caption{Location of the SDSS-like \textsc{Gaea-lc} galaxy populations in the $\rm [\ion{O}{iii}]/H\beta$ versus $\rm[\ion{N}{ii}]/H\alpha$ BPT diagram with the same layout as the top row in Fig. \ref{fig:BPT lowz}, but now with individual data points colour-coded by their BHAR/SFR ratio. Shown are galaxies above the SDSS flux cut in the redshift range 0.4--2.5.}
    \label{fig:BPT BHAR/SFR 3x3}
\end{figure*}

\subsection{[\ion{N}{ii}]-based emission-line ratios as new tracers for BHAR/SFR}
 \label{sec:BHAR_SFR}
The $\rm BHAR/SFR$ ratio measures the relative strength of the star formation and black hole activity, which defines the balance of energetic processes driving both the ionisation of the ISM, as well as the evolution of the entire galaxy.

In this work, we have utilised the $\rm BHAR/SFR$ ratio as a theoretical criterion to distinguish between dominant ionising sources in galaxies. Both Fig. \ref{fig:BPT lowz} and Fig. \ref{fig:BPT Euclid} have demonstrated that this distinction is successful at visually separating the different galaxy types in standard BPT diagrams. The [\ion{N}{ii}] BPT diagrams show a particular dependence of the three populations on the $\rm [\ion{N}{ii}]/H\alpha$ ratio, which we aim further explore in this Section.

In Fig. \ref{fig:BPT BHAR/SFR 3x3}, we show the location of SDSS-like \textsc{Gaea-lc} populations between redshift 0.4 and 2.5 in the $\rm [\ion{O}{iii}]/H\beta$ versus $\rm [\ion{N}{ii}]/H\alpha$ BPT diagram. We follow the same layout as in the top row in Fig. \ref{fig:BPT lowz}, but plot individual data points for the SF-dominated, composite and AGN-dominated galaxies, coloured according to their $\rm BHAR/SFR$ ratio. For reference, we indicate the $\rm BHAR/SFR$ boundaries between the three types (black vertical lines in colour bar). As we aim to explore the physical relationship between these quantities, we employ no dust correction or \Euclid-like flux cuts. 

As seen before, the three galaxy populations are well-separated by the $\rm BHAR/SFR$ criteria, with some overlap at the edges of the SF-dominated and AGN-dominated populations with the area occupied by composite galaxies. We further point out that even across galaxies of the same type there is a clear $\rm BHAR/SFR$-dependence of the $\rm [\ion{N}{ii}]/H\alpha$ ratio. The $\rm [\ion{N}{ii}]/H\alpha$ is mostly driven by changes in [\ion{N}{ii}], while the H$\alpha$ line provides a roughly constant baseline. Thus we conclude that it is the [\ion{N}{ii}] emission which depends on the relative strength of star-forming and accretion processes. This can be explained with the high excitation energy of this $\rm N^+$ state, which traces the hardness of the ionising radiation from AGNs.
As a result, we further explore [\ion{N}{ii}]-based emission-line ratios as potential tracers for the $\rm BHAR/SFR$ ratio.

\begin{figure*}[htbp!]
\centering
	\includegraphics[width=\textwidth]{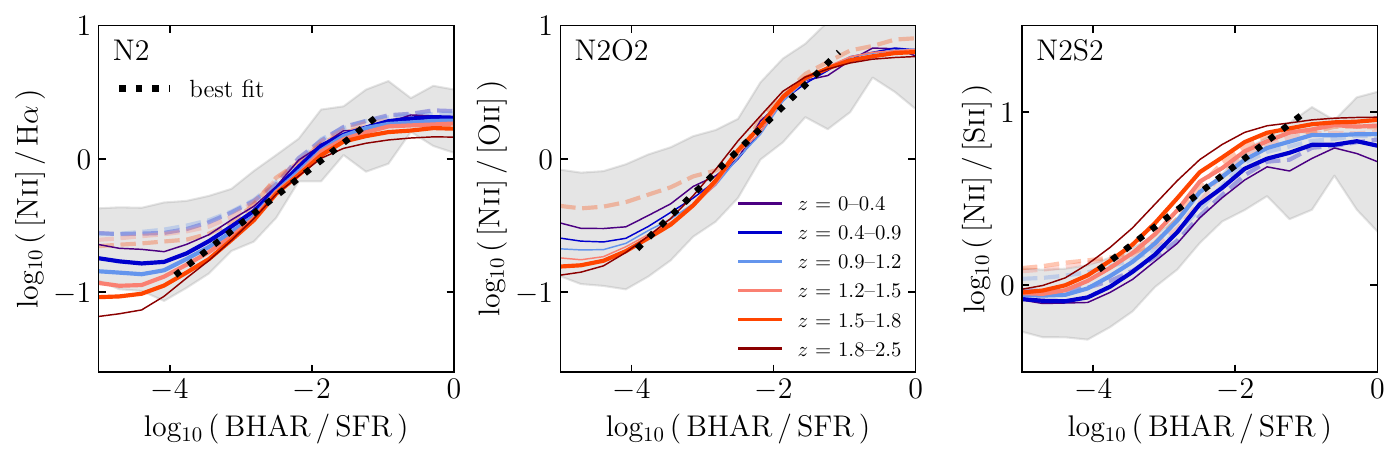}
    \caption{Average ratios of $\rm [\ion{N}{ii}]/H\alpha$ (left panel), $\rm [\ion{N}{ii}]/[\ion{O}{ii}]$ (middle panel), and $\rm [\ion{N}{ii}]/[\ion{S}{ii}]$ (right panel) versus $\rm BHAR/SFR$ in different redshift intervals (same key as in Fig. \ref{fig:SFR+LAGN highz}, with additional redshift 0--0.4 line in indigo). Overplotted are the best linear fits between $\mathrm{-4 \leq \logten(BHAR/SFR) \leq -1}$ (dotted lines). }
    \label{fig:BHAR SFR EL ratio}
\end{figure*}

Thus far, determining the $\rm BHAR/SFR$ ratio of a galaxy relied on combining separate estimates for the SFR and the BHAR. Results are heavily dependent on AGN selection and derivation methods, which results in large combined uncertainties, especially with increasing redshift \citep[see][]{McDonald2021ObservationalGalaxies}.  
In Fig. \ref{fig:BHAR SFR EL ratio}, we propose three novel [\ion{N}{ii}]-based emission-line calibrations to the $\rm BHAR/SFR$ for intermediate redshifts. We show the average $\rm [\ion{N}{ii}]/H\alpha$ (N2, left panel), $\rm [\ion{N}{ii}]/[\ion{O}{ii}]$ (N2O2, middle panel), and $\rm [\ion{N}{ii}]/[\ion{S}{ii}]$ (N2S2, right panel) at fixed $\rm BHAR/SFR$ in the same redshift intervals as in Fig. \ref{fig:SFR+LAGN highz}. We additionally include the 0--0.4 redshift interval (indigo) with a scatter of 1 standard deviation (shaded area). As before, we vary line thickness according to the observability in the EDS and show flux-limited populations in dashed lines. We omit dust-attenuated populations, as, since we are considering line ratios, these do not present any significant differences.

Additionally, we found linear best fits (dotted lines) in the $\logten(\mathrm{BHAR/SFR})$ range between $-4$ and $-1$, which can be used to derive the $\rm BHAR/SFR$ for \Euclid spectra in which $\rm [\ion{N}{ii}]$ can be de-blended from $\rm H\alpha$.

For all three emission-line ratios, we find a clear positive correlation. Over the given range, the N2 ratio increases by 1.25\,dex, represented by:
\begin{equation}
     \mathrm{N2} = 0.42\;\mathrm{\logten(BHAR/SFR)} + 0.77.
\end{equation}
N2O2 covers a range as large as 1.6\,dex, with a slightly steeper best fit:
\begin{equation}
    \mathrm{N2O2} = 0.52 \; \mathrm{\logten(BHAR/SFR)} + 1.36. 
\end{equation}
This strong correlation with N2O2 arises due to additionally decreased [\ion{O}{ii}] emission for high $\rm BHAR/SFR$ due to a greater abundance of multiply-ionised oxygen at the expense of singly-ionised oxygen, caused by harder ionising radiation from AGN.
Both the N2- and N2O2-$\rm BHAR/SFR$ relation appear to be almost redshift-invariant. The N2S2 ratio covers a range of 0.9\,dex with a less steep correlation of  
\begin{equation}
    \mathrm{N2S2} = 0.31 \; \mathrm{\logten(BHAR/SFR)} + 1.32.
\end{equation}

Additionally, N2S2 exhibits a slight rise of around 0.2--0.3\,dex with increased redshift, making it a less robust tracer than N2 and N2O2. However, average relations derived from flux-limited populations appear completely unbiased for N2S2, while N2 and N2O2 start to differ from the intrinsic relations below $\logten(\mathrm{BHAR/SFR})$ of -3. This biasing becomes significant at the low end of the linear fit, with an increased line ratio of at most 0.3\,dex with respect to the intrinsic population. While the N2O2-$\rm BHAR/SFR$ correlation is the steepest, we predict the [\ion{O}{ii}] line to be relatively faint over \Euclid's sensitivity range. 
This makes the N2 and N2S2 ratios the most favourable tracers. However, due to their strong correlations with the $\rm BHAR/SFR$ ratio, we consider all three of these line ratios to be effective tracers, valid across SF-dominated, composite, and AGN-dominated galaxies. This could allow for the characterisation of galaxies using just two emission line measurements.

\begin{figure*}[htbp!]
\centering
	\includegraphics[width=0.95\textwidth]{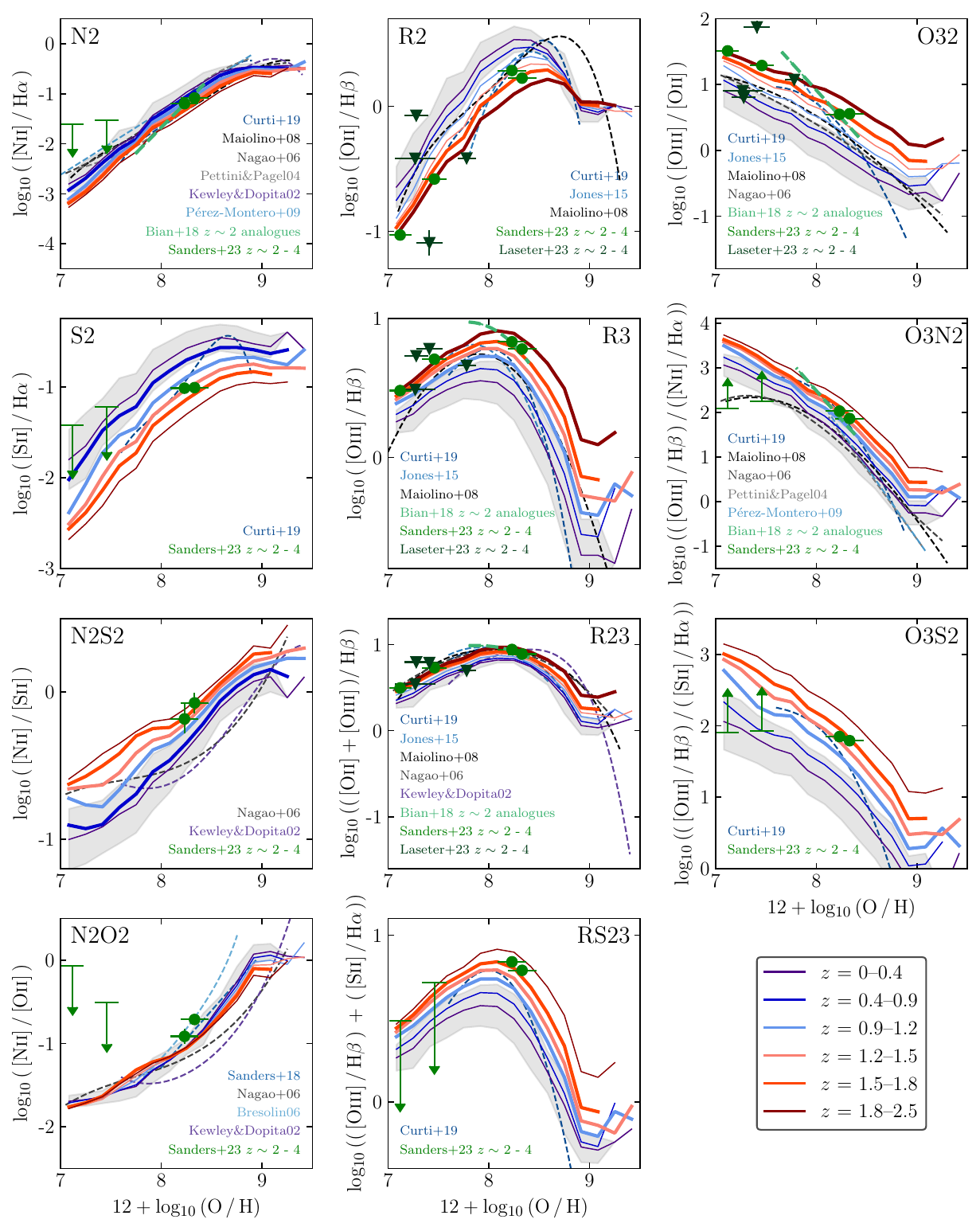}
    \caption{Average ratios of strong emission lines against interstellar O/H abundance. We consider in the left column: N2, $\rm [\ion{S}{ii}]/H\alpha$ (hereafter S2), N2S2, and N2O2; in the middle column: $\rm [\ion{O}{ii}]/H\alpha$ (R2), $\rm [\ion{O}{iii}]/H\alpha$ (R3), $\rm ([\ion{O}{iii}]+[\ion{O}{ii}])/H\alpha$ (R23), and $\rm ([\ion{O}{iii}]/H\alpha)+([\ion{S}{ii}]/H\alpha$) (RS23); and in the right column: $\rm [\ion{O}{iii}]/[\ion{O}{ii}]$ (O32), ($\rm [\ion{O}{iii}]/H\alpha$)/($\rm [\ion{N}{ii}]/H\alpha$) (O3N2), and ($\rm[\ion{O}{iii}]/H\alpha)/([\ion{S}{ii}]/H\alpha$) (O3S2). Solid lines show each relation for the \textsc{Gaea-lc} SF-dominated populations in different redshift bins, using the same colours as in Fig. \ref{fig:BHAR SFR EL ratio}, with thick lines indicating overlap of the relevant emission lines with the EDS sensitivity range. Overplotted are empirical and theoretical calibrations at redshift 0 \citep[blue and purple dashed lines, respectively; grey indicates calibrations using a combination of both;][]{Curti2019TheGalaxies,Jones2015TEMPERATURE-BASEDREDSHIFT,Perez-Montero2009TheLines,Bresolin2006TheIndicators,Tremonti2004TheSurvey,Kewley2002Galaxies,Marino2013TheData,Maiolino2008AMAZE3,Nagao2006GasGalaxies,Pettini2004ORedshift}. Additionally shown are recent estimates for higher redshifts \citep[green dashed lines and data points][]{Bian2018DirectGalaxies,Sanders2023DirectNoon,Laseter2023JADES:9.5}. }
    \label{fig:Z ratios}
\end{figure*}

\section{Deriving interstellar metallicity}
\label{sec:deriving Z}
The interstellar metallicity provides valuable insight into the chemical processes which have enriched the galaxy and, as a result, imprinted into the spectrum by influencing the relative intensities of various emission lines \citep[see][and references therein]{Kewley2019UnderstandingLines,Maiolino2019DeGalaxies}. 

The direct temperature method uses collisionally-excited auroral lines, like [\ion{O}{iii}]$\lambda$4363 and [\ion{N}{ii}]$\lambda$5755, to estimate the electronic temperature $T_\mathrm{e}$ and the underlying chemical abundances. Unfortunately, auroral lines are fairly faint and thus primarily detectable for nearby galaxies. However, they have been used to calibrate relationships between the metallicity and ratios of various strong emission lines, which are more easily detectable \citep[e.g.][]{Bresolin2006TheIndicators,Perez-Montero2009TheLines,Jones2015TEMPERATURE-BASEDREDSHIFT,Curti2019TheGalaxies}. This approach is particularly successful for metal-poor galaxies, while in metal-rich galaxies the lack of collisional excitation due to temperatures lowered by metal-line cooling makes the auroral lines even fainter. Additionally, direct $T_\mathrm{e}$ metallicities may be biased towards lower values if the ISM region is subject to temperature fluctuations or outside of thermal equilibrium.

For this reason, some studies resort to photoionisation models to establish relationships between metallicity and strong-line ratios, either in combination with direct $T_\mathrm{e}$ estimates for the metal-poor end \citep{Pettini2004ORedshift,Nagao2006GasGalaxies,Maiolino2008AMAZE3, Marino2013TheData} or to cover the entire range of metallicities  \citep{Kewley2002Galaxies,Tremonti2004TheSurvey}. While these are able to cover a wide range of metallicities and line ratios, their parameter space is often poorly constrained \citep[see][]{Chevallard2016ModellingBEAGLE, Vidal-Garcia2024BEAGLE-AGNGalaxies} and thus limited in their predictive power. 

As with the emission-line relations for SFR and AGN luminosity in Sect. \ref{sec:SFR+AGN}, most line-ratio calibrations have been derived and validated at redshift 0 and it is unclear if they provide robust estimates at higher redshifts. In fact, \citet{Bian2018DirectGalaxies} found that for local redshift 2 analogues, direct $T_\mathrm{e}$ estimates disagree with calibrations at redshift 0. New JWST measurements at redshift 2--9 show similar discrepancies \citep[e.g.][]{Curti2022The8,Laseter2023JADES:9.5,Sanders2023DirectNoon}. 

In Fig. \ref{fig:Z ratios}, we thus explore if commonly used emission-line ratios (as detailed in caption) for the interstellar metallicity (usually expressed as the oxygen abundance O/H) evolve across the \Euclid-detectable redshift range. As, thus far, most studies have focused on SF galaxies, we only show relations for SF-dominated \textsc{Gaea-lc} galaxies \citep[for specific calibrations to AGN narrow-line regions, see][]{Dors2017NewNuclei,Dors2021ChemicalShocks,Carvalho2020ChemicalSDSS}. We use the same colour code as in Fig. \ref{fig:BHAR SFR EL ratio}. Thicker lines indicate that all emission lines making up the ratio fall into the detectable range of the EDS. For visual clarity, we do not include average relations for flux-limited populations in this case, but comment on biasing in the final paragraph of the Section.

Alongside our predictions, we plot empirical and theoretical calibrations for redshift 0 (direct $T_\mathrm{e}$ method: blue dashed lines, photoionisation models: purple dashed lines, combination of both: grey lines). Additionally shown are the metallicity estimates for the redshift 2 analogues \citep[light green dashed lines]{Bian2018DirectGalaxies} and recent JWST measurements of galaxies at redshift 2--4 from the CEERS and JADES surveys. For two CEERS galaxies from \citet[][medium green data points]{Sanders2023DirectNoon}, there are only upper limits for [\ion{N}{ii}] and [\ion{S}{ii}], which is reflected in the one-sided error bars of length 0.5\,dex for any emission line ratio including them. For N2S2, we thus omit these two galaxies. \citet[dark green data points]{Laseter2023JADES:9.5} recovered measurements for the [\ion{O}{iii}], [\ion{O}{ii}], and H$\beta$ lines for four JADES galaxies at redshift 2--4.

Overall, we find that the average relations between line ratios and the oxygen abundance broadly agree with the local calibrations, especially at low redshifts. While N2O2 and R23 appear fairly constant across all redshift bins, other relations show some evolution with redshift. These results are in excellent agreement with \citet{Hirschmann2023High-redshiftSimulations}, who find the same broad evolution in their emission-line coupled IllustrisTNG galaxies between redshift 2 and 0. We note that this evolution presents itself as a change in normalisation; metallicity estimators are progressively shifted toward lower or higher values, while the shape generally remains the same. For each estimator, we thus determined the shifts at $\mathrm{12 + \logten(O/H) = 8}$ between our predicted average relation at redshift 0--0.4 and the relations at redshift 1, 1.5, 2, and 2.5. The results are shown in Table \ref{tab:shifts}. Depending on metallicity, the magnitude of the shifts can vary, and as such, we provide approximate ranges within which the specified shifts are applicable. In the following paragraphs, we will assess each estimator in more detail and explain the evolutionary shifts between redshift 0--0.4 and 2.5.

Looking at the N2, S2, N2S2, and N2O2 estimators, we note that they all show a positive correlation with oxygen abundance as metal-rich galaxies contain elevated nitrogen and sulfur abundances. S2, and to a much lesser degree N2, show a drop with increasing redshift. This can be attributed to the elevated ionisation parameter at fixed metallicity and stellar mass (see Fig. \ref{fig:logU relation} and \ref{fig:logU evolution Mstar}), which causes nitrogen and sulfur to favour the higher ionisation states $\mathrm{N}^{2+}$ and $\mathrm{S}^{2+}$. For N2, the resulting drop of $-0.46$\,dex lies within the scatter of local literature calibrations. S2 shows a greater evolution of $-0.87$\,dex, which translates into a +0.41\,dex rise with redshift in the N2S2 estimator. The changing ionisation parameter has little influence on the N2O2 ratio, as nitrogen and oxygen have similar ionisation energies, which explains its near redshift-invariance.

The R2, R3, R23, and RS23 all show a peak in their average relations around $\mathrm{12+\logten\,(O\,/\,H)} \approx 8$. They are all oxygen-based and thus initially increase with oxygen abundance. However, at high metallicities, low temperatures caused by metal-line cooling make collisional excitation less likely. At high redshift, the increased ionisation parameter due to higher SFRs causes oxygen to favour $\mathrm{O}^{2+}$ over $\mathrm{O}^{+}$, which is why R2 shifts toward lower and R3 and RS23 toward higher values. This amounts to changes of $-0.52$, +0.37, and +0.35\,dex. The R23 ratio is able to compensate some of this evolution by summing the [\ion{O}{ii}] and [\ion{O}{iii}] line contributions, which results in a comparatively small shift of +0.16\,dex in the metallicity estimator.

The combined effects of stronger [\ion{O}{ii}], [\ion{N}{ii}], and [\ion{S}{ii}], as well as decreased [\ion{O}{iii}] emission with rising metallicity cause O32, O3N2, and O3S2 to drop steeply. Increased abundances of doubly-ionised atoms at higher redshifts, at the detriment of $\mathrm{O}^{+}$, $\mathrm{N}^{+}$, and $\mathrm{S}^{+}$, cause up to 1.24\,dex higher line-ratios at higher redshifts.

In general, the CEERS galaxies at redshift 2--4 from \citet{Sanders2023DirectNoon} consistently agree better with our intermediate-redshift prediction from the \textsc{Gaea-lc} galaxies than the local calibrations. Upper and lower limits for the [\ion{N}{ii}]- and [\ion{S}{ii}]-based ratios deviate from our predictions in the expected direction. Similarly, the calibrations to redshift 2 analogues from \citet{Bian2018DirectGalaxies} tend to lie closely to our predictions between redshift 1.5 to 2.5. This close agreement is encouraging and further highlights that metallicity estimators already start evolving at intermediate redshifts. The R3 and R23 line ratios for four JADES galaxies at redshift 2--4 determined by \citet{Laseter2023JADES:9.5} show a similar trend to \citet{Sanders2023DirectNoon}. However, the R2 and O32 estimates exhibit a scatter of roughly 1\,dex across the low-metallicity range between $\mathrm{12 + \logten(O/H)} = 7.27$ and 7.78. This scatter is also present in their measurements for nine additional JADES and twelve additional CEERS galaxies with redshift above 4 from \citet[not shown here, see Figures 4--8 in \citealt{Laseter2023JADES:9.5}]{Sanders2023DirectNoon}, from which they conclude that the metallicity dependence of R2 and O32 breaks down in the high-redshift Universe. No distinction is made for the intermediate-redshift range.
Given the apparent agreement of our predictions with \citet{Bian2018DirectGalaxies} and \citet{Sanders2023DirectNoon} below and around redshift 4, the small number statistics of these samples and the systematics resulting from varying assumptions, data reduction and metallicity precriptions (see Sect. 3 in \citealt{Laseter2023JADES:9.5}), it is however difficult to determine if and when exactly this breakdown might occur. This further illustrates the need for additional spectroscopic data at intermediate redshifts.
 
Overall, we conclude that across the redshift range 0.4--2.5, most commonly used emission-line ratios evolve away from their local O/H calibrations. The N2O2 and R23 relations evolve the least, making them, in theory, the most robust to extend to higher redshifts without modifications. For all relations we observe a shift in the normalisation, while the overall shape is preserved. Not accounting for this evolution could result in false metallicity estimates. For instance, \citet[][]{Hirschmann2023High-redshiftSimulations} found that classical calibrations at redshift 0 underestimate the mass-metallicity relation by up to 1\,dex when applied above redshift 4. We expect a similar trend to occur already up to redshift 2.5.

For the purpose of deriving metallicity in \Euclid-observable galaxies specifically, the detectability of lines according to the grism sensitivity ranges and the observability according to flux limits should also be considered. We expect \Euclid's flux limit to bias the observable galaxy population to the brightest, most massive and metal-rich galaxies, which could deviate from predicted relations. In order to test the robustness of various metallicity estimators in a \Euclid-like sample, we thus computed the average relations for EDS-observable populations. For all average relations except R3, no metal-poor galaxies passed the flux cut, resulting in a truncated $\mathrm{12 + \logten(O/H)}$ range of 8--9. If the EDS recovers metal-poor galaxies outside this truncated range, they will most likely be extreme objects. Inside the metallicity range, R2, R3, R23, RS23, O32, O3N2, and O3S2 did not deviate significantly from the predictions in Fig. \ref{fig:Z ratios}. N2, S2, N2S2, and N2S2 showed an 0.5--1\,dex overestimation of the line ratio below $\mathrm{12 + \logten(O/H) \approx 8.5}$. Additionally, the deconvolution of blended [\ion{N}{ii}] and H$\alpha$ lines (method detailed in Sect. \ref{sec:discussion_inst}) might not always be possible. We thus recommend the use of [O{\sc ii}]- and [O{\sc iii}]-based metallicity estimators for \Euclid data. However, for line combinations that are widely spaced in wavelength, the line ratio is more likely to suffer from differential dust extinction. Therefore, when applying calibrations for N2O2, O3N2, O3S2, O32, R23, and RS23 some form of dust correction should be considered.

\begin{table*}[h]
\centering
\caption{\label{tab:shifts} Average offsets in $\mathrm{\logten(Line \, ratio)}$ at different redshifts with respect to the average relation at redshift 0--0.4, determined at $\mathrm{12 + \logten(O/H) = 8}$ and applicable in the given $\mathrm{12 + \logten(O/H)}$ range.}
\begin{tabular}{|l|c|c|c|c|c|}
\hline
  & & & & & \\[-9pt]
\begin{tabular}[c]{@{}c@{}}Line ratio\end{tabular} & 12 + $\rm \logten(O/H)$ range & \textbf{$z=1$} & \textbf{$z=1.5$} & \textbf{$z=2$} & \textbf{$z=2.5$} \\[3pt] \hline 
 & & & & & \\[-9pt]
N2   & 7.5--8.7 & $-0.15 \pm 0.32$ & $-0.28 \pm 0.34$ & $-0.38 \pm 0.38$ & $-0.46 \pm 0.38$ \\
S2   & 7.5--8.7 & $-0.34 \pm 0.28$ & $-0.56 \pm 0.26$ & $-0.74 \pm 0.25$ & $-0.87 \pm 0.24$ \\
N2S2 &      8      & $0.19 \pm 0.29$ & $0.28 \pm 0.31$ & $0.36 \pm 0.34$ & $0.41 \pm 0.34$ \\
N2O2 &      -      & $0.04 \pm 0.27$ & $0.04 \pm 0.31$ & $0.06 \pm 0.35$ & $0.06 \pm 0.36$ \\
R2   &  7.5--8.5  & $-0.18 \pm 0.19$ & $-0.32 \pm 0.18$ & $-0.44 \pm 0.16$ & $-0.52 \pm 0.15$ \\
R3   &  7.5--8.5  & $0.17 \pm 0.17$ & $0.26 \pm 0.16$ & $0.32 \pm 0.16$ & $0.37 \pm 0.15$ \\
R23  &      -      & $0.05 \pm 0.05$ & $0.09 \pm 0.05$ & $0.13 \pm 0.05$ & $0.16 \pm 0.06$ \\
RS23 &  7.5--8.5  & $0.16 \pm 0.16$ & $0.24 \pm 0.14$ & $0.3 \pm 0.14$ & $0.35 \pm 0.13$ \\
O32  &  7.5--8.5  & $0.35 \pm 0.35$ & $0.58 \pm 0.33$ & $0.76 \pm 0.31$ & $0.89 \pm 0.3$ \\
O3N2 &    7--8    & $0.31 \pm 0.44$ & $0.53 \pm 0.44$ & $0.7 \pm 0.46$ & $0.83 \pm 0.46$ \\
O3S2 &   7--8.7   & $0.50 \pm 0.42$ & $0.82 \pm 0.39$ & $1.06 \pm 0.38$ & $1.24 \pm 0.36$ \\ \hline

\end{tabular}
\end{table*}

\section{Discussion}
\label{sec:discussion}
Based on our self-consistent modelling framework, we have provided forecasts for line-emitting galaxies in the intermediate redshift range 0.4--2.5, thereby addressing the lack of theoretical guidance for spectroscopic diagnostics in this regime. In anticipation of the Euclid Wide and Deep Surveys, we indicated the expected biasing of \Euclid's future spectral catalogues as well as made recommendations for the use of a wide range of spectroscopic diagnostics to characterise observed galaxies.
We consider these predictions robust, as in Sect. \ref{sec:validation}, we validated our approach by showing that the emission-line coupled \textsc{Gaea-lc} framework successfully reproduces a range of key observations and calibrations. However, we acknowledge that our study may be affected by several caveats related to the treatment of interstellar dust, modelling details, and \Euclid's instrumental effects. We discuss these in sections below. Caveats that are not directly related to \textsc{Gaea} are largely similar to those discussed in Sect. 5.2 of \cite{Hirschmann2023High-redshiftSimulations}, which we refer the reader to for additional details.
 
\subsection{Treatment of interstellar dust}
\label{sec:discussion_dust}
We expect the presence of dust to affect \Euclid observations, primarily by obscuring emitted line fluxes along the line-of-sight to objects. While dust attenuation is treated self-consistently within ionised regions using \texttt{Cloudy}, estimating the impact of interstellar dust is more challenging, especially beyond redshift 0, due to limited data coverage. \textsc{Gaea} also does not incorporate a self-consistent prescription of dust processes, like dust grain formation, evolution, and depletion within the interstellar medium. However, the limited data available suggests that the locally found stellar mass-dependent scaling of dust attenuation seems to hold for higher redshifts. Thus, we used the \citet{Calzetti2000TheGalaxies} curve based on mass-dependent $A_\mathrm{V}$ from the \citet{Garn2010PredictingGalaxy} relation to estimate the attenuation effects on our modelled emission-line catalogue.

We find that interstellar dust leads to substantial reduction in line fluxes, particularly for bluer wavelengths, where observable percentages are reduced by up to 30\%. Consequently, intrinsically fainter lines, such as H$\beta$, [\ion{O}{ii}], and [\ion{O}{i}], may become difficult to observe. This effect is more pronounced at higher redshifts, where predominantly massive galaxies produce intrinsic line fluxes above \Euclid's nominal flux limit. Due to the mass-dependence of the scaling, they are also more strongly attenuated, which would decrease many estimated line fluxes below the threshold. However, due to large uncertainties involved, these estimates should be understood as an indication of how the presence of dust might affect observed samples rather than precise predictions. We further note that the \citet{Garn2010PredictingGalaxy} relation is purely empirical and is not derived from first principles. 

There are a variety of complicating factors. The environmental conditions at higher redshifts are largely unknown and variations in dust grain composition, size, shape, and distribution can directly affect the resulting attenuation \citep[for a review, see][]{Draine2003InterstellarGrains}. Additionally, we assume galaxies of the same stellar mass also have the same dust mass, disregarding other potential dependencies, such as a positive correlation between extinction and the star-formation rate \citep[i.e.][]{Hopkins2001TowardRate,Zahid2013THEGALAXIES}. Further, we expect the presence of a dusty torus around AGN to affect line emission differently compared to interstellar dust \citep[][]{Urry1995UnifiedNuclei,Hasinger2008AbsorptionNuclei}. This would introduce additional complexities in estimating the overall attenuation effects, especially for composites. 

Lastly, we note that during the coupling process we adopted a dust-to-metal mass ratio $\xi_{\mathrm{d}}$ of 0.3 for all galaxies (see Sect. \ref{sec:coupling}), which sets the depletion of metals onto dust grains. This is inspired by the Solar-neighbourhood value $\xi_{\mathrm{d}, \odot}$ of 0.36.
Increasing $\xi_{\mathrm{d}}$ would result in greater depletion of metal coolants and, thus, higher temperatures and greater probability of collisional excitation. 
Significantly different or evolving values for $\xi_{\mathrm{d}}$ could thus have an effect on the resulting line emission. 
However, \citet{Hirschmann2017SyntheticRatios}
explored the influence of setting different $\xi_{\mathrm{d}}$ during the coupling process and found only a negligible influence on the cosmic evolution of simulated emission-line ratios, in agreement with observational and other theoretical works \citep{Remy-Ruyer2014Gas-to-DustRange,Popping2017The9}.

\subsection{Caveats related to photoionisation models}
\label{sec:discussion_models}
The model grids for SF, AGN, and PAGB contributions used in this study are based on the \texttt{Cloudy} photoionisation code. It follows the non-equilibrium ionisation and the thermal and chemical state of a gas element under incident radiation, calculating ionisation, recombination, collision, emission, and absorption, as well as accounting for various other physical processes, like radiation pressure on dust grains, metal depletion on dust, and attenuation. 
Computations are performed in one dimension, assuming spherical geometry and constant density. This represents a simplification, as real galaxies exhibit complex 3D gas distributions, which can not be captured by the models.

We further assume that all ionised regions are ionisation-bounded, as opposed to density-bounded. Density-bounded regions are optically thin to Lyman Continuum (LyC) photons, allowing them to leak out and affect emission-line intensities. This process appears particularly important at higher redshifts \citep[i.e.][]{DeBarros2019TheJWST,Shapley2016Q1549-C25:Z=3.15,Shapley2015TheGalaxies,Bian2017HIGH2.5} and generally decreases the intensities of low-ionisation relative to high-ionisation lines \citep{Jaskot2013THEGALAXIES,Plat2019ConstraintsGalaxies}. However, LyC leakage remains poorly understood and it is unclear to what extent it might affect diagnostics at intermediate redshifts, like the ones presented in this work. 

In contrast to recent works with similar methodology \citep[i.e.][]{Hirschmann2023Emission-lineJWST,Hirschmann2023High-redshiftSimulations}, we did not include any emission-line models for fast radiative shocks. Shocks can be produced by galactic outflows caused by supernovae, stellar winds, and AGN \citep[e.g.][]{Rich2010NGCSuperwind,Sharp2010Three-DimensionalBlows,Rich2011Galaxy-wideGalaxies,Soto2012TheOutflows,Weistrop2012Characteristics4194} and have been observed out to redshift 3 \citep[e.g.][]{Steidel2010THEGALAXIES,Genzel2011THECLUMPS, Kornei2012THESTRIP,Newman2012TheProperties}.
These shocks can cause excitation in the ISM gas, which contributes to the overall line emission.

In order to model line emission from shocks self-consistently, \citet{Hirschmann2023Emission-lineJWST,Hirschmann2023High-redshiftSimulations} take advantage of a shock finder which has been applied to IllustrisTNG on the fly. Fast radiative shock models from the \texttt{Mappings V}-based grids by \citet{Alarie2019ExtensiveDatabase} were then matched to the averaged quantities of shocked regions. As our modelling framework is based on \textsc{Gaea}, a semi-analytic model, we do not have access to shocks and other internal gas kinematics (see also Sect. \ref{sec:discussion_coupling}). \citet{Hirschmann2023Emission-lineJWST} predict that at low redshifts, shock-dominated galaxies produce similar signatures to AGN on the BPT-diagram, making them hard to distinguish. However, since shocks are usually caused by stellar and AGN-driven outflows, purely shock-dominated galaxies are rare. They find that at redshifts above 1, fractions of shock-dominated galaxies drop below 1\% and produce a negligible contribution to the overall line emission. In light of this, we conclude that the inclusion of fast radiative shocks would not influence our results significantly.

\subsection{Caveats related to the \textsc{Gaea} semi-analytic model and the coupling methodology}
\label{sec:discussion_coupling}
While semi-analytic models, like \textsc{Gaea}, differ in their modelling methodology to cosmological simulations, they are subject to similar uncertainties. Complex physical processes and mechanisms, like star formation and stellar and AGN feedback, have to be simplified into model prescriptions in order to represent the complex evolution of baryonic components. As for cosmological simulations, resulting galaxy properties can depend substantially on the specific model chosen.
However, one advantage of semi-analytic models is that, due to their computation speed, it is possible to run the model with various implementations. This allows the exploration of large parameter spaces, ultimately enabling the choice of the most successful scheme and the subsequent fine tuning of parameters to reproduce observational benchmarks. This has been done for both stellar and AGN feedback \citep[see][respectively]{Hirschmann2016GalaxyModel,Fontanot2020TheModel}.
The resulting combined model we make use of in this work reproduces key observational constraints out to high redshift. The galaxy mass function is robust out to approximately redshift 7 and the cosmic star-formation rate density out to redshift 10 \citep{Fontanot2017STRONGDAWN}. \textsc{Gaea} also agrees well with the observed cold gas fractions out to around redshift 2, as well as the mass-metallicity relation at redshift 0 and its evolution for increasing redshifts \citep{Hirschmann2016GalaxyModel}. \citet[][]{Fontanot2020TheModel} have further shown that the bolometric AGN luminosity function is in agreement with observations up to roughly redshift 4.
This forms a robust foundation for our emission-line predictions at intermediate redshifts. We note however, that a preliminary comparison to observations \citep{DeLucia2018NatureGalaxies} indicates that \textsc{Gaea} tends to underpredict quiescent fractions for massive galaxies at high redshift, which could introduce a bias in our results.

Given that \textsc{Gaea} does not explicitly treat internal gas distribution and the related dynamics, additional assumptions and simplifications were necessary in the coupling process, which are detailed in Sect. \ref{sec:coupling}. Following preceding works, we fix the hydrogen gas density within ionised regions to $10^2 \rm \, cm^{-3}$ for \ion{H}{ii} regions, $10^3 \rm \,cm^{-3}$ for narrow-line regions, and $10 \rm \,cm^{-3}$ for line-emitting regions ionised by post-AGB stars. As discussed in  Sect. \ref{sec:discussion_dust}, the dust-to-metal mass ratio $\xi_{\mathrm{d}}$ has been set to 0.3. The impact of varying these parameters on predicted line emission is relatively small and has been discussed in  \citet{Hirschmann2017SyntheticRatios,Hirschmann2019SyntheticSources}. For the computation of the ionisation parameter, we initially calibrate the volume filling factor at redshift 0 to reproduce the \citet{Carton2017InferringApproach} relation. At higher redshifts, the filling factor for the SF component evolves according to the average global gas density within galaxies sourced from IllustrisTNG, whereas for the AGN component we use the average central gas density. In order to couple the AGN models, an additional estimate of the central metallicity is necessary, which we have set to twice that of the global metallicity of each galaxy. 
Our results remain largely unaffected by varying this assumption by a factor of a few.

We further cannot model individual \ion{H}{ii} regions in \textsc{Gaea} and instead adopt the methodology introduced by \citet[][as discussed in Sect. \ref{sec:coupling}]{Charlot2001NebularGalaxies}, in which the temporal evolution of a typical \ion{H}{ii} region is convolved with the star-formation history of each galaxy. That means that, even though we assume that a galaxy comprises multiple \ion{H}{ii} regions with varying star cluster ages, we do not account for potential variations in gas properties across the galaxy, such as density, filling factor, and metallicity. 

Lastly, we stress that we only model the narrow-line regions for all AGN and exclude broad-line regions (BLR), thus not distinguishing between type-I and type-II AGN. In reality, a fraction of future observed AGN, will be of type-I and thus will exhibit increased fluxes due to the contribution from the BLR in addition to the narrow-line region. At the high densities of BLRs ($n_\mathrm{H}>10^9 \,\mathrm{cm}^{-3}$, see \citealt{Peterson2006TheNuclei}), forbidden optical transitions like [\ion{O}{iii}], [\ion{N}{ii}], [\ion{S}{ii}], [\ion{O}{ii}], and [\ion{O}{i}] are disfavoured at the expense of collisional de-excitation, meaning that we expect only the Balmer lines to be strengthened.

\subsection{Caveats related to instrumental and environmental effects}
\label{sec:discussion_inst}
In this work, we focus our predictions on the physical evolution of emission-line properties for galaxies at intermediate redshifts, without modelling instrumental and environmental effects. Effects which could cause deviations from our predictions include low signal-to-noise ratios (SNR) and blending of lines due to the spectrometer's spectral resolution limit.

For our analysis, we have assumed all galaxies with predicted line fluxes greater than $ 2 \times 10^{-16}\, \operatorname{erg}\, \mathrm{s}^{-1}\, \mathrm{cm}^{-2}$ to be observable in the EWS (and by analogy greater than $6 \times 10^{-17}\,\mathrm{erg} \, \mathrm{s}^{-1}\, \mathrm{cm}^{-2}$ in the EDS). This is a simplification of the formal requirement that in the EWS, the NISP spectrometer is expected to detect line emission with a sensitivity greater than $2 \times 10^{-16} \,\operatorname{erg}\, \mathrm{s}^{-1}\, \mathrm{cm}^{-2}$ and a SNR of 3.5 for a typical source of size \ang{;;0.5}. We expect these additional specification to reduce the size of the overall observable galaxy population, meaning that our estimate represents an upper limit. \cite{Gabarra-EP31} have assessed the performance of the NISP red grism in detail for star-forming galaxies and discuss the influence of the SNR, source size, and morphological effects. Their dataset is based on simulated spectra from the Pilot simulation, a \Euclid legacy science project, which models the instrument output resulting from observing a galaxy spectrum on a patch of simulated sky, including instrumental and astrophysical noise. \citep{EuclidCollaboration:Lusso2023EuclidNISP} have performed a similar analysis assessing NISP performance for active galaxies, using mock AGN spectra created from a library of empirical templates. We also refer the reader to ongoing work by Mancini et al. (in prep.), which will further assess the impact of \Euclid's instrumental effects on emission-line forecasts, with an empirical approach based on the Millennium Mambo catalogues, complementary to ours.

Our approach further assumes that \Euclid can recover separable line fluxes for the H$\alpha$ and [\ion{N}{ii}]$\lambda6584$ line. Given the resolution of the NISP spectrometer, the $\rm H\alpha + [\ion{N}{ii}]\lambda6584 + [\ion{N}{ii}]\lambda6548$ complex will be blended in \Euclid spectra (see also \citealt{EuclidCollaboration:Lusso2023EuclidNISP}). However, in the OU-SPE spectroscopic pipeline (Le Brun et al., in prep) two different approaches are implemented to measure the line fluxes: a direct integration (DI) and a Gaussian-fit (GF) method. 
In the GF method, the $\rm H\alpha+[\ion{N}{ii}]$ is modeled as three Gaussians, where the free parameters are the amplitude of H$\alpha$ and of [\ion{N}{ii}]$\lambda6584$, the width of the line, the position of H$\alpha$, and the value of the continuum. The flux ratio of the [\ion{N}{ii}] lines is set to 1/3 and their positions are fixed to that of the H$\alpha$ line. As discussed in Le Brun et al. (in prep), the GF method provides an estimate of the deconvolved H$\alpha$ and [\ion{N}{ii}] fluxes.

\section{Conclusion}
\label{sec:conclusion}
In this work, we presented optical emission-line predictions at intermediate redshifts for the upcoming Euclid Wide and Deep Surveys, addressing the lack of comprehensive theoretical guidance in this regime.
We followed a methodology adapted from \citet{Hirschmann2017SyntheticRatios,Hirschmann2019SyntheticSources,Hirschmann2023Emission-lineJWST,Hirschmann2023High-redshiftSimulations} to construct emission-line catalogues by coupling galaxies from a mock light cone based on the \textsc{Gaea} semi-analytic model to state-of-the-art photoionisation models. This enabled us to self-consistently compute the emission lines of galaxies at different cosmic epochs originating from young star clusters \citep{Gutkin2016ModellingGalaxies}, AGN narrow-line regions \citep{Feltre2016NuclearWavelengths}, and post-asymptotic giant branch stellar populations \citep{Hirschmann2017SyntheticRatios}. As a last step, we validated the resulting emission-line catalogues by comparing its predictions to observational data and well-calibrated theoretical predictions for low redshifts (Sect. \ref{sec:validation}). This framework represents the first emission-line catalogue based on a semi-analytic model which contains self-consistent modelling of the ionising contributions not only from young star clusters, but also from other ionising sources, such as AGN, and post-AGB stellar populations.  

We then focused on spectroscopic diagnostics in the redshift range 0.4--2.5 based on the seven optical emission lines: H$\alpha$, H$\beta$, [S{\sc ii}], [N{\sc ii}],  [O{\sc i}], [O{\sc iii}], and [O{\sc ii}]. In order to make targeted predictions for their observability with \Euclid, we further computed observer-like fluxes based on the location and redshift of each galaxy in the light cone and modelled attenuation due to interstellar dust using the \citet{Calzetti2000TheGalaxies} relation, with attenuation $A_{V}$ modelled according to the mass-dependent \citet{Garn2010PredictingGalaxy} relation.

Our main results from the analysis can be summarised as follows:
\begin{enumerate}
 \setlength\itemsep{0.3em}
 
\item 
We tested how emission lines trace scaling relations in two observing scenarios: observing H$\alpha$ in the EWS and the BPT lines, namely H$\alpha$, H$\beta$, [\ion{O}{iii}], and [\ion{N}{ii}], in the EDS (Fig. \ref{fig:selection}). In both cases, the resulting observable populations bias standard scaling relation towards high stellar and halo masses, high specific SFR, and high metallicities, meaning \Euclid will predominantly observe line-emitting galaxies that are massive ($M_{\star} \gtrsim 10^{9}\,\si{\solarmass}$), star-forming ($\mathrm{sSFR} > 10^{-10}\,\mathrm{yr^{-1}}$), and metal-rich ($\mathrm{\logten(O/H)+12} > 8$). We predict that both survey configurations will recover galaxies containing AGN with black hole masses between $10^{6}$--$10^{9.5} \si{\solarmass}$ and bolometric luminosities of $10^{39}$--$10^{46}\rm \, erg\,s^{-1}$ (Fig. \ref{fig:MBH LAGN}). We estimate that the influence of interstellar dust could reduce observable percentages by an additional 20--30\% with respect to the intrinsic population, which may pose challenges in measuring fainter lines, especially toward higher redshifts (Fig. \ref{fig:no count hists}). If accounting for dust attenuation, we anticipate that at redshift less than 1, \Euclid will successfully capture approximately 30--70\% of both SF and active galaxies emitting H$\alpha$, [\ion{N}{ii}], [\ion{S}{ii}], and [\ion{O}{iii}], given our \textsc{Gaea-lc} galaxy sample with a mass resolution limit of $10^{9} \, \si{\solarmass}$ and $H$-band magnitude cut of 25. At higher redshifts, these percentages decline to below 10\%. H$\beta$, [\ion{O}{ii}], and [\ion{O}{i}] exhibit particularly faint signatures and, consequently, we expect observable percentages to be limited to below 5\% for both SF and active galaxies at low redshifts, which are reduced to below 1\% with increasing redshift. 

\item For EDS-observable galaxies, we expect [\ion{O}{iii}]/H$\beta$ versus [\ion{N}{ii}]/H$\alpha$ BPT diagrams to continue to distinguish between SF-dominated, composite, and AGN-dominated galaxies up to at least redshift 1.8 (Fig. \ref{fig:BPT Euclid}). This can be attributed to the bias toward metal-rich systems, introduced by requiring the observability of all four emission lines. After including the impact of dust attenuation, we expect this to be the case for 1.4\% of \textsc{Gaea-lc} galaxies, with an upper limit of 11.8\% in the no dust scenario.

\item We find that relationships between H$\alpha$ and $\rm [\ion{O}{iii}]+H\beta$ luminosities and the SFR show only a negligible, if any, evolution with increasing redshift, when compared to the local calibrations from \citet{Kennicutt2012} and \citet[Fig. \ref{fig:SFR+LAGN highz}]{Osterbrock2006AstrophysicsNuclei}. This indicates, that they could be applied in the analysis of future \Euclid data. The $L_\mathrm{AGN}$-$L_{[\ion{O}{iii}]}$ relationship shows a redshift evolution of up to +0.5\,dex with respect to the local \citet{Lamastra2009TheSources} calibration, but generally retains the same slope. As a result, we find that \citet{Lamastra2009TheSources} still provides a reasonable estimate for the AGN luminosity, especially at the lower end of \Euclid's target redshift range. We further indicate up to which luminosity thresholds these tracers appear largely unbiased for EDS-observable and dust-attenuated galaxy populations at different redshifts. It is important to note that these calibrations are only valid for the dominant ionisation mechanism in the respective galaxy types, meaning SF-dominated galaxies for the SFR relations and AGN-dominated galaxies for the $L_\mathrm{AGN}$ relation. As a result, a pre-sorting of galaxies according to their type is necessary.

\item We find that [\ion{N}{ii}] emission strongly depends on the $\rm BHAR/SFR$ ratio (Fig. \ref{fig:BPT BHAR/SFR 3x3}). As a result, we explored various [\ion{N}{ii}]-based emission-line tracers for the $\rm BHAR/SFR$ ratio and discover strong positive correlations with $\rm [\ion{N}{ii}]/H\alpha$, $\rm [\ion{N}{ii}]/[\ion{O}{ii}]$, and $\rm [\ion{N}{ii}]/[\ion{S}{ii}]$ (Fig. \ref{fig:BHAR SFR EL ratio}). These relationships appear to be largely redshift-invariant in the 0--2.5 range, are valid across all galaxy types and show only slight biasing when applying \Euclid flux cuts. We propose these as novel tracers and provide linear best fits between $\mathrm{-4 \leq \logten(BHAR/SFR) \leq -1}$, allowing for the derivation of the relative strength of star-forming and accretion processes from combinations of only two emission lines.

\item  We examined the potential evolution between redshift 0 and 2.5 of the strong emission-line ratios commonly used to estimate oxygen abundance from observed spectra (Fig. \ref{fig:Z ratios}): $\rm [\ion{N}{ii}]/H\alpha$ (N2), $\rm [\ion{S}{ii}]/H\alpha$ (S2), $\rm [\ion{N}{ii}]/[\ion{S}{ii}]$ (N2S2), $\rm [\ion{N}{ii}]/[\ion{O}{ii}]$ (N2O2), $\rm [\ion{O}{ii}]/H\alpha$ (R2), $\rm [\ion{O}{iii}]/H\alpha$ (R3), $\rm ([\ion{O}{iii}]+[\ion{O}{ii}])/H\alpha$ (R23),  $\rm ([\ion{O}{iii}]/H\alpha)+([\ion{S}{ii}]/H\alpha)$ (RS23), $\rm [\ion{O}{iii}]/[\ion{O}{ii}]$ (O32), $\rm ([\ion{O}{iii}]/H\alpha)/([\ion{N}{ii}]/H\alpha)$ (O3N2), and $\rm ([\ion{O}{iii}]/H\alpha)/([\ion{S}{ii}]/H\alpha)$ (O3S2). We found that, in general, they evolve away from their locally established calibrations. This evolution manifests as a shift in normalisation, with metallicity estimators gradually moving towards either lower or higher values, while the overall shape remains consistent with local calibrations. This is in tentative agreement with current, although sparse, JWST data. Notably, the N2O2 and R23 ratios exhibit the weakest evolutions, suggesting that they are the most robust to extend to higher redshifts without adjustments. We assessed the robustness of various metallicity estimators for EDS-like samples and found that [\ion{O}{ii}]- and [\ion{O}{iii}]-based estimators are reliable within the observable $\rm 12 + \logten(O/H)$ range of approximately 8--9. However differential dust extinction should be considered when using widely spaced wavelength line combinations, like N2O2, O3N2, O3S2, O32, R23, and RS23 estimators.

\end{enumerate}
In summary, the comprehensive predictions presented in this paper offer valuable insights into emission-line properties of galaxy populations at intermediate redshifts, and their relationship to the ionising source properties and local gas conditions. This represents theoretical guidance for a redshift range that has seen limited spectroscopic coverage thus far, and we expect it to serve as a reference for interpreting results from the upcoming \Euclid surveys, as well as other spectroscopic surveys with instruments like DESI and VLT/MOONs.

\begin{acknowledgements}
  
\AckEC  
L.S., M.H., and A.P. acknowledge funding from the Swiss National Science Foundation (SNF) via a PRIMA Grant PR00P2 193577 “From cosmic dawn to high noon: the role of black holes for young galaxies”. We also acknowledge the computing centre of INAF-OATs, under the coordination of the CHIPP project \citep{Taffoni2020CHIPP:HPDA}, for the availability of computing resources and support. C.S. acknowledges support by NASA ROSES grant 12-EUCLID12-0004. A.F. acknowledges the support from project "VLT-MOONS" CRAM 1.05.03.07, INAF Large Grant 2022 "The metal circle: a new sharp view of the baryon cycle up to Cosmic Dawn with the latest generation IFU facilities" and INAF Large Grant 2022 "Dual and binary SMBH in the multi-messenger era”. V.A. acknowledges support from INAF-PRIN 1.05.01.85.08

\end{acknowledgements}

\section*{Data availability}
An introduction to {\sc Gaea}, a list of our recent work, as well as data file containing published model predictions, can be found at \url{https://sites.google.com/inaf.it/gaea/home}.

%
%

\bibliography{references}

%

\begin{appendix}


\section{Redshift evolution of the ionisation parameter}
\label{app}
Throughout this study, we have explained the evolution of emission-line properties and their relationships to galaxy properties, like the gas-phase metallicity and the location of star-forming galaxies in the BPT diagrams, using the redshift evolution of the ionisation parameter. We demonstrate this evolution at fixed metallicity and fixed stellar mass.

\begin{figure}
	\includegraphics[width=\columnwidth]{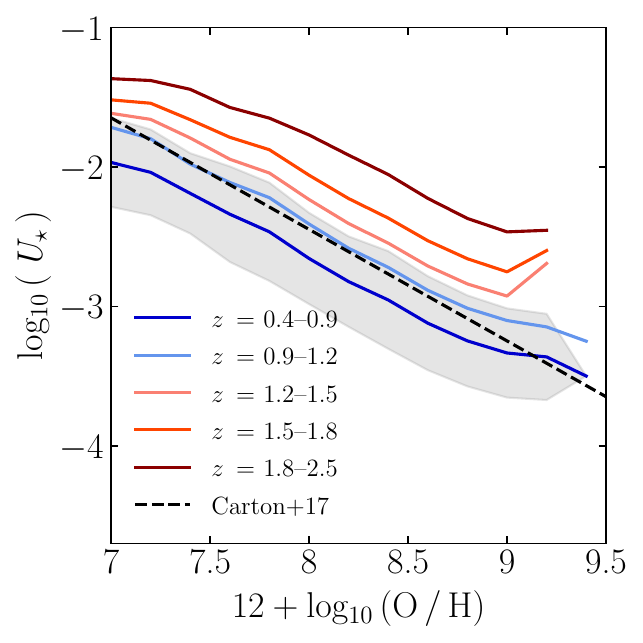}
    \caption{Average ionisation parameter $U_{\star}$ for SF models against O/H abundance in different redshift bins (same colour coding as in Fig. \ref{fig:SFR+LAGN highz}). Alongside, we show the relation from \citet{Carton2017InferringApproach}.}
    \label{fig:logU relation}
\end{figure}

In Fig. \ref{fig:logU relation}, we show in different redshift bins (using the colour coding as in Fig. \ref{fig:SFR+LAGN highz}) the average ionisation parameter $U_{\star}$ for SF photoionisation models against global O/H abundance. We compare this against the empirical relation from \citet[dashed line]{Carton2017InferringApproach}, which has been derived from local star-forming SDSS galaxies.
Across all redshift bins, our prediction from \textsc{Gaea} presents an anti-correlation between the ionisation parameter and the metallicity, in agreement with \citet{Carton2017InferringApproach}. At fixed redshift, the ionisation parameter decreases by around 1.5\,dex between  $\mathrm{12 + \logten(O/H) = 7}$ and 9.5.
We further note that the agreement is best in the lower redshift bins, while higher redshifts exhibit a systematically increased average ionisation parameter at fixed metallicity. This is caused by increased star-formation rates and gas densities at increased redshift.

\begin{figure}
	\includegraphics[width=\columnwidth]{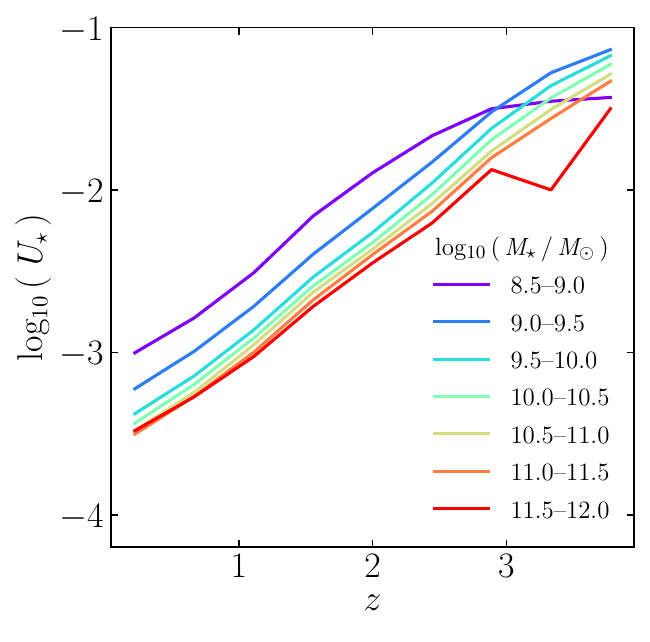}
    \caption{Redshift evolution of average ionisation parameter $U_{\star}$ for SF models in different stellar mass bins (colour coding as indicated in legend).}
    \label{fig:logU evolution Mstar}
\end{figure}

The increase of the ionisation parameter with redshift persists when keeping the stellar mass fixed. In Fig. \ref{fig:logU evolution Mstar}, we present the redshift evolution of average SF ionisation parameter in different stellar mass bins (see colour code in legend). Due to the anti-correlation of metallicity and ionisation parameter, and the correlation of mass and metallicity (see Fig. \ref{fig:selection} and \citealt{Maiolino2008AMAZE3}), the ionisation parameter is most elevated at the lowest stellar masses.

\end{appendix}

\end{document}